\documentclass[conference]{IEEEtran-3}

\usepackage{graphicx}
\usepackage{float}
\usepackage{times}
\usepackage{amsfonts}
\usepackage[square,sort&compress,comma,numbers]{natbib}

\usepackage[small,bf]{caption}
\usepackage{subcaption}
\usepackage[marginal,hang]{footmisc}
\renewcommand{\footnoterule}{%
  \kern -3pt
  \hrule width 1in 
  \kern 2pt
}

\usepackage{xcolor}
\usepackage{flushend}
\usepackage{paralist}
\usepackage{booktabs}
\usepackage{dsfont}

\usepackage[
	n,
	operators,
	advantage,
	sets,
	adversary,
	landau,
	probability,
	notions,	
	logic,
	ff,
	mm,
	primitives,
	events,
	complexity,
	asymptotics,
	keys]{cryptocode}

\definecolor{darkgreen}{RGB}{47,109,79}
\definecolor{darkblue}{RGB}{57,79,99}
\usepackage[bookmarks=false, colorlinks=true, plainpages = false,  linkcolor = darkgreen,   citecolor = darkgreen, urlcolor = darkblue, filecolor = blue]{hyperref}

\newcommand{\setsample}{\ensuremath{\subset_{\$}}}

\newcommand{\descr}[1]{\vspace{0.05cm} \noindent \textbf{#1}}
\newcommand{\descrit}[1]{\smallskip \noindent \textit{#1}}

\usepackage[all]{xy}

\newif\ifcomment
\commentfalse

\ifcomment
	\newcommand{\edc}[1]{\textbf{\em\color{red}#1}}
	
\else  
    \newcommand\edc[1]{}
\fi

\makeatletter
\def\@copyrightspace{\relax}
\makeatother
\begin{document} 

\sloppy
\title{Knock Knock, Who's There?\\ Membership Inference on Aggregate Location Data$^*$\IEEEoverridecommandlockouts\thanks{$^*$To appear in the Proceedings of the 25th Network and Distributed System Security Symposium (NDSS 2018).}}

\author{\IEEEauthorblockN{Apostolos Pyrgelis}
\IEEEauthorblockA{University College London\\
apostolos.pyrgelis.14@ucl.ac.uk}
\and
\IEEEauthorblockN{Carmela Troncoso}
\IEEEauthorblockA{IMDEA Software Institute\\
carmela.troncoso@imdea.org}
\and
\IEEEauthorblockN{Emiliano De Cristofaro}
\IEEEauthorblockA{University College London\\
e.decristofaro@ucl.ac.uk}}

\maketitle

\begin{abstract}
Aggregate location data is often used to support smart services and applications, e.g., generating live traffic maps or predicting visits to businesses. In this paper, we present the first study on the feasibility of membership inference attacks on aggregate location time-series. We introduce a game-based definition of the adversarial task, and cast it as a classification problem where machine learning can be used to distinguish whether or not a target user is part of the aggregates.

We empirically evaluate the power of these attacks on both raw and differentially private aggregates using two mobility datasets. We find that membership inference is a serious privacy threat, and show how its effectiveness depends on the adversary's prior knowledge, the characteristics of the underlying location data, as well as the number of users and the timeframe on which aggregation is performed. Although differentially private mechanisms can indeed reduce the extent of the attacks, they also yield a significant loss in utility. Moreover, a strategic adversary mimicking the behavior of the defense mechanism can greatly limit the protection they provide. Overall, our work presents a novel methodology geared to evaluate membership inference on aggregate location data in real-world settings and can be used by providers to assess the quality of privacy protection before data release or by regulators to detect  violations.
\end{abstract}

\vspace{-0.1cm}
\section{Introduction}
\label{sec:intro}
The ability to model the context in which users and applications operate enables the development of intelligent applications and pervasive personalized services. Naturally, location information plays a crucial role in shaping such context, motivating the continuous collection of users' location data by applications and service providers.
In some cases, %
entities may be interested in only collecting or releasing {\em aggregate location statistics}, %
which, for instance, can be used to calculate the average speed along a road and generate live traffic maps~\cite{googletraffic,waze}, or to estimate the number of people at a restaurant and predict availability and waiting times~\cite{popa2011privacy}.
Apple also lets iOS as well as third-party app developers collect differentially private aggregate statistics about emojis, deep links, as well as locations, via dedicated APIs~\cite{apple}. 
Moreover, aggregate location information is relied upon by companies like factual.com to offer statistics to advertisers, or like Telefonica, which provides consultancy services around footfall measures calculated from network events~\cite{smartsteps}.

Aggregation is often considered as a way to hinder the exposure of individual users' data~\cite{popa2011privacy}; however, access to, or release of, aggregate location statistics might ultimately violate privacy of the individuals that are part of the aggregates~\cite{xu2017trajectory, pyrgelis2017does}. In this paper, we focus on {\em membership inference attacks}, whereby an adversary attempts to determine whether or not location data of a target user is part of the aggregates. 

\descr{Motivation.} The ability of an adversary to ascertain the presence of an individual in aggregate location time-series constitutes an obvious privacy threat if the aggregates relate to a group of users that share a sensitive characteristic. For instance, learning that an individual is part of a dataset aggregating movements of Alzheimer's patients implies learning that she suffers from the disease. Similarly, inferring that statistics collected over a sensitive timeframe or sensitive locations include a particular user also harm the individual's privacy.

Recent work~\cite{pyrgelis2017does} also shows that an adversary with some prior knowledge about a user's mobility profile can exploit aggregate information to improve this knowledge, %
or even localize her. Also, users' ``trajectories'' can in some cases be extracted from aggregate mobility data, even without prior knowledge~\cite{xu2017trajectory}. 
However, in order to mount these attacks, the adversary needs to know that the user is part of the aggregate dataset, which further motivates our research objectives. %

Membership inference can also be leveraged by providers to evaluate the quality of privacy protection on the aggregates {\em before} releasing them, and by regulators, to support enforcement of individual's rights (e.g., the right to be forgotten) %
or detect violations.
For instance, if a service provider is not allowed to release location data, or make it available to third-parties even in aggregate form, one can use membership inference attacks to verify possible misuse of the data. 

\descr{Approach \& Results.} In this paper, we present the first formalization of membership inference in the context of location data. We model the problem as a game in which an adversary aims at distinguishing location aggregates that include data of a target user from those that do not. We instantiate the distinguishing task using a machine learning classifier trained on the prior knowledge of the adversary (e.g., past users' locations, aggregates of groups including and excluding the target user), and use it to infer the target's membership in unseen aggregate statistics. %

We evaluate our approach on two mobility datasets with different characteristics, and find that releasing raw aggregates poses a significant privacy threat. In particular, our results show that membership inference is very successful when the adversary knows the locations of a small subset of users including her target -- the classifier achieves up to 0.83 Area Under Curve (AUC) with 100 users per aggregation group -- or when she has prior information for user groups on which she attempts to infer membership (up to 1.0 AUC even with 9,500 users per group). In weaker adversarial knowledge settings, membership inference is less effective but still yields non-negligible privacy leakage. %
Overall, we find that the number of users as well as the timeframe used to compute the aggregation have a profound effect on the accuracy of the attacks. Interestingly, certain characteristics of the data, like regularity and sparseness, also affect the power of the membership inference adversarial task.

We also study membership inference on statistics protected using defense mechanisms based on differential privacy. %
We find that they are generally effective at preventing inference, although at the cost of a non-negligible reduction in utility. Moreover, we show that a strategic adversary mimicking the behavior of the mechanisms -- i.e., training the classifier on noisy aggregates -- can reduce their protection (up to 83\%).

\descr{Contributions.} In summary, this paper makes the following contributions: 
(i) we introduce a generic methodology to study membership privacy in location aggregates that formalizes membership inference on aggregate location data as a distinguishability game and instantiates the distinguishing task with a machine learning classifier; (ii) we deploy our methods to quantify privacy leakage on raw aggregates, using two real-world mobility datasets; and (iii) we illustrate how our techniques can be used to study the effectiveness of defense mechanisms aimed at preventing these attacks.
\descr{Paper Organization.} The rest of this paper is organized as follows. 
The next section reviews related work. 
Then, we formalize the problem of membership inference on aggregate location time-series in Section~\ref{sec:problem}
and present the methodology used to evaluate it in Section~\ref{sec:methodology}.
In Section~\ref{sec:setup}, we introduce our experimental setup and, in Section~\ref{sec:results}, present the results
of our experiments on raw aggregates. 
After evaluating differentially private defense mechanisms in Section~\ref{sec:counter}, the paper concludes in 
Section~\ref{sec:conclusion}.

\section{Related Work}
\label{sec:related}
In this section, we review previous work on membership inference, differentially private release of location data, as well as location privacy.

\descr{Membership inference attacks.} Such attacks aim to determine the presence of target individuals within a dataset. This is relevant in many settings, e.g., in the context of {\em genomic research}, where data inherently tied to sensitive information, such as health stats or physical traits, is commonly released in aggregate form for Genome Wide Association Studies (GWAS)~\cite{welter2013nhgri}.
Homer et al.~\cite{homer2008resolving} show that one can learn whether a target individual was part of a case-study group associated to a certain disease by comparing the target's profile against the aggregates of the case study and those of a reference population obtained from public sources. This attack has then been extended by Wang et al.~\cite{wang2009learning} to use correlations within the human genome, reducing the need for prior knowledge about the target. 
Also, Backes et al.~\cite{backes2016membership} show that membership inference can be mounted against individuals contributing their microRNA expressions to scientific studies.

Another line of work focuses on membership inference in {\em machine learning} models. Shokri et al.~\cite{shokri2017membership} show that such models may leak information about data records on which they were trained. 
Hitaj et al.~\cite{hitaj2017deep} present active inference attacks on deep neural networks in collaborative settings, while Hayes et al.~\cite{hayes2017logan} focus on privacy leakage from generative models in Machine Learning as a Service applications.
Moreover, Buscher et al.~\cite{bucher2017} recently evaluate membership inference in the context of data aggregation in {\em smart metering}, studying how many household readings need to be aggregated in order to protect privacy of individual profiles in a smart grid.

Overall, our research differs from these works in that we focus on membership inference over aggregate location time-series, which present different characteristics and challenges than the other domains. Despite the importance of location data in terms of its availability, frequency of collection, and the amount of sensitive information it carries~\cite{reza}, this problem, to the best of our knowledge, has not been examined before.

\descr{Differentially private mechanisms.} Differential privacy (DP)~\cite{dwork2008differential} can be used to mitigate membership inference, as its indistinguishability-based definition guarantees that the presence or the absence of an individual does not significantly affect the output of the data release. 
Li et al.~\cite{li2013membership} introduce a framework geared to formalize the notion of Positive vs Negative Membership Privacy, considering an adversary parameterized by her prior knowledge.
However, to the best of our knowledge, no specific technique has been presented to instantiate their framework in our setting.
Common mechanisms to achieve DP include using noise from the Laplacian~\cite{dwork2008differential} or the Gaussian distribution~\cite{dwork2006our} (see Section~\ref{sec:counter}).

Specific to the context of spatio-temporal data are the techniques proposed by Machanavajjhala et al.~\cite{machanavajjhala2008privacy}, who use synthetic data generation to release differentially private mobility patterns of commuters in Minnesota. 
Also, Rastogi and Nath~\cite{rastogi2010differentially} propose an algorithm based on Discrete Fourier Transform to privately release aggregate time-series, while Acs and Castelluccia~\cite{acs2014case} improve on~\cite{rastogi2010differentially} and present a differentially private scheme tailored to the spatio-temporal density of Paris. %
Finally, To et al.~\cite{to2016differentially} release the entropy of certain locations with DP guarantees, %
and show how to achieve better utility although with weaker privacy notions.

\descr{Location privacy.} Previous location privacy research taking into account traces or profiles of {\em single users}~\cite{golle2009anonymity,shokri2010unraveling,zang2011anonymization,ji2015your,rossi2014s,de2013unique} does not apply to our problem, which focuses on  aggregate location statistics.
Closer to our work is our own PETS'17 paper~\cite{pyrgelis2017does}, which shows that aggregate location time-series can be used by an adversary to improve her prior knowledge about users' location profiles.
Also, Xu et al.~\cite{xu2017trajectory} present an attack that exploits the uniqueness and the regularity of human mobility, and extracts location trajectories from aggregate mobility data. %
As opposed to these efforts, which attempt to learn data about individuals (e.g., mobility profiles, trajectories) from the aggregates, we focus on inferring their membership to datasets, which, to the best of our knowledge, has not been studied before.

\section{Defining Membership Inference on\\ Aggregate Locations }
\label{sec:problem}
In this section, we formalize the problem of membership inference on aggregate location time-series. We consider the case where one or more entities periodically release the number of users in some Regions Of Interest (ROIs), within a given time interval (e.g., 99 taxis in Union Square on Fri, Aug 11th between 9--10am).
By relying on this data, as well as some prior knowledge, an adversary tries to infer if a target individual contributed to the aggregates, i.e., whether or not she is a \emph{member} of the group yielding the released aggregates. 

\begin{table}[t]
\centering
\footnotesize
  \begin{tabular}{ r  l }
  \toprule
    {\bf Symbol} & {\bf Description} \\ 
    \midrule
	Adv, Ch & Adversary, Challenger \\ %
	$\mathcal{P}$ & Adv's prior knowledge\\		   
    $U$ & Set of mobile users \\ %
    $S$ & Set of locations (ROIs) \\ %
    $T$ & Time period considered \\ %
	$T_O$ & Observation period \\ %
	$T_I$ & Inference period \\ %
	$L_{u}$ & User $u$'s location time-series, where \\
	        & ~~$L_{u}[s,t]$ = 1 if $u$ is in $s$ at time $t$, 0 otherwise\\
	$\mathcal{L}$ & Location time-series of all users in $U$\\
	$\Upsilon\setsample U$ & Random subset $\Upsilon \subset U$\\
	$A_{X}$ & Aggregate location time-series of users in $X \subset U$\\
	& ~~ where $A_X[s,t]=\sum_{j\in X} L_j[s,t]$\\
	$m$ & Variable representing size of aggregation group\\
	\bottomrule
	  \end{tabular}
  \vspace{-0.1cm}
	\caption{Notation.} 
  \label{table:notation}
  \vspace{-0.3cm}
  \end{table}

\subsection{Notation}
\label{sec:notation}
The notation used throughout the paper is summarized in Table~\ref{table:notation}.
We denote the set of all users as $U = \{u_1, u_2, \cdots, u_{|U|} \}$, and the set of regions of interest as $S = \{s_1, s_2, \cdots, s_{|S|} \}$. We also use $T = \{t_1, t_2, \cdots, t_{|T|} \}$ to denote the set of time intervals on which aggregate locations are collected, although, without loss of generality, the problem can be extended to infinite intervals.
We model the location of a user $u \in U$ over time as a binary matrix $L_{u}$ of size $| S | \times | T |$, where $L_u[s, t]$ is 1 if $u$ is in location $s \in S$, at time $t \in T$, and 0 otherwise. That is, $L_u$ contains the location time-series of $u$, while those of {\em all} users are stored in a matrix $\mathcal{L}$, which is of size $|U| \times |S| \times |T|$.

Also, %
$A_{X}$ denotes the {\em aggregate} location time-series over the users in $X \subset U$. $A_{X}$ is modeled as a matrix of size $|S| \times |T|$, where each element $A_{X}[s, t]$ represents the number of users in $X$ %
that are in ROI $s$ at time $t$. 

Finally, we denote the {\em prior knowledge} of an adversary (Adv) about users as $\mathcal{P}$, which is %
built %
during an {\em observation period}, denoted as $T_O \subset T$ (see Section~\ref{sec:prior}). 
The prior knowledge is used by Adv to perform membership inference during the {\em inference} period, denoted as $T_I \subset T$, for which aggregates are available.%

\subsection{Membership Inference as a Distinguishability Game}
\label{sec:game}

We model membership inference by means of a distinguishability game (DG), played by the adversary Adv and a challenger Ch, which generates the location aggregates over various user groups. 
The former, having some prior knowledge about the users ($\mathcal{P}$), tries to infer whether data of a particular user ($u^*$) is included in the aggregates. Naturally, Adv could be interested in multiple target users, however, to ease presentation, we describe the case of a single target user. 

The game is parameterized by the set of users $U$, the number of users included in the aggregation group ($m$), and the inference period $T_I$. Note that $m$ and $T_I$ inherently affect Adv's performance, as we discuss in our experimental evaluation (Section~\ref{sec:results}).

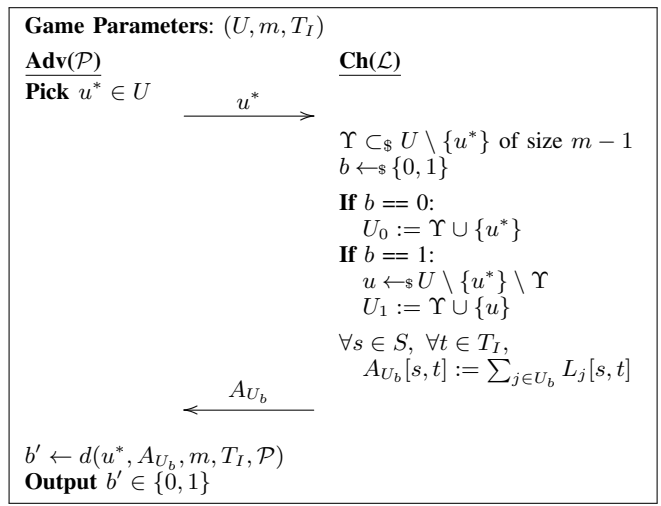
\begin{figure}[t]
\resizebox{.475\textwidth}{!}{
\fbox{\small
\begin{minipage}{0.95\columnwidth}
\hspace{0.1cm}{\bf Game Parameters}: $(U, m, T_I)$\\[1ex]
\begin{tabular}{lcl}
\hspace*{-0.1cm}\textbf{\underline{Adv($\mathcal{P}$)}} & & \textbf{\underline{Ch($\mathcal{L}$)}}\\[0.5ex]
\hspace*{-0.1cm}$\text{\bf Pick~}u^* \in U \hspace{-0.05cm} $ & & \vspace{-0.3cm}\\
	&  $\xymatrix@1@=50pt{\ar[r]^*{u^*}&}$\hspace*{-0.15cm}&\\
 & & $\Upsilon \setsample U \setminus \{u^*\} ~ \text{of size} ~ m-1$ \\
 & & $b \sample \bin $\\[1ex]
 & & $\text{\textbf{If} } b \text{ == 0:}$\\
 & & $ \pcind U_0 := \Upsilon \cup \{u^*\} $\\
 & & $\text{\textbf{If} } b \text{ == 1:}$\\
 & & $\pcind u \sample U \setminus \{u^*\} \setminus \Upsilon $\\
 & & $\pcind U_1 := \Upsilon \cup \{u\} $\\[1ex]
 & & $\forall s\in S,~\forall t\in T_I,$\\ 
 & & $\pcind A_{U_b}[s, t] := \sum_{j\in U_b} L_{j}[s,t]$\\[-1.25ex]
 & $\xymatrix@1@=50pt{& \ar[l]_*{A_{U_b}}}$\hspace*{-0.15cm}&\\[2ex]
 \end{tabular}
 \hspace*{0.1cm}$b' \leftarrow d(u^*, A_{U_b}, m, T_I, \mathcal{P})$\\ %
  \hspace*{0.1cm}$\text{\bf Output~} b' \in \{0, 1\} $
\end{minipage}
}}
\caption{Distinguishability Game (DG) between adversary Adv and challenger Ch, capturing membership inference over aggregate location time-series. The game is parameterized by the set of users ($U$), the aggregation group size ($m$) and the inference period ($T_I$).}
\label{fig:dg}
\vspace{-0.3cm}
\end{figure}

We present the game in Fig.~\ref{fig:dg}. 
First, Adv selects the target user $u^*$ %
and sends it to Ch.
The latter randomly selects a subset $\Upsilon \subset U$ of size $m - 1$, excluding $u^*$,
and draws a random bit $b$. If $b=0$, she aggregates the location matrices of all users in $\Upsilon$ along with that of $u^*$; whereas, if $b=1$, she selects another random user $u \neq u^*$ not in $\Upsilon$ and adds her data to the aggregates instead. The resulting matrix $A_{U_b}$, computed over all timeslots of $T_I$, is sent back to Adv, which attempts to guess $b$. 
Adv wins if $b'=b$, i.e., she successfully distinguishes whether $u^*$ is part of the aggregates or not;
naturally, her goal is to win the game, over multiple iterations, with probability higher than 1/2 (i.e., a random guess). 

We model Adv's guess as a \textit{distinguishing function}, $d$, with input $(u^*, A_{U_b}, m, T_I, \mathcal{P})$. How to instantiate the function is discussed in Section~\ref{sec:distinguishing}.
Observe that the parameters of the DG game include the set of users $U$, but this information is {\em not} used in the distinguishing function. In other words, we only assume that Adv knows that $u^*$ is in the universe of possible users, but not that she knows all users in $U$.

\section{Methodology}
\label{sec:methodology}

We now introduce our methodology to evaluate membership inference on aggregate location time-series, modeled by the DG game in Fig.~\ref{fig:dg}. Specifically, we discuss ways to build Adv's prior knowledge ($\mathcal{P}$) during the observation period $T_O$, how to instantiate the distinguishing function (i.e., deciding the bit $b'$), and measure the performance of the inference.

\subsection{Adversarial Knowledge}
\label{sec:prior}

Our generic game-based definition of the adversarial goal enables the consideration of adversaries of variable strength, modeled by their prior knowledge, $\mathcal{P}$. We consider two possible priors, discussed next.

\descr{(1) Subset of Locations.} We start with a setting in which Adv knows the real locations of a subset of users $Y \subset U$, {\em including} the target user (i.e., $u^* \in Y$), during the inference period $T_I$. Thus, in this case observation and inference periods coincide (i.e., $T_O=T_I$). We consider $|Y| = \alpha \cdot |U|$, where $\alpha \in [0, 1]$ models the percentage of users for which Adv knows their actual location.
Formally, we define it as: \vspace{-0.1cm}
\begin{equation}
\mathcal{P}: ~ L_u[s,t] ~~\forall u \in Y~\forall s \in S~\forall t \in T_I \vspace{-0.1cm}
\end{equation}

This prior knowledge represents the case of an adversary that has access to location information of some users at a point in time, e.g., a telecommunications service provider getting locations from cell towers, or a mobile app provider collecting location data. Using this information she attempts to infer membership of her target to an aggregate dataset published by another entity.

\descr{(2) Participation in Past Groups.} We then consider an adversary that knows aggregates computed during an observation period $T_O$, disjoint from the inference period $T_I$ (i.e., $T_O \cap T_I = \emptyset$) for $\beta$ groups $W_i$ of size $m$, which may or may not include $u^*$. For each group $W_i$, we assume that Adv knows: (i) the aggregates of the observation period, i.e., $A_{W_{i}}[s, t],\forall s \in S ~\text{and} ~ t \in T_O$, and (ii) $u^*$'s membership to the group. More formally: \vspace{-0.1cm}
\begin{equation}
\mathcal{P} : A_{W_{i}} ~ \wedge ~ \mathds{1}_{W_i}(u^*) ~ \forall ~ i \in \{1, \cdots, \beta\}\vspace{-0.1cm}
\end{equation}
where $\mathds{1}_{W_i}(u^*)$ is the indicator function modeling the membership of the target user to the group $W_i$.
In our experiments, we consider two different ``flavors'' of this prior:
\begin{itemize}
\item[--] \textbf{(2a) {\em Same Groups as Released}},  Adv knows the target user's participation in past groups which {\em are} also used to compute the aggregates released by Ch during the inference period;
\item[--] \textbf{(2b) {\em Different Groups than Released}}, Adv knows the user's participation in past groups that {\em are not} used to compute aggregates released in the inference period.
\end{itemize}  
Observe that (2a) simulates the case of continuous data release related to particular groups, where users are stable over time (e.g., statistics about a neighborhood), and with the adversary (e.g., a group member) having observed the participation of the target user in past aggregates of the same groups. Prior (2b) is less restrictive, as it only assumes that the adversary has some aggregates of groups in which the target was previously included, but does not require these groups to be fixed over time -- e.g., if the target user moves to a new neighborhood and her data is mixed with other users, Adv attempts to infer membership using past information.

\vspace{-0.1cm}
\subsection{Distinguishing Function}
\label{sec:distinguishing}
Recall from Section~\ref{sec:problem} that, in the DG game (Fig.~\ref{fig:dg}), the adversary tries to guess whether or not the target user is part of the aggregates using a distinguishing function, which we denoted as $d$. 
This function takes as input the target user $u^*$, the ``challenge'' $A_{U_b}$, parameters of the game $m$ and $T_I$, and the prior knowledge $\mathcal{P}$.

We opt to instantiate $d$ with a {\em supervised machine learning classifier}, trained using data included in the adversarial prior knowledge. Our intuition is that the adversary's distinguishing goal can be modeled as a binary classification task, i.e., categorizing observations into two classes corresponding to whether or not the data of target user $u^*$ is part of the location aggregates under examination.
\vspace{-0.1cm}
\subsection{Privacy Metric}
\label{sec:metric}

Given our game-based definition, we reason about privacy leakage %
in terms of the adversarial performance in distinguishing whether or not $u^*$'s data is included in the aggregates. %
In particular, we introduce a {\em privacy loss} metric, capturing Adv's advantage in winning the DG game over a random guess (assuming that the adversary plays the distinguishability game for a specific user multiple times), while relying on the Area Under the Curve (AUC) to measure Adv's performance. %

\descr{AUC Score.} %
For a series of instances of the game for $u^*$, we count the Adv's guesses $b'$ regarding the presence of $u^*$'s data in the released aggregate location time-series as:
\begin{itemize}
\item[--] True Positive (TP) when $b=0$ and $b'=0$;
\item[--] True Negative (TN) when $b=1$ and $b'=1$;
\item[--] False Positive (FP) when $b=1$ and $b'=0$;
\item[--] False Negative (FN) when $b=0$ and $b'=1$.
\end{itemize}
We then calculate True Positive and False Positive Rates, as
TPR $=$ TP/(TP+FN) and FPR $=$ FP/(FP+TN), respectively. From these, we 
derive the Receiver Operating Characteristic (ROC) curve, which represents the TPR and FPR obtained at various discrimination classification thresholds, and compute the Area Under Curve (AUC). The AUC captures a classifier's overall performance in the distinguishability game. %

\descr{Privacy Loss (PL).} As mentioned, we measure the privacy loss of $u^*$ as the adversary's \textit{improvement} over a random guess baseline (AUC $=$ 0.5). %
Formally, we define PL as:
\begin{equation}\label{eq:pl}
\text{PL} = \begin{cases}
\frac{ \text{AUC} - 0.5}{0.5}~~~~~~~~\text{if} ~~\text{AUC} > 0.5 \\
0 ~~~~~~~~~~~~~~~~\text{otherwise}
\end{cases}
\end{equation}
Hence, PL is a value between 0 and 1 that captures the adversary's advantage over random guessing when distinguishing whether the target user's data is part of the  aggregates. 

\section{Experiment Design}
\label{sec:setup}
In this section, we present our experimental setup as well as the datasets used in our evaluation.
(Results are given later in Sections~\ref{sec:results} and~\ref{sec:counter}).
\vspace{-0.1cm}
\subsection{Datasets}
\label{sec:datasets}

We use two real-world datasets that capture different mobility characteristics, obtained, respectively, from the Transport for London (TFL) authority and the San Francisco Cab (SFC) network. Both datasets contain about one month of location data, and have been used often in ubiquitous computing~\cite{silva2015predicting,ceapa2012avoiding} and location privacy~\cite{shokri2011quantify,pyrgelis2017does} research. 
We choose these datasets primarily because of their different characteristics: data from public transport (TFL) includes more users and is more ``predictable'' (due to commuting patterns) than the SFC dataset, which, on the other hand, is less sparse (i.e., it involves more data points per user per day).

\descr{Transport For London (TFL).} The TFL dataset consists of trips made by passengers on the TFL network in March 2010 using the Oyster Card -- an RFID pre-paid card. Each record in the data describes a unique trip and includes the (anonymized) oyster card id, start time, touch-in station id, end time, and touch-out station id. We discard trips from March 29--31 to have exactly four weeks of data, which contain 60M trips made by 4M unique Oyster cards, visiting 582 train/tube stations. We select the top 10K oyster ids per total number of trips, which account for about 6M trips. Considering oyster trips start/end stations as ROIs, the top 10K users report, on average, 728 $\pm$ 16 ROIs in total, out of which 20 $\pm$ 9 are unique. Setting the time granularity to {\em one hour}, the top 10K oysters are in the ``system'' for 115 $\pm$ 21 out of the 672 slots (28 days). We consider each Oyster card as a user $u$, and compute the matrix $L_u$ setting $L_u[s,t]$ to 1 if the user $u$ touched-in or out at station $s$, during time slot $t \in T$, and 0 otherwise. When a card does not report any station at a particular time slot, we assign it to a special ROI denoted as \textit{null}. For this dataset, $L_u$ is a matrix of size $|S| \times |T| = 583 \times 672$. %

\descr{San Francisco Cabs (SFC).} This dataset includes mobility traces recorded by San Francisco taxis from May 17 to June 10, 2008~\cite{epfl-mobility-20090224}. Each record consists of a cab identifier, latitude, longitude, and a time stamp. The dataset includes approximately 11 million GPS coordinates generated by 536 taxis. We select 3 weeks (Monday May 19 to Sunday June 8) worth of data and discard traces outside downtown San Francisco (i.e., those of 2 taxis). To generate ROIs, we split the city into a 10 $\times$ 10 grid, whose cells are 0.18 sq. miles. Setting the time granularity to one hour, the 534 cabs report over 2M ROIs, on average 3,827 $\pm$ 1,069 locations per taxi, out of which 78 $\pm$ 6 ROIs are unique. The SFC data is less sparse than the TFL one, as cabs report locations more frequently -- specifically 340 $\pm$ 94 out of the 504 time slots in the 21 considered days. For each cab $u$, we populate $L_u$ by setting $L_u[s,t] $ to 1 if $u$ was in the $s$ cell at time $t \in T$, and 0 otherwise. If a cab does not report any location, we assign it to the special \textit{null} ROI. For this dataset $L_u$ is a matrix of size $|S| \times |T| = 101 \times 504$.

\descr{Sampling Users.} For both datasets, we perform a basic analysis of the number of ROIs reported by their users. We observe that for TFL (resp., SFC), the median is 727 (resp., 4,111), with a maximum of 881 (resp., 8,136) and a minimum of 673 (resp., 504). We sort the users in each dataset per total number of ROI reports and split them in 3 groups of equal size, capturing their mobility patterns as: \textit{highly}, \textit{mildly}, and \textit{somewhat} mobile. To avoid bias, to %
select target users, we sample 50 users from each mobility group {\em at random}. Thus, we run membership inference attacks against a total 150 users for each dataset.

\vspace{-0.1cm}
\subsection{Experimental Setup}
\label{sec:exp-design}
Our experiments aim to evaluate the effectiveness of the distinguishing function $d$, used in the DG game, to guess whether the target user $u^*$ is in the aggregates or not.
As mentioned, we instantiate $d$ using a machine learning classifier. 

We train the classifier on a {\em balanced} dataset of {\em labeled} aggregates over user groups that include and groups that exclude $u^*$, so that it learns patterns that distinguish its participation in the aggregates. The training dataset is generated using data from the prior knowledge $\mathcal{P}$. We then play the game, i.e., we use the trained classifier to infer membership on a {\em balanced} testing set of aggregates previously \textit{unseen}.%

More specifically, we go through three phases: aggregation, feature extraction, and classification, which we describe in high-level. The concrete details of each phase depend on the adversarial prior knowledge, as we discuss later in Section~\ref{sec:results} (where we evaluate membership inference attacks with different priors). The three phases are discussed next.

\descrit{Aggregation.} We create a dataset $D$ %
by repeating these steps:
\begin{enumerate}
\item Randomly generate a group $U_{0}$ of $m$ users, which {\em includes} $u^*$;
\item Aggregate the location matrices of users in $U_{0}$, for $|T_I|$ intervals;
\item Append a row with the aggregates $A_{U_{0}}$ to dataset $D$, and attach the label \textit{in};
\item Randomly generate a group $U_{1}$ of $m$ users, which {\em excludes} $u^*$;
\item Aggregate the location matrices of users in $U_{1}$, for $|T_I|$ intervals;
\item Append a row with the aggregates $A_{U_{1}}$ to the dataset $D$, and attach the label \textit{out}.
\end{enumerate}

\descrit{Feature Extraction.} For each row of the dataset, corresponding to the aggregates of a group with/without $u^*$, we extract statistics that are given as input to the classifier. Such statistics are calculated per location (ROI) and include variance, minimum, maximum, median, mean, standard deviation, as well as the sum of values of each location's time-series.

\descrit{Classification.} We first split the dataset $D$ into the non-overlapping balanced training and testing sets mentioned above. We then train the classifier on the features extracted from the {\em training} set. Finally, we play the distinguishability game on the aggregates of the {\em testing} set (data previously unseen by the classifier), classifying them as including or excluding $u^*$. 

\descr{Implementation.} Our experiments are implemented in Python using the {\em scikit-learn} machine learning suite.\footnote{\url{http://scikit-learn.org/stable/}} Source code is available upon request. We instantiate the following classifiers: i) Logistic Regression (LR), for which we employ a linear solver using a coordinate descent optimization algorithm suitable for binary classification~\cite{fan2008liblinear}; 
ii) Nearest Neighbors (k-NN), configured to use Euclidean distance, with k set to 5, i.e., to predict the output class based on the votes of the 5 nearest samples;
iii) Random Forests (RF), set up to train 30 decision trees and to consider all the features during the node splits using the Gini criterion to measure their quality; and iv) Multi-Layer Perceptron (MLP), consisting of 1 hidden layer with 200 nodes, whose weights are calculated via a stochastic gradient-based optimizer. For more details about the classifiers, we refer to Appendix~\ref{sec:appendix}.

For the feature extraction, we use the {\em tsfresh} Python package.\footnote{\url{http://tsfresh.readthedocs.io/en/latest/}}  
For both datasets, and for all groups, we extract the 7 statistical features mentioned above, for each ROI. 
We obtain 4081 features for TFL (583 ROIs) and 707 features for SFC (101 ROIs). To avoid overfitting, we use Recursive Feature Elimination (RFE) to reduce the number of features to the number of samples we create for each user's dataset $D$. We then feed the features in their original form to all classifiers, except for MLP where we standardize them to have mean of 0 and variance 1.

\section{Evaluating Membership Inference on Raw Aggregate Locations}
\label{sec:results}
We now present the results of our experimental evaluation, measuring the performance of different classifiers in instantiating the distinguishing function (i.e., performing membership inference) on raw aggregates. 
We do so vis-\`a-vis the different priors discussed in~Section~\ref{sec:prior}, using the experimental methodology and the datasets described above. Recall that, in each experiment, we perform attacks against 150  users sampled from high, mild, and somewhat mobility profiles (50 each).
\begin{figure*}[t]
  \centering
    \begin{subfigure}[b]{0.245\textwidth}
        \includegraphics[width=\textwidth]{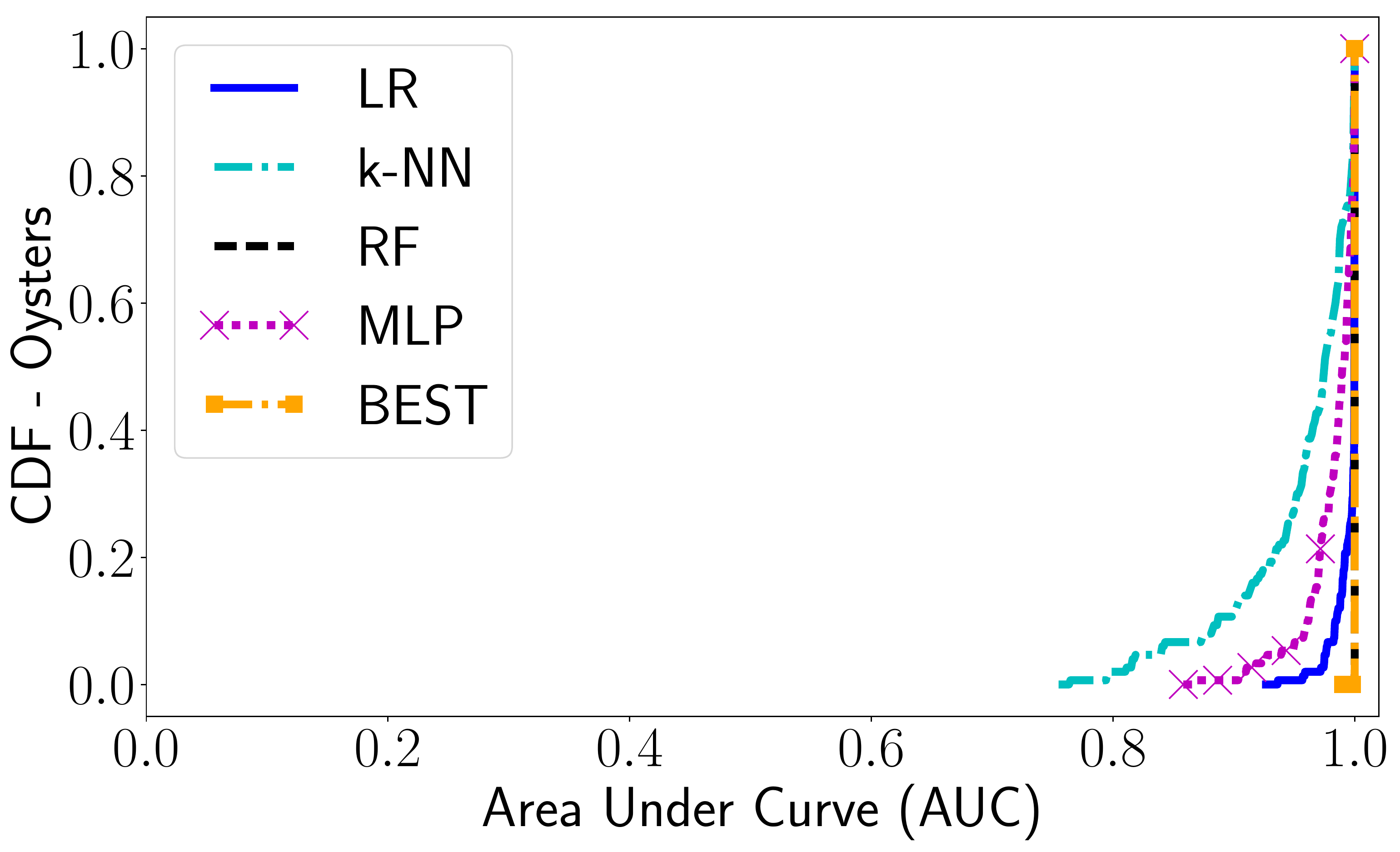}
        \caption{$m=5$}
        \label{fig:tfl-10perc-gr5}
    \end{subfigure}
    ~
    \begin{subfigure}[b]{0.245\textwidth}
        \includegraphics[width=\textwidth]{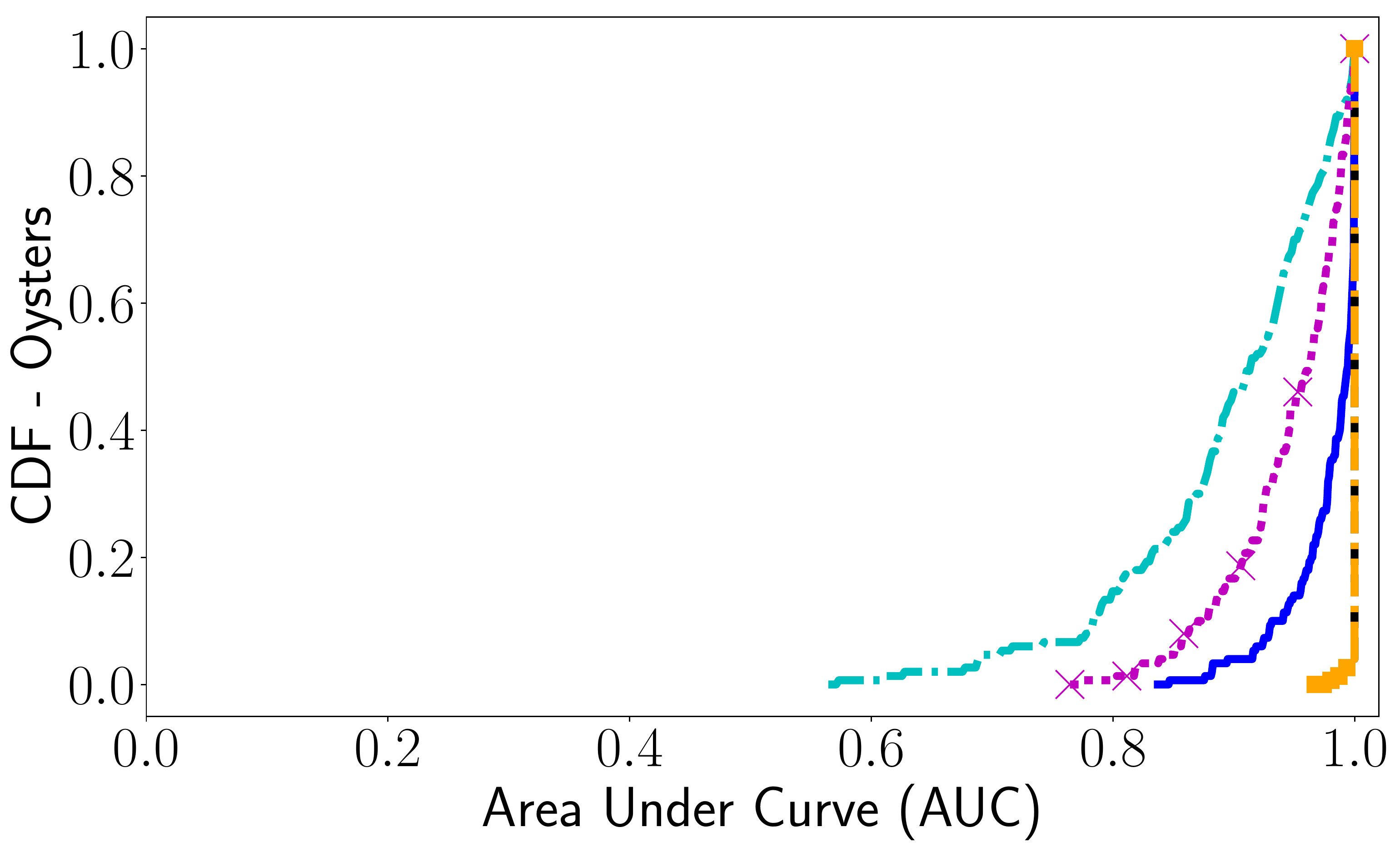}
        \caption{$m=10$}
        \label{fig:tfl-10perc-gr10}
    \end{subfigure}
    ~
    \begin{subfigure}[b]{0.245\textwidth}
        \includegraphics[width=\textwidth]{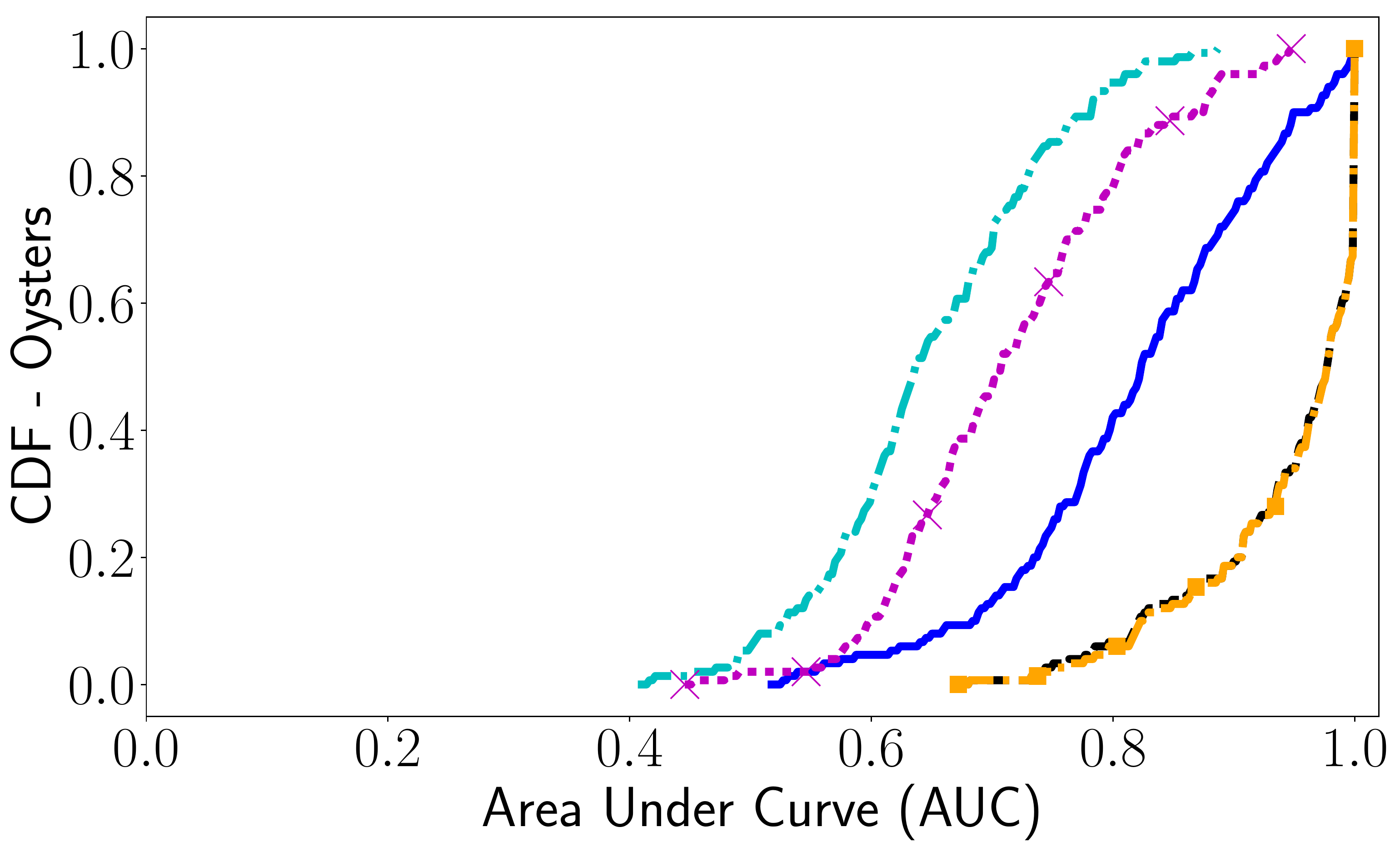}
        \caption{$m=50$}
        \label{fig:tfl-10perc-gr50}
    \end{subfigure}
    \\
    \begin{subfigure}[b]{0.245\textwidth}
        \includegraphics[width=\textwidth]{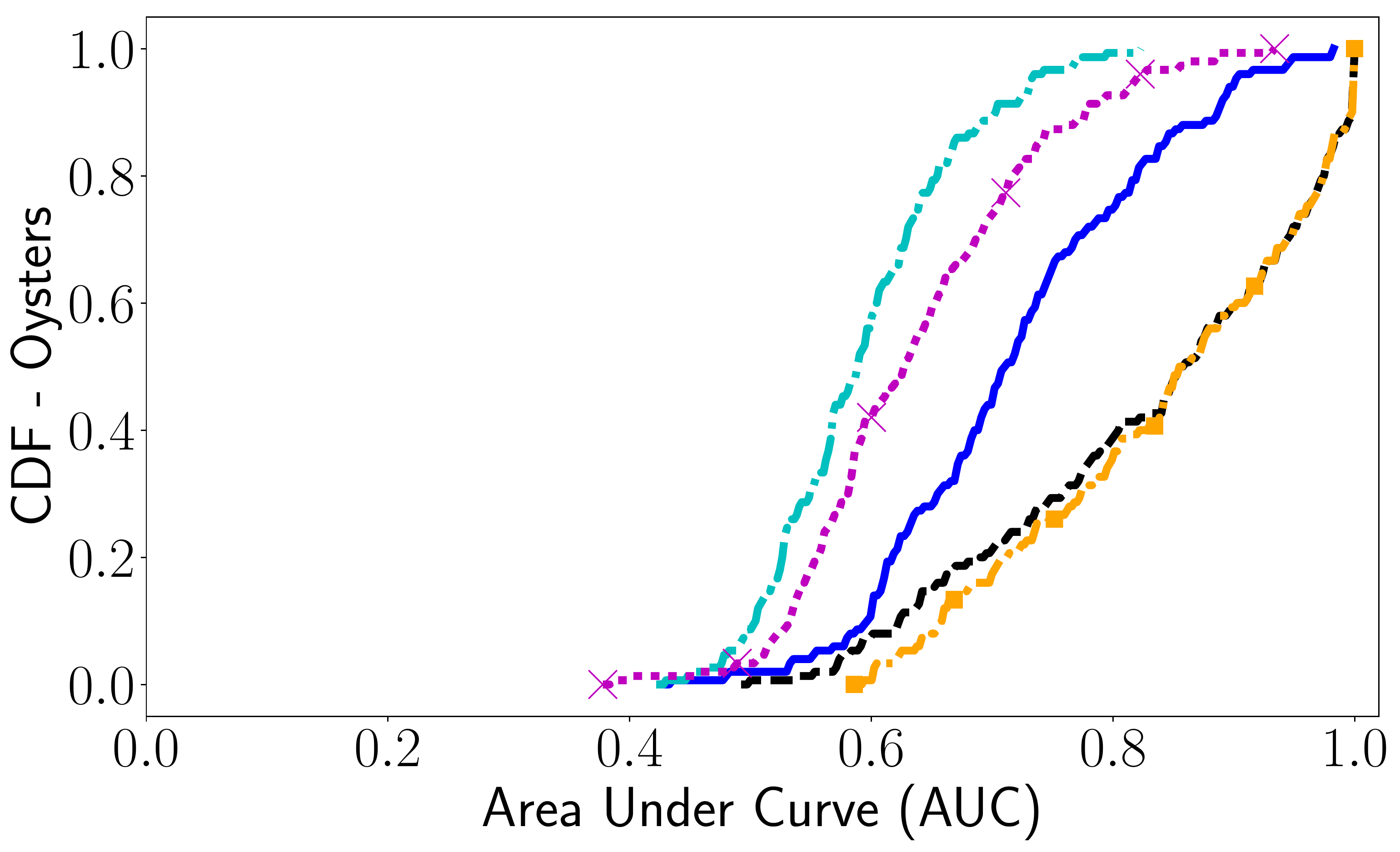}
        \caption{$m=100$}
        \label{fig:tfl-10perc-gr100}
    \end{subfigure}
    ~
    \begin{subfigure}[b]{0.245\textwidth}
        \includegraphics[width=\textwidth]{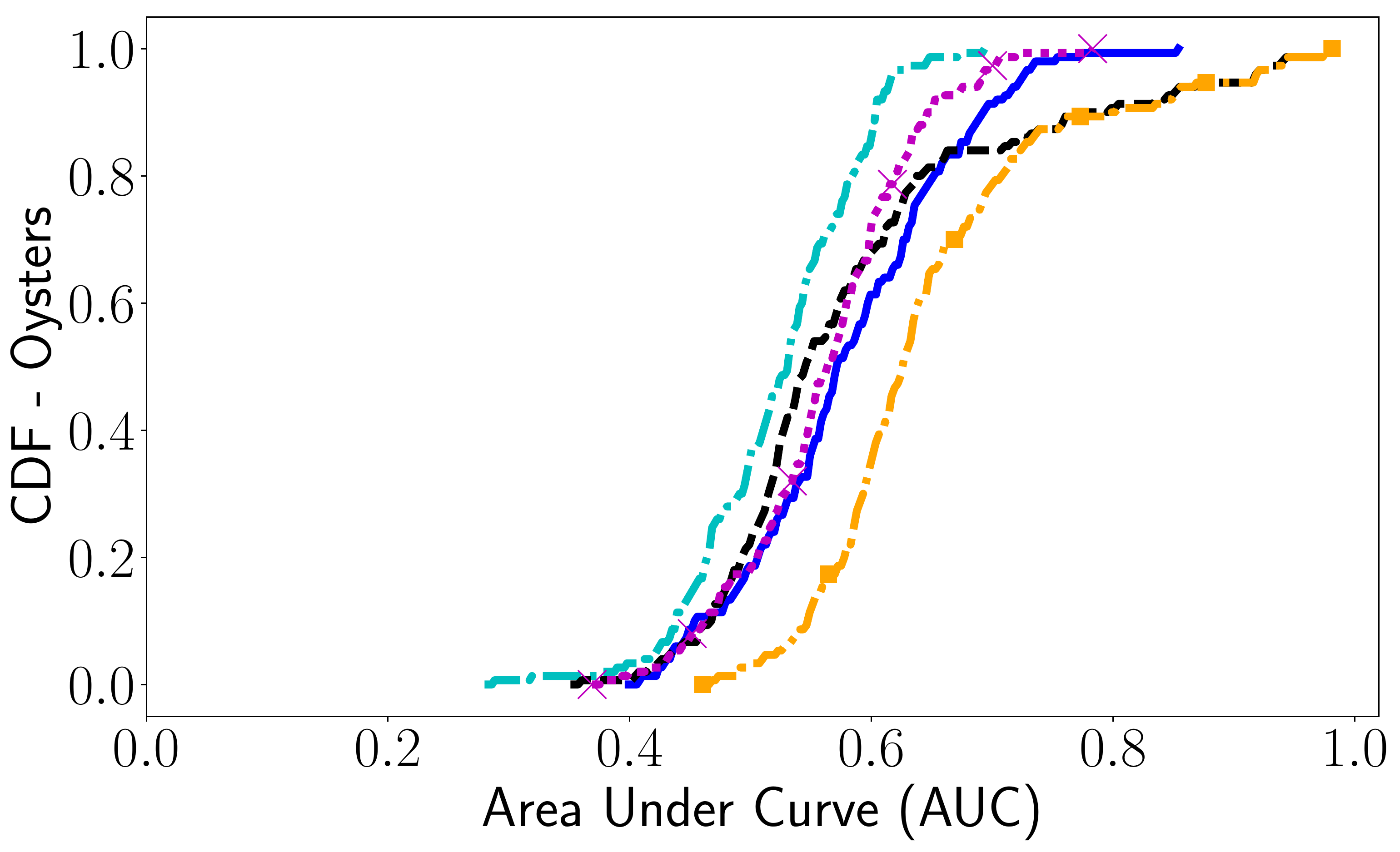}
        \caption{$m=500$}
        \label{fig:tfl-10perc-gr500}
    \end{subfigure}
    ~
    \begin{subfigure}[b]{0.245\textwidth}
        \includegraphics[width=\textwidth]{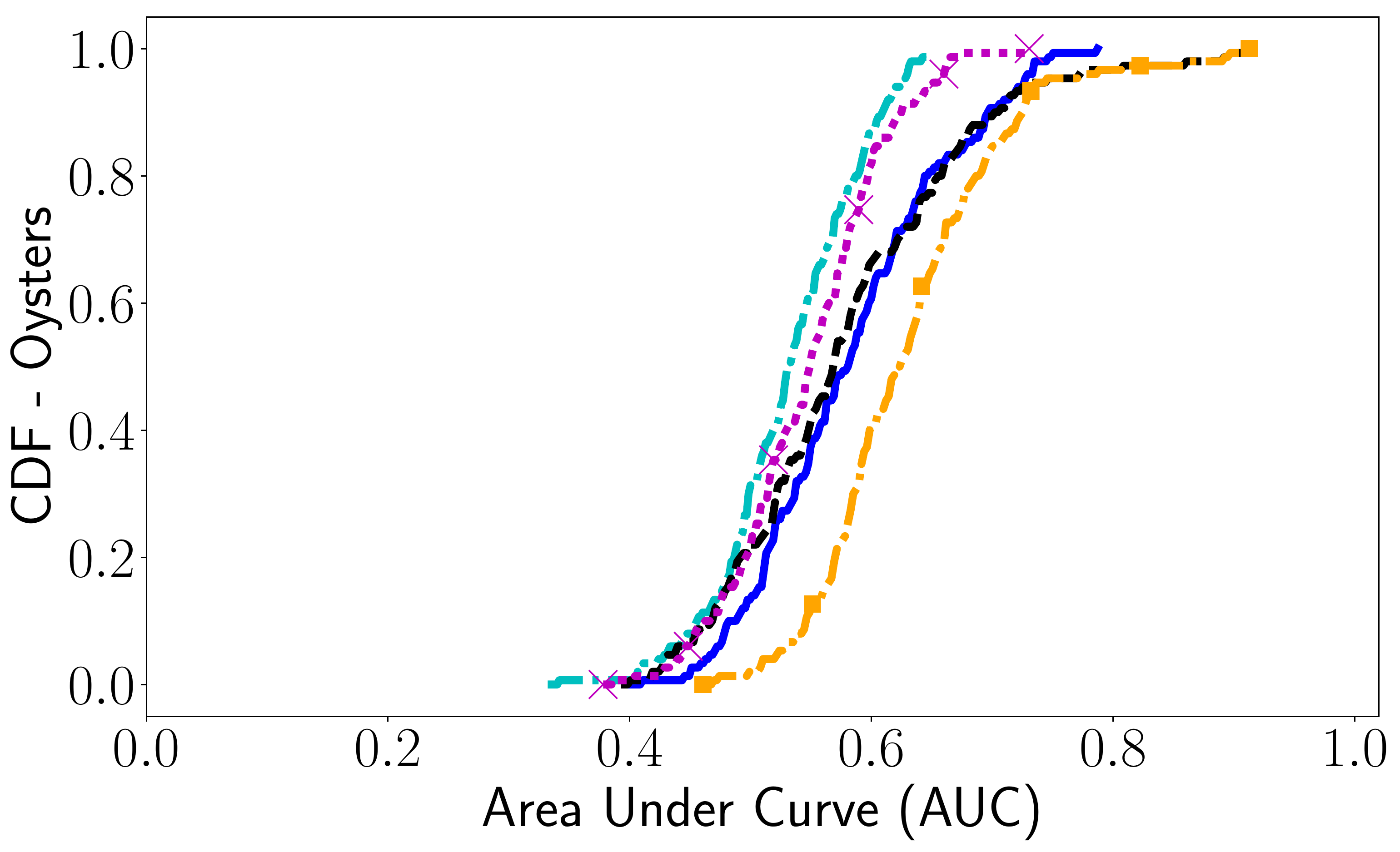}
        \caption{$m=1,000$}
        \label{fig:tfl-10perc-gr1000}
    \end{subfigure}
    \vspace{-0.3cm}    
    \caption{{\em Subset of Locations} prior (TFL,  $\alpha = 0.11$, $|T_I| = 168$) -- Adv's performance for different values of $m$.}
    \label{fig:tfl-10perc}
    \vspace{-0.2cm}
\end{figure*}

\begin{figure*}[t]
\center
    \begin{subfigure}[b]{0.33\textwidth}
\includegraphics[width=1\textwidth]{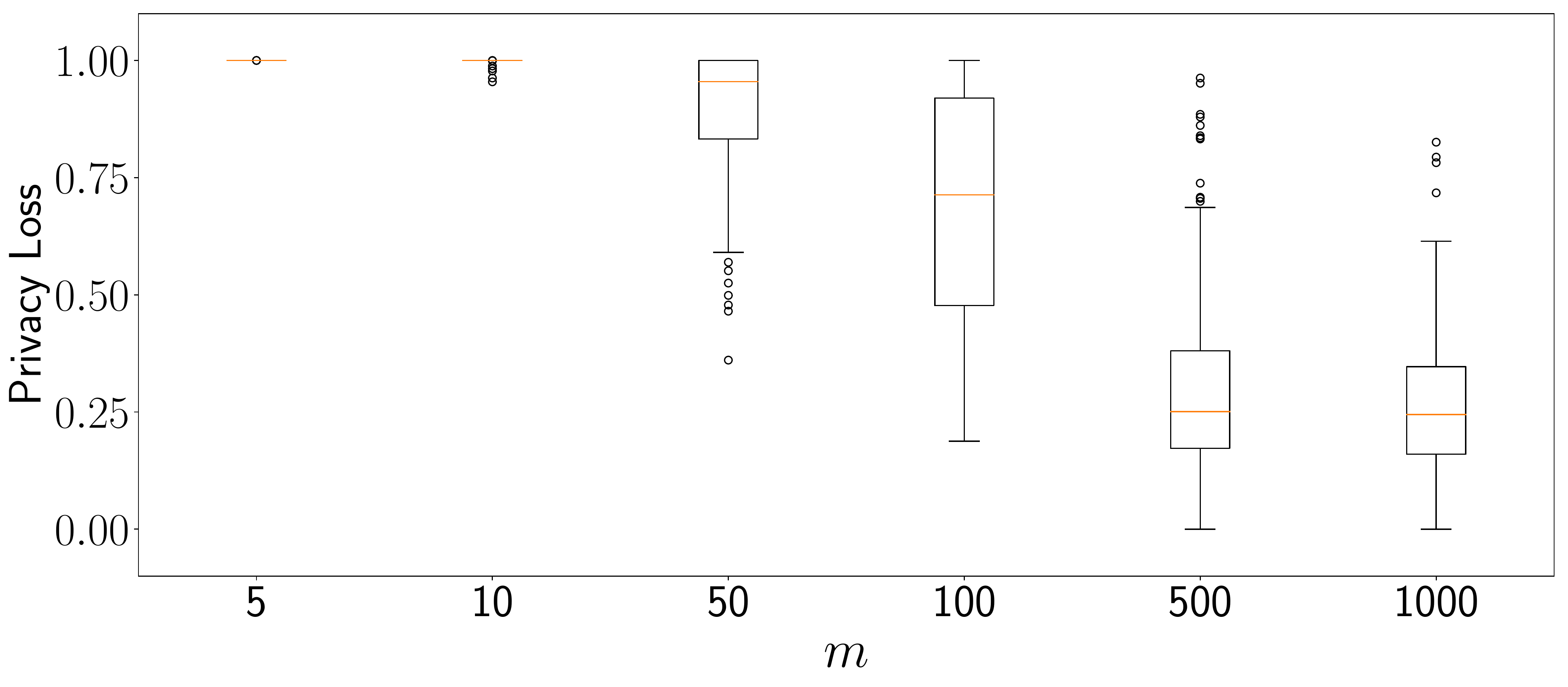}
\caption{TFL ($\alpha = 0.11$, $|T_I| = 168$)}
\label{fig:tfl-10perc-pl}
    \end{subfigure}
    ~~
    \begin{subfigure}[b]{0.33\textwidth}
\includegraphics[width=1\textwidth]{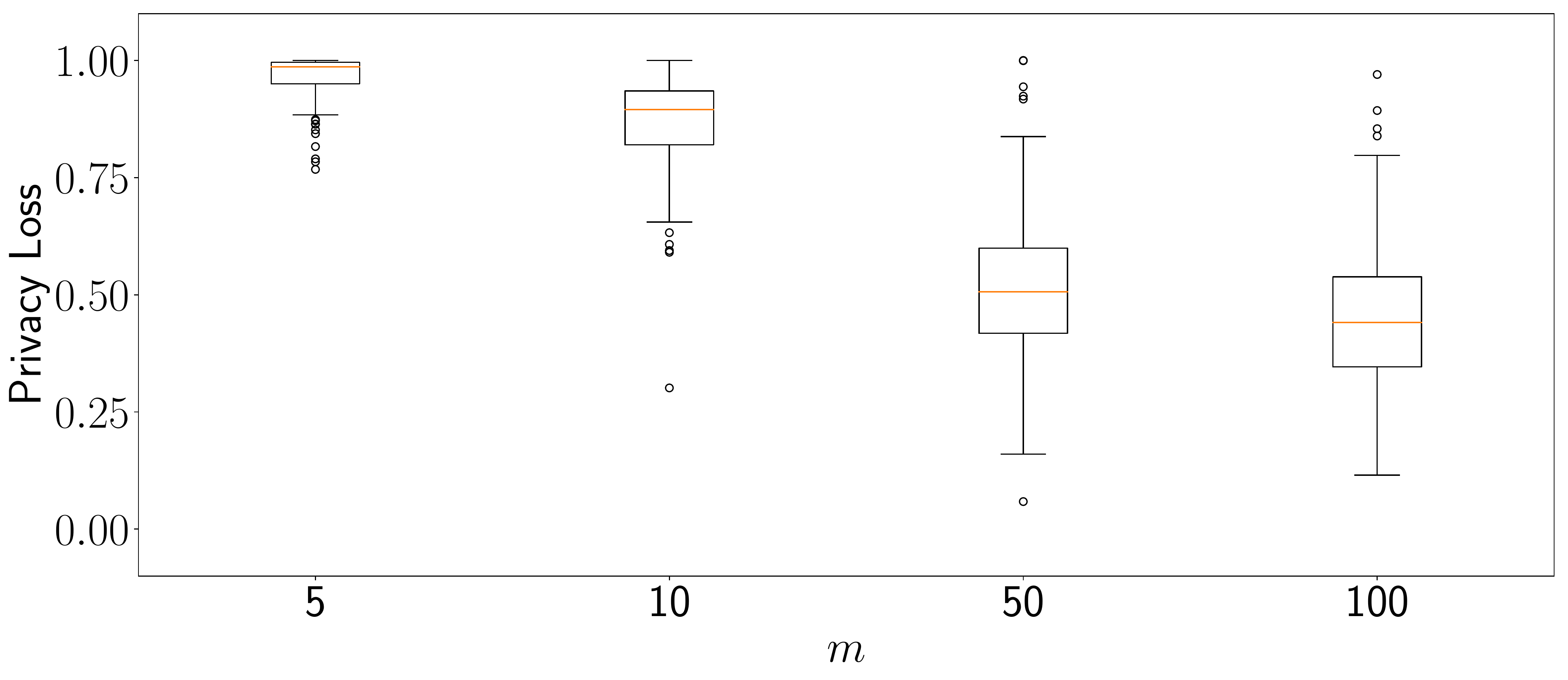}
\caption{SFC ($\alpha = 0.2$, $|T_I| = 168$)}
\label{fig:sfc-20perc-pl}
    \end{subfigure}
    \vspace{-0.3cm}
   \caption{{\em Subset of Locations} prior - Privacy Loss (PL) for different values of $m$.} 
    \vspace{-0.2cm}
\end{figure*}

\begin{figure*}[t]
  \centering
    \begin{subfigure}[b]{0.23\textwidth}
        \includegraphics[width=\textwidth]{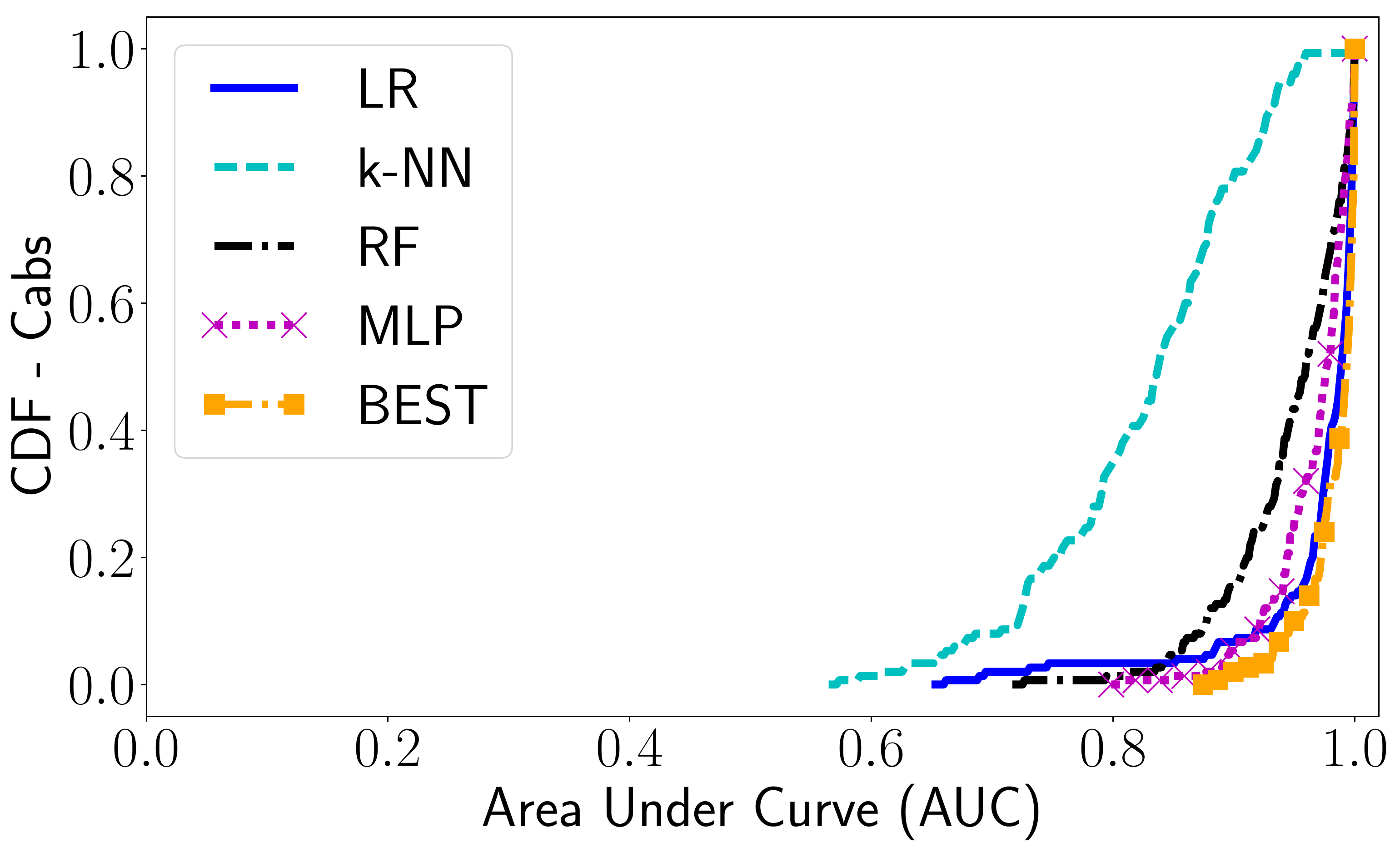}
        \caption{$m=5$}
        \label{fig:sfc-20perc-gr5}
    \end{subfigure}
	~
    \begin{subfigure}[b]{0.23\textwidth}
        \includegraphics[width=\textwidth]{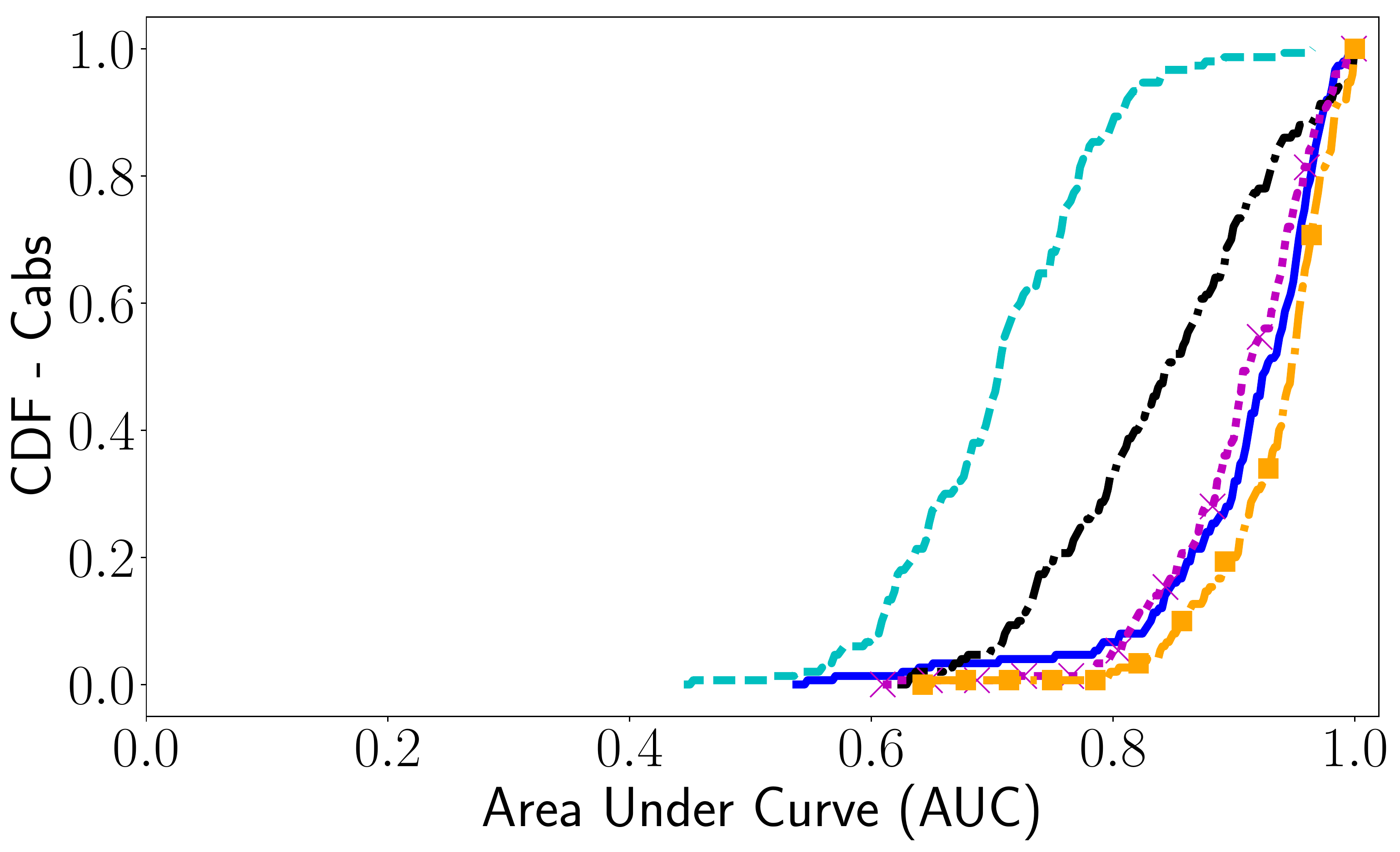}
        \caption{$m=10$}
        \label{fig:sfc-20perc-gr10}
    \end{subfigure}
    ~
    \begin{subfigure}[b]{0.23\textwidth}
        \includegraphics[width=\textwidth]{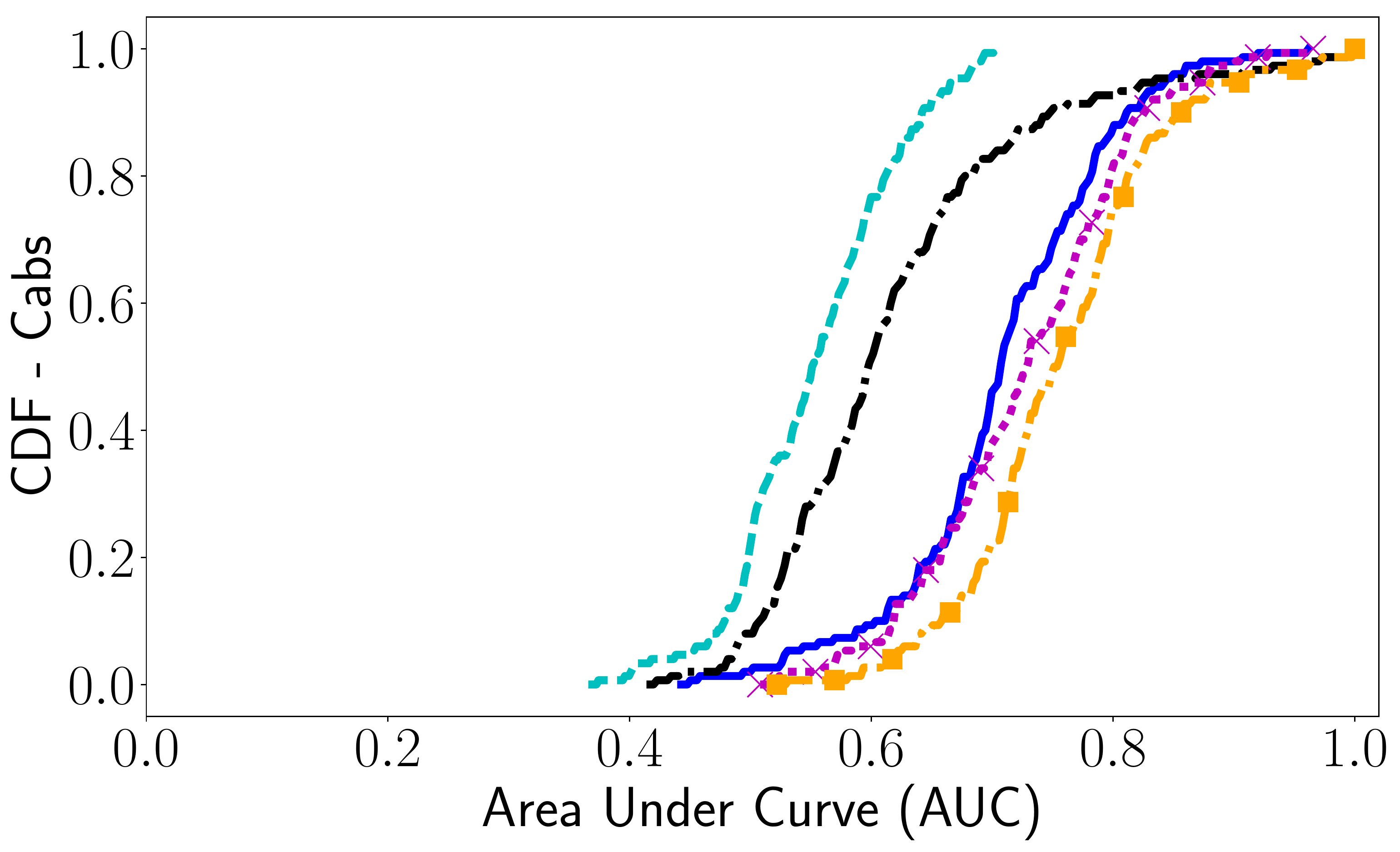}
        \caption{$m=50$}
        \label{fig:sfc-20perc-gr50}
    \end{subfigure}
	~
    \begin{subfigure}[b]{0.23\textwidth}
        \includegraphics[width=\textwidth]{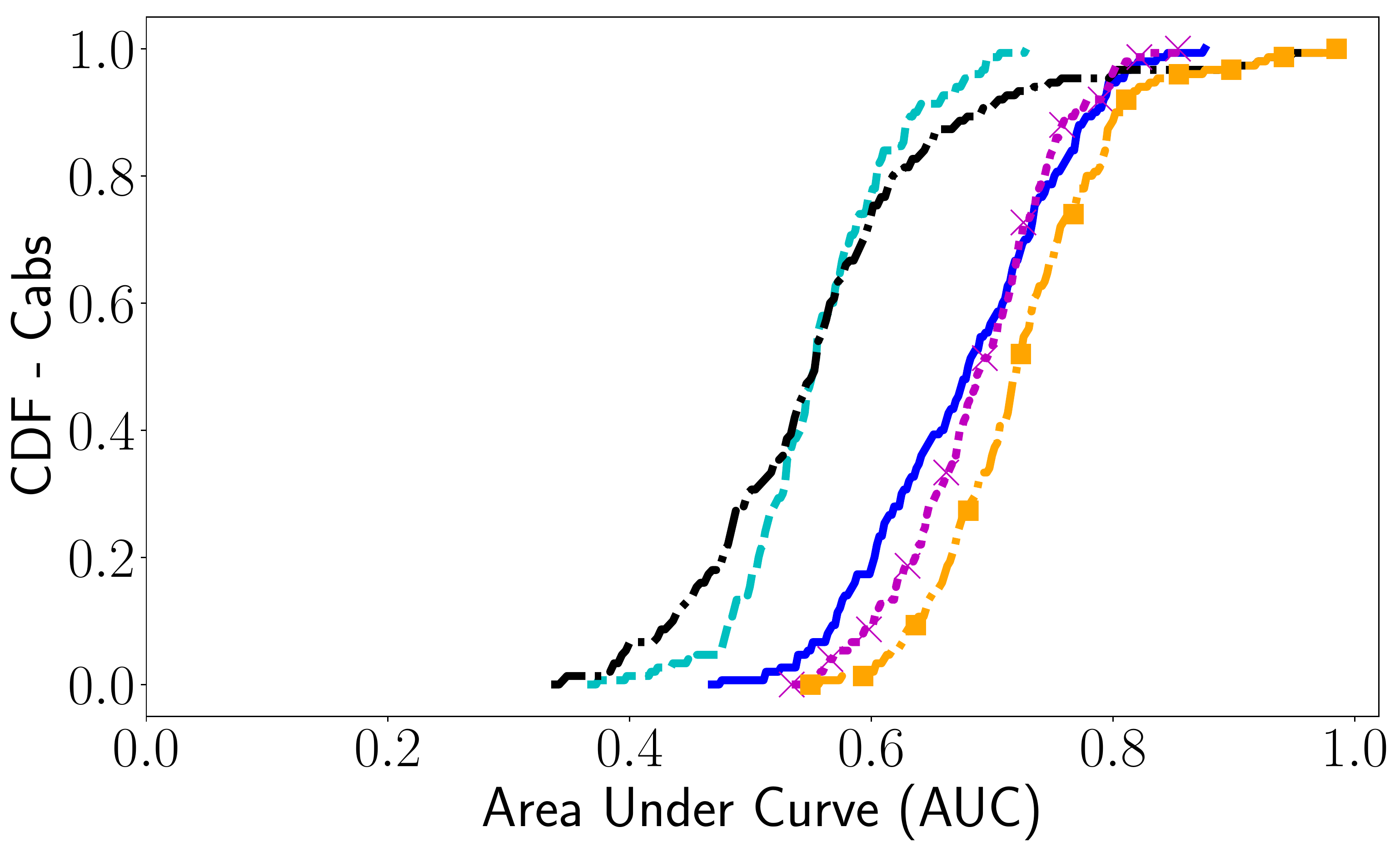}
        \caption{$m=100$}
        \label{fig:sfc-20perc-gr100}
    \end{subfigure}
    \vspace{-0.3cm}    
	\caption{{\em Subset of Locations} prior (SFC,  $\alpha = 0.2$, $|T_I|=168$) -- Adv's performance for different values of $m$.}
    \label{fig:sfc-20perc}
    \vspace{-0.5cm}    
\end{figure*}

\vspace{-0.1cm}
\subsection{Subset of Locations} 
We start with the setting where the observation and inference periods coincide, and the adversary knows the time-series for a subset of users, including the target, during this period.
This information can then be used by Adv to create groups, with and without her target, and train a classifier. %
We consider, as the observation/inference period, the first week of both TFL and SFC datasets -- that is, $|T_O|=|T_I|=$ 24 $\cdot$ 7 $=$ 168 hourly timeslots. We build Adv's prior by setting $\alpha=$ 0.11 for TFL and $\alpha=$ 0.2 for SFC, i.e, we randomly choose 1,100 out of 10,000 TFL users and 106 out of 534 SFC cabs. This represents a setting where Adv (e.g., a telecommunications service provider) knows location information for a small subset of users, including her target. %

We then generate (i) a {\em balanced} training dataset by randomly sampling 400 \textit{unique} user groups from Adv's prior knowledge, whereby half include the target user and half exclude it; and (ii) a {\em balanced} testing set by sampling 100 \textit{unique} user groups from the set of users {\em not} in the prior knowledge (apart from the target user). 
Our choice for training/testing sizes (400 and 100, resp.) is so that the datasets are large enough to enable learning and evaluation, and experiments run in reasonable time. 
Finally, we extract features from the aggregates of both training and testing groups, labeling them as per the participation of the target in the group, and perform experiments with different values of $m$ in order to evaluate the effect of aggregation group size. %

\descrit{TFL Dataset.} %
Fig.~\ref{fig:tfl-10perc} plots the CDF, computed over the 150 target TFL users, of the AUC score achieved by the classifiers for different values of $m$. Limited by the adversarial knowledge (1,100 users), we examine aggregation group sizes up to 1,000. 
Note that the orange line labeled as `BEST' represents a hypothetical best case in which Adv chooses the classifier that yields the highest AUC score for each target user.%

When groups are small, i.e., $m=$ 5 or 10, all classifiers achieve very high AUC scores. For instance, with $m=$ 10, Linear Regression (LR) and Random Forest (RF) achieve a mean AUC score of 0.97 and 0.99, respectively. This indicates that for such small groups, where users' contribution to the aggregates is very significant, membership inference is very effective. As the size of the aggregation groups increases to $m=$ 50 or 100, %
the performance only slightly decreases, with RF outperforming LR, Nearest Neighbors (k-NN), and Multi-Layer Perceptron (MLP), yielding 0.94 mean AUC for groups of 50 users, and 0.83 for 100. %
With larger aggregation sizes, $m=$ 500 or 1,000, performance drops closer to the random guess baseline (AUC $=$ 0.5).
Nonetheless, even for groups of 1,000 users, Adv can still infer membership of 60\% of the target population with an AUC score higher than 0.6.

We also measure the impact of the effectiveness of the distinguishing function on privacy using the Privacy Loss metric (PL, cf.~Eq.~\ref{eq:pl}). More specifically, in Fig.~\ref{fig:tfl-10perc-pl}, we report a box-plot with the PL for different aggregation group sizes, when the adversary picks the best classifier for each target user (orange line in Fig.~\ref{fig:tfl-10perc}). For small groups, mean PL is very large, e.g., 0.99 for $m=$ 10, 0.89 for 50, and 0.68 for 100. Unsurprisingly, PL decreases as the group size increases, i.e., users enjoy better privacy when their data is aggregated in larger groups. Yet, even then they experience a 25\% reduction of privacy vs a random guess ($m=$ 1,000).

\descrit{SFC Dataset.} In Fig.~\ref{fig:sfc-20perc}, we plot the classifiers' performance %
on the SFC dataset for groups of up to 100 users, as we are limited by the adversarial knowledge (106 cabs).
As in the previous case, for small groups ($m=$ 5, 10) Adv  can infer membership with high accuracy. 
For instance, for groups of 10 users, LR and MLP achieve mean AUC of 0.9, followed by RF (0.84) and k-NN (0.7). Again, performance decreases as group size increases: for groups of 50 cabs (resp., 100) MLP and LR yield mean AUC scores of 0.72 (resp., 0.68) and 0.7 (resp., 0.67). Nonetheless, when Adv picks the best classifier for each cab (orange line), mean AUC score is still 0.72 even with 100 users per group.

PL over the different values of $m$ is explored in Fig.~\ref{fig:sfc-20perc-pl}, using the best classifier for each target. Similar to the TFL case, the loss is very large for small groups (e.g., PL $=$ 0.86 when $m=$ 10), and remains significant in larger ones (e.g., PL $=$ 0.44 when $m=$ 100). %
Interestingly, for groups up to 100 users, PL is larger on TFL than on SFC data (e.g., PL $=$ 0.68 on TFL vs 0.44 on SFC, for $m=$ 100), indicating that membership inference is easier on sparse data.

\begin{figure*}[h]
  \centering
    \begin{subfigure}[b]{0.23\textwidth}
        \includegraphics[width=\textwidth]{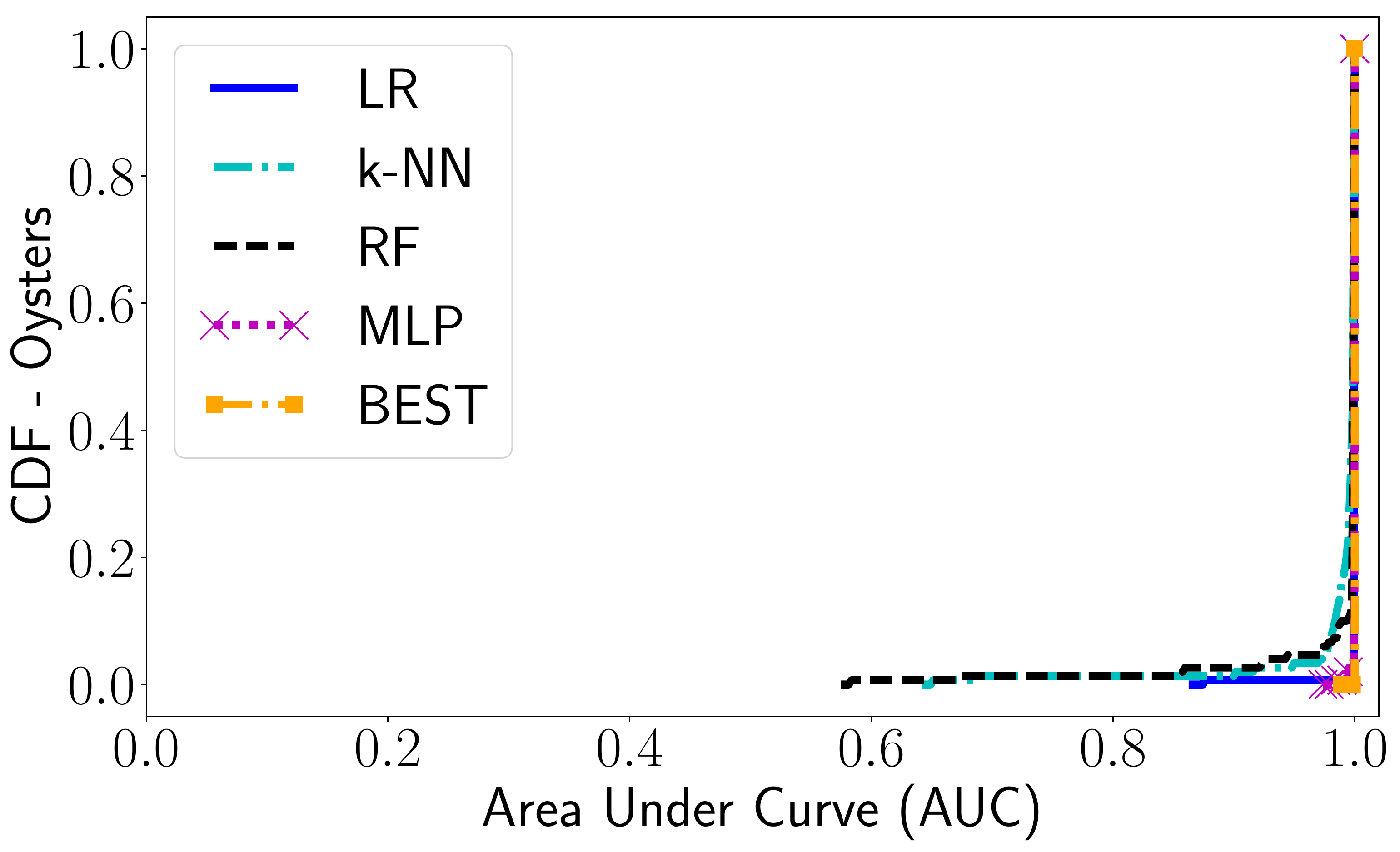}
        \caption{$m=5$}
        \label{fig:tfl-gr5}
    \end{subfigure}
	~
    \begin{subfigure}[b]{0.23\textwidth}
        \includegraphics[width=\textwidth]{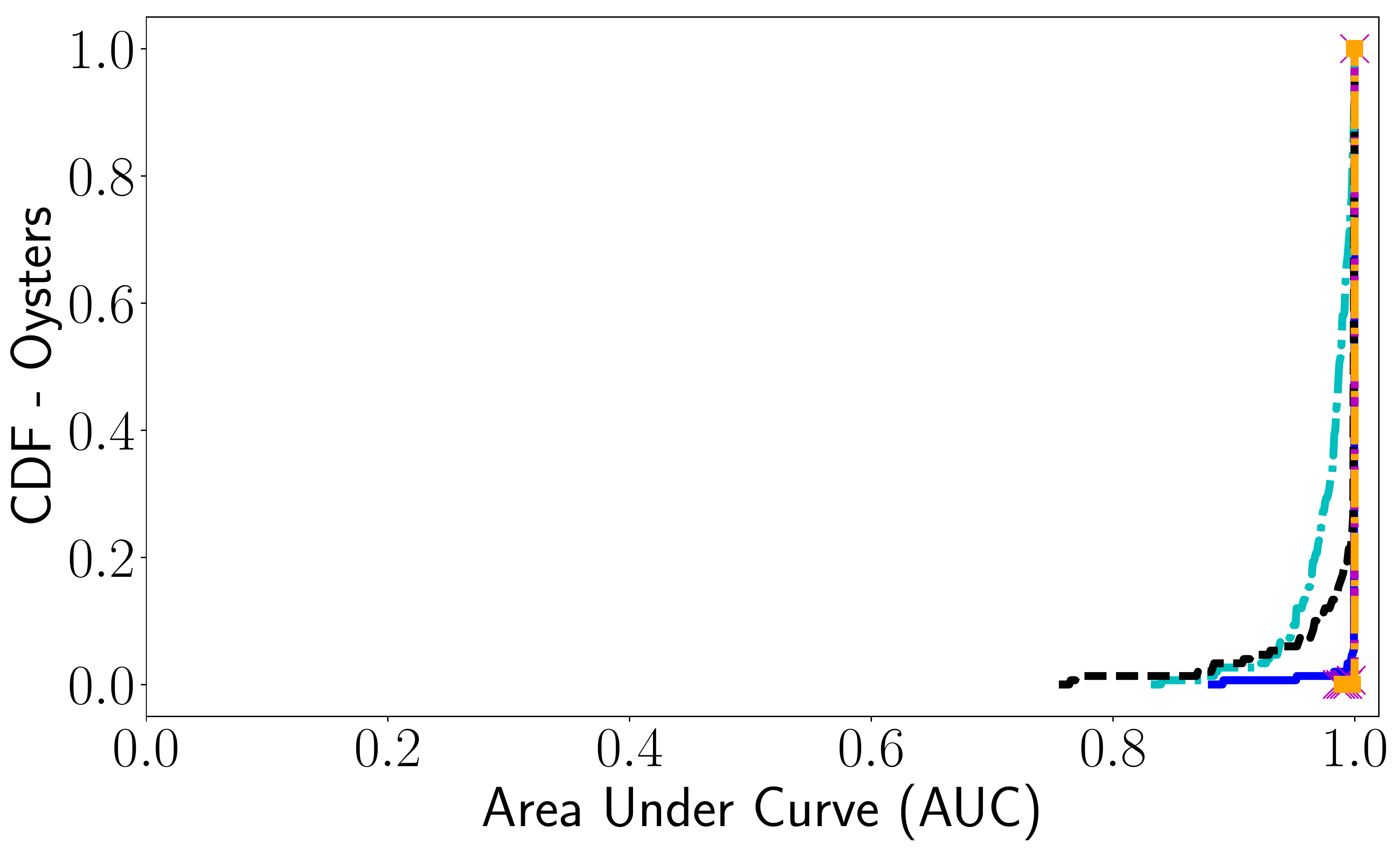}
        \caption{$m=10$}
        \label{fig:tfl-gr10}
    \end{subfigure}
    ~
    \begin{subfigure}[b]{0.23\textwidth}
        \includegraphics[width=\textwidth]{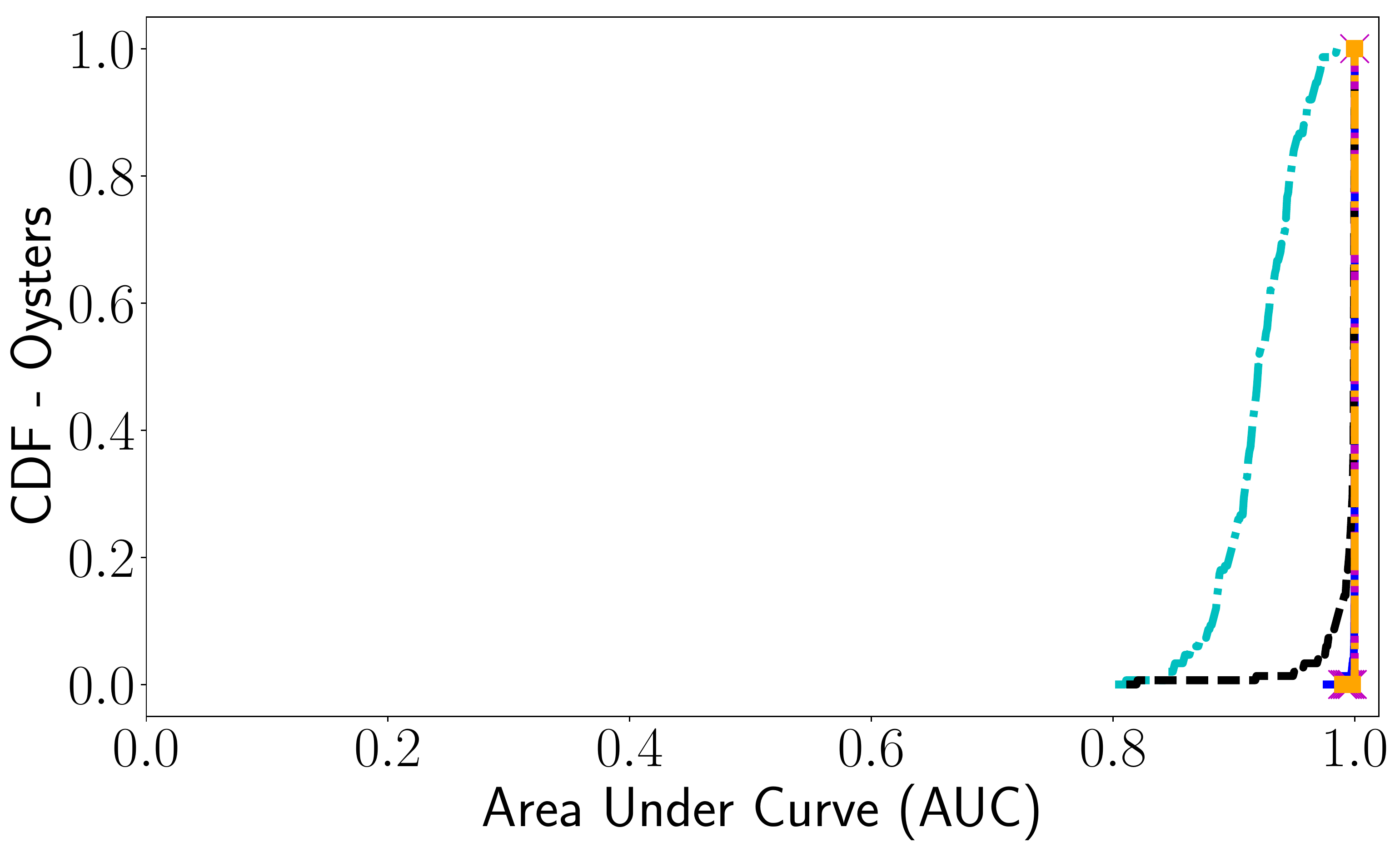}
        \caption{$m=50$}
        \label{fig:tfl-gr50}
    \end{subfigure}
	\\
    \begin{subfigure}[b]{0.23\textwidth}
        \includegraphics[width=\textwidth]{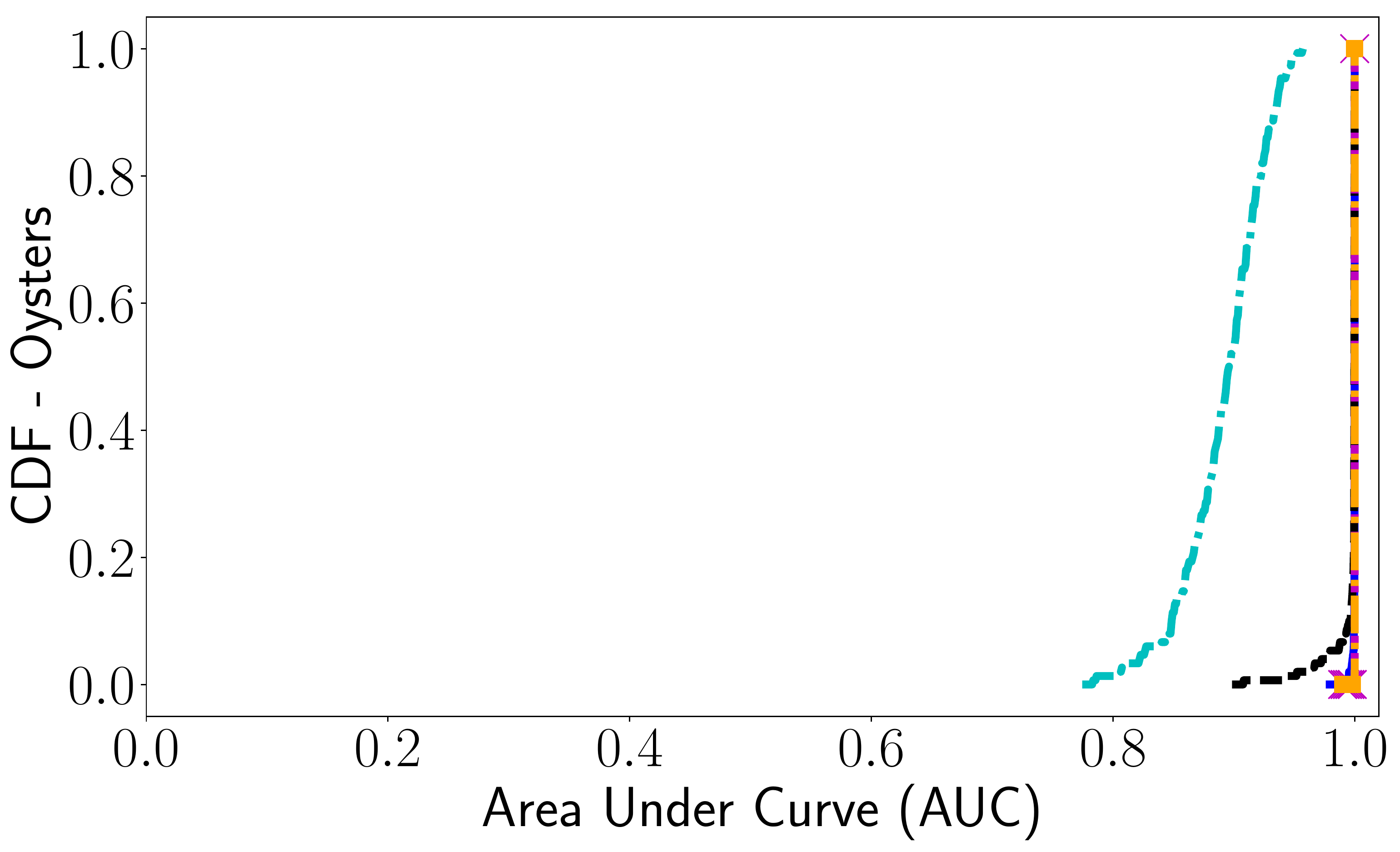}
        \caption{$m=100$}
        \label{fig:tfl-gr100}
    \end{subfigure}    
    ~
    \begin{subfigure}[b]{0.23\textwidth}
        \includegraphics[width=\textwidth]{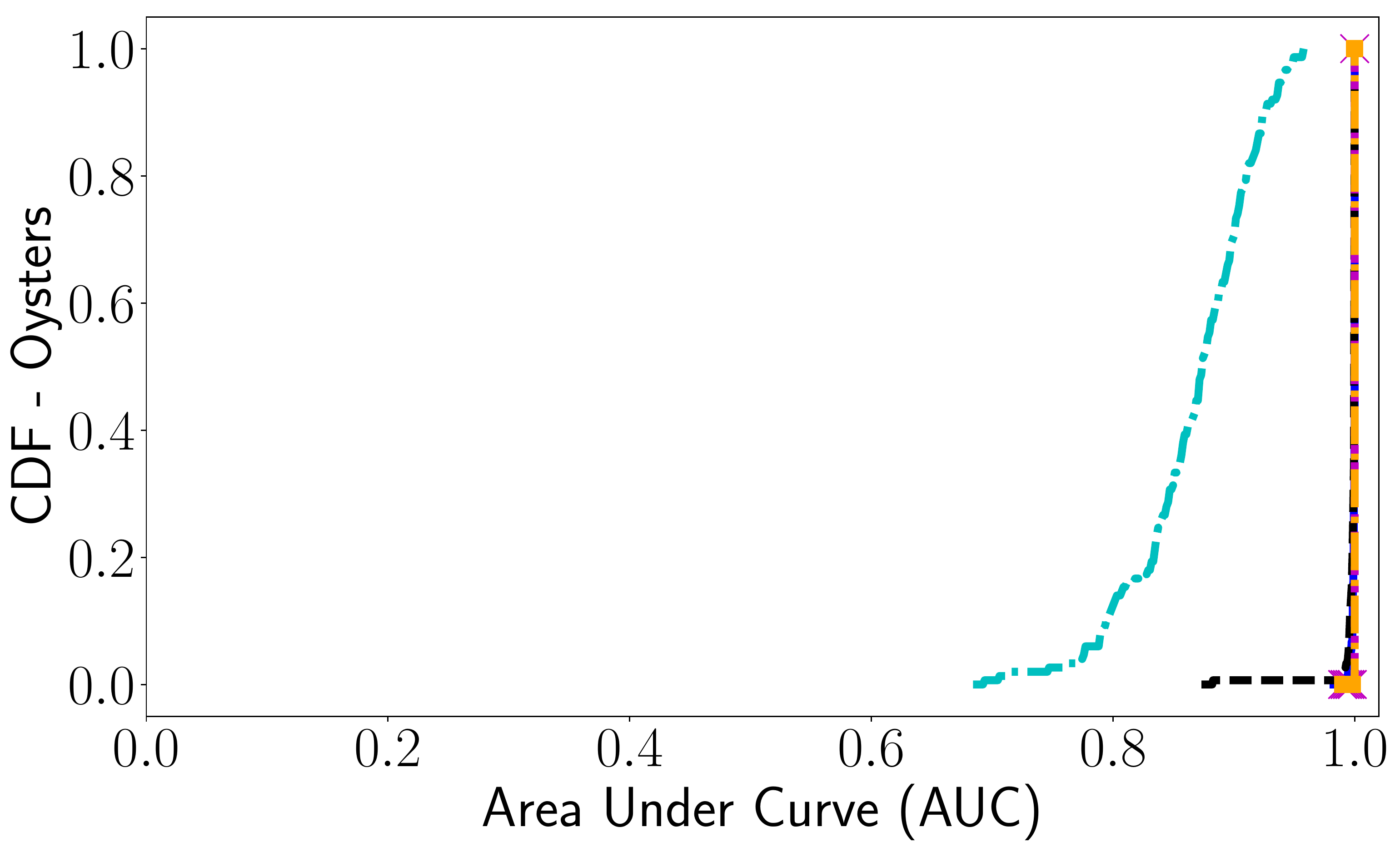}
        \caption{$m=1,000$}
        \label{fig:tfl-gr1000}
    \end{subfigure}
    ~
    \begin{subfigure}[b]{0.23\textwidth}
        \includegraphics[width=\textwidth]{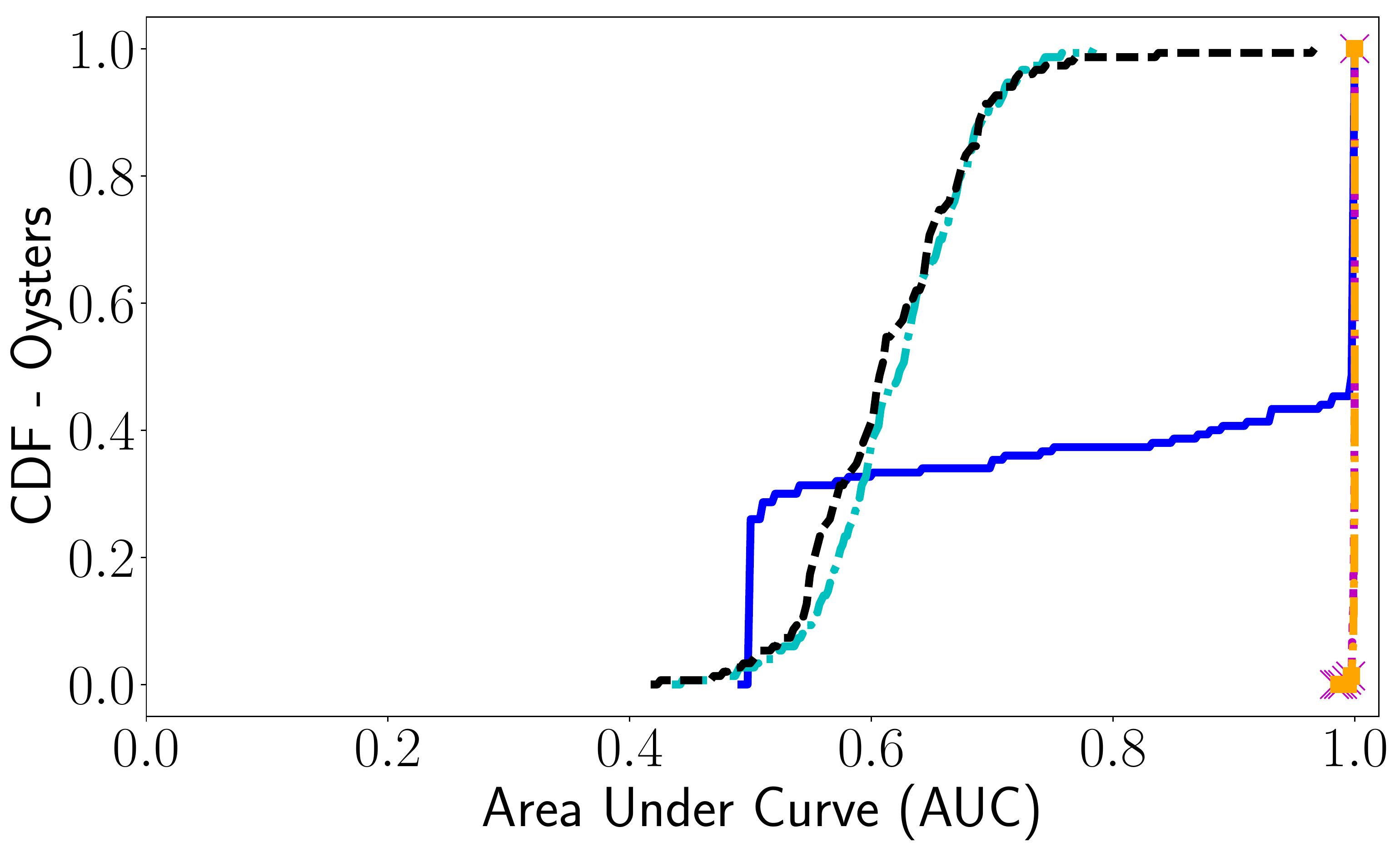}
        \caption{$m=9,500$}
        \label{fig:tfl-gr9500}
    \end{subfigure}
    \vspace{-0.3cm}
 \caption{{\em Same Groups as Released} prior (TFL, 75\%-25\% split, $\beta=150$, $|T_I|=168$) -- Adv's performance for different values of $m$.}   
    \label{fig:tfl-1week-time}
    \vspace{-0.2cm}
\end{figure*}

\begin{figure*}[t]
\centering
\begin{subfigure}[b]{0.4\textwidth}
\centering
\includegraphics[width=0.82\textwidth]{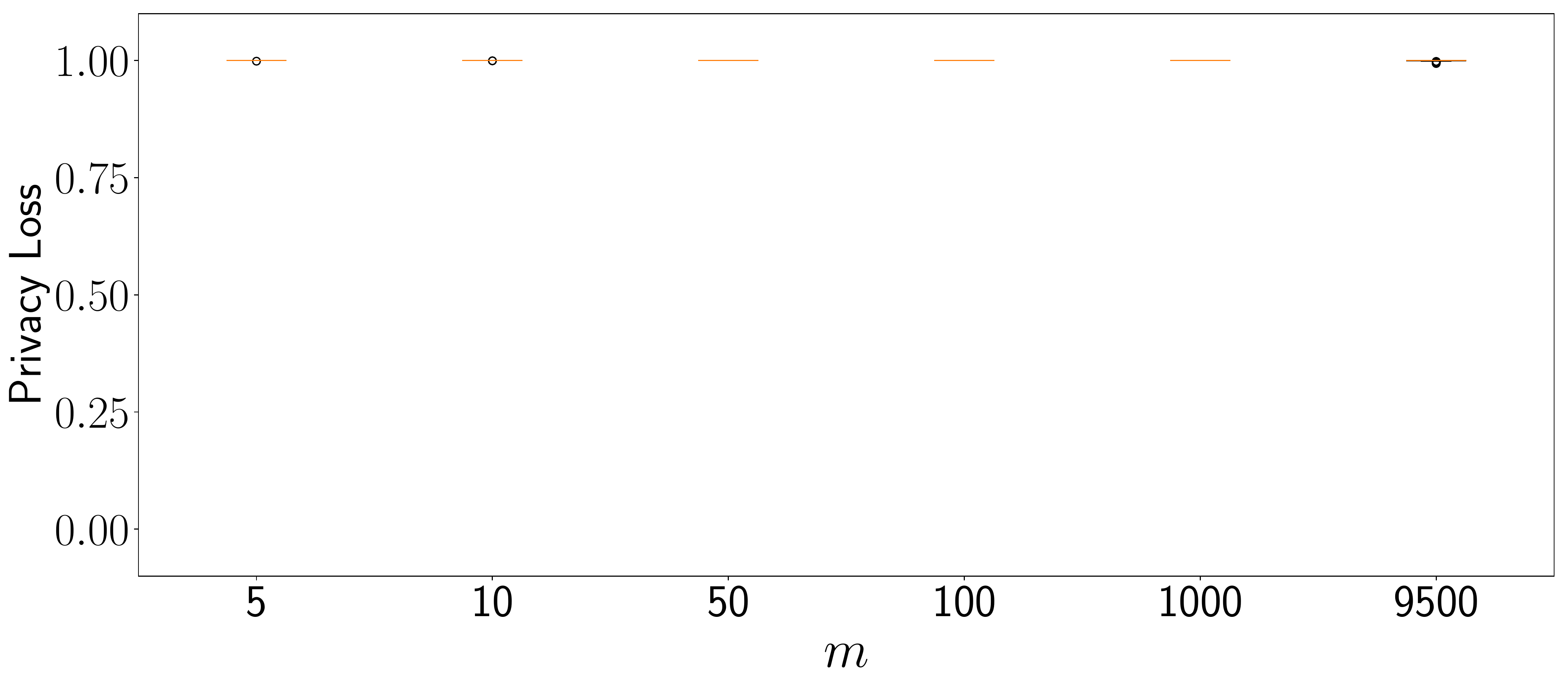}
\caption{TFL, $75\%-25\%$ split, $\beta=150$, $|T_I|=168$}
\label{fig:tfl-1week-pl}
\end{subfigure}
~
\begin{subfigure}[b]{0.4\textwidth}
\centering
\includegraphics[width=0.82\textwidth]{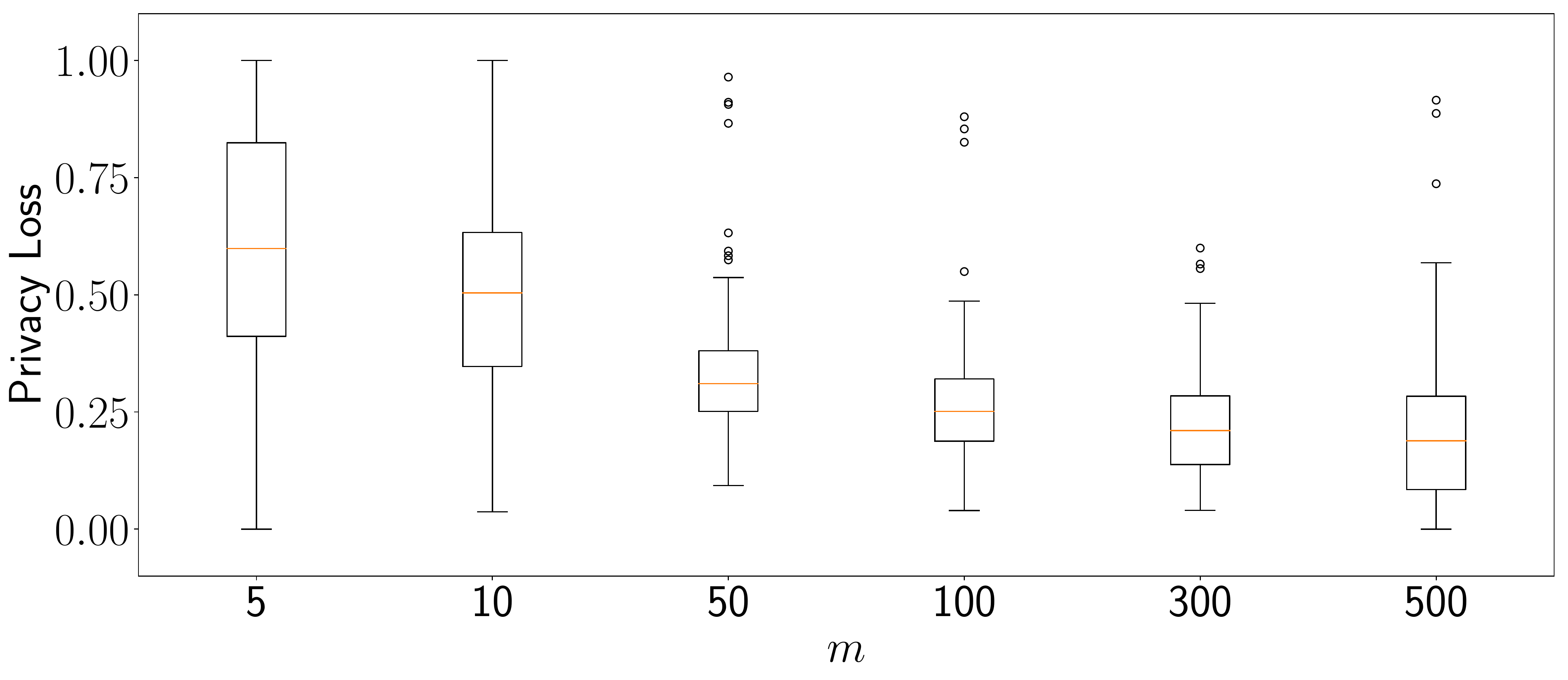}
\caption{SFC,  $67\%-33\%$ split, $\beta=150$, $|T_I|=168$}
\label{fig:sfc-1week-time-pl}
\end{subfigure}
\vspace{-0.3cm}
\caption{{\em Same Groups as Released} prior - Privacy Loss (PL) for different values of $m$.} 
\vspace{-0.2cm}
\end{figure*}

\begin{figure*}[t]
  \centering
    \begin{subfigure}[b]{0.23\textwidth}
        \includegraphics[width=\textwidth]{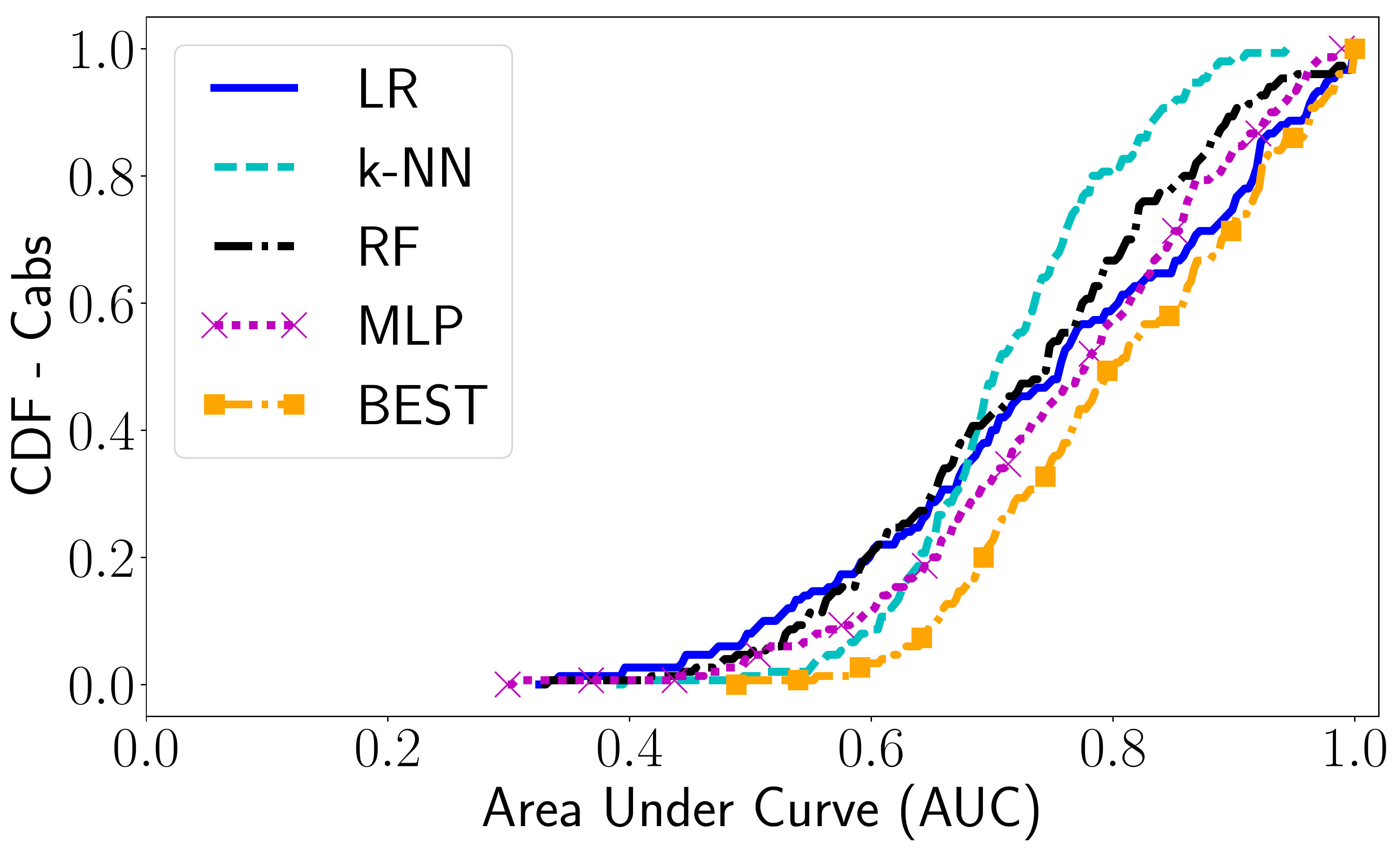}
        \caption{$m=5$}
        \label{fig:sfc-gr5-time}
    \end{subfigure}
	~
    \begin{subfigure}[b]{0.23\textwidth}
        \includegraphics[width=\textwidth]{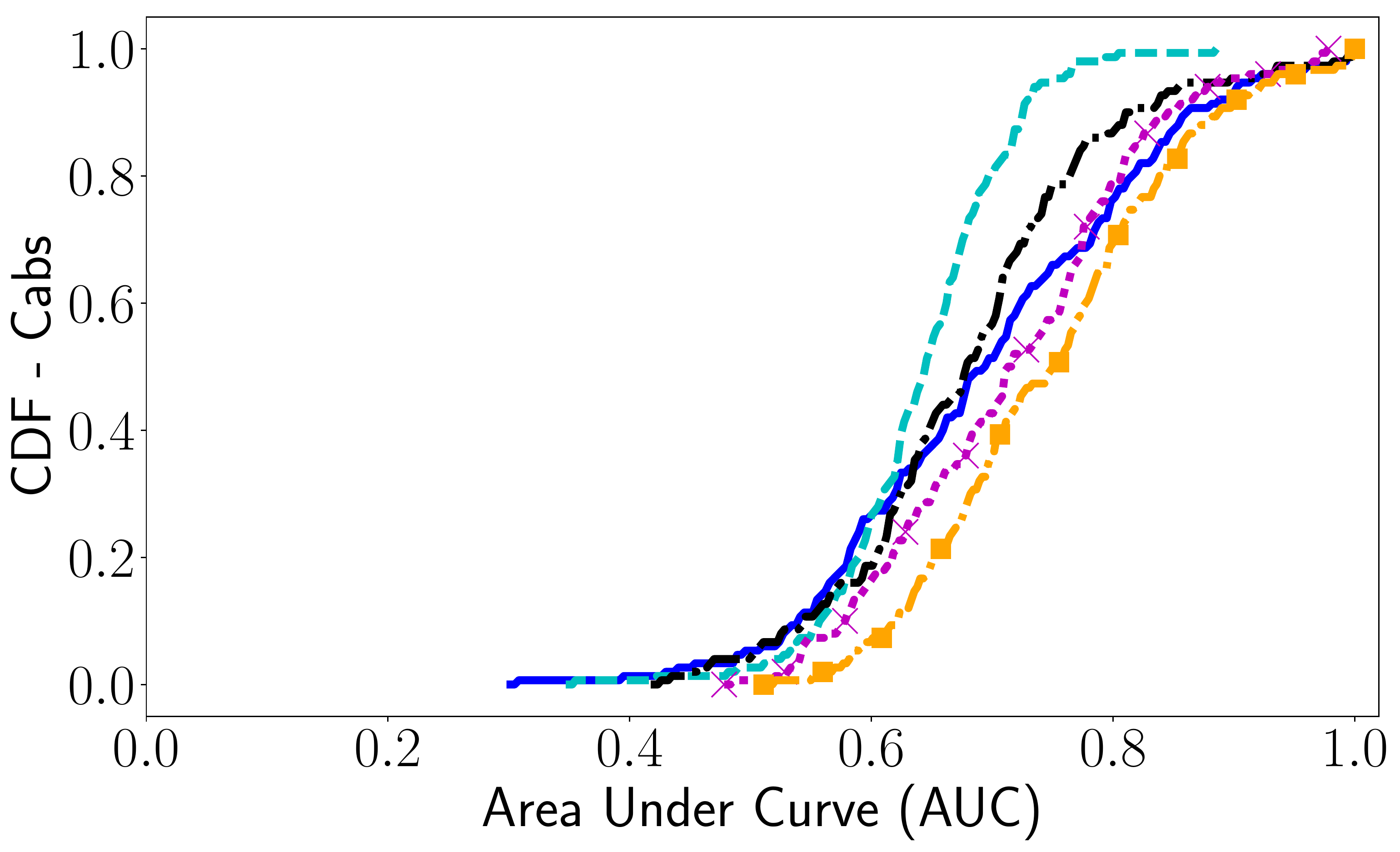}
        \caption{$m=10$}
        \label{fig:sfc-gr10-time}
    \end{subfigure}
    ~
    \begin{subfigure}[b]{0.23\textwidth}
        \includegraphics[width=\textwidth]{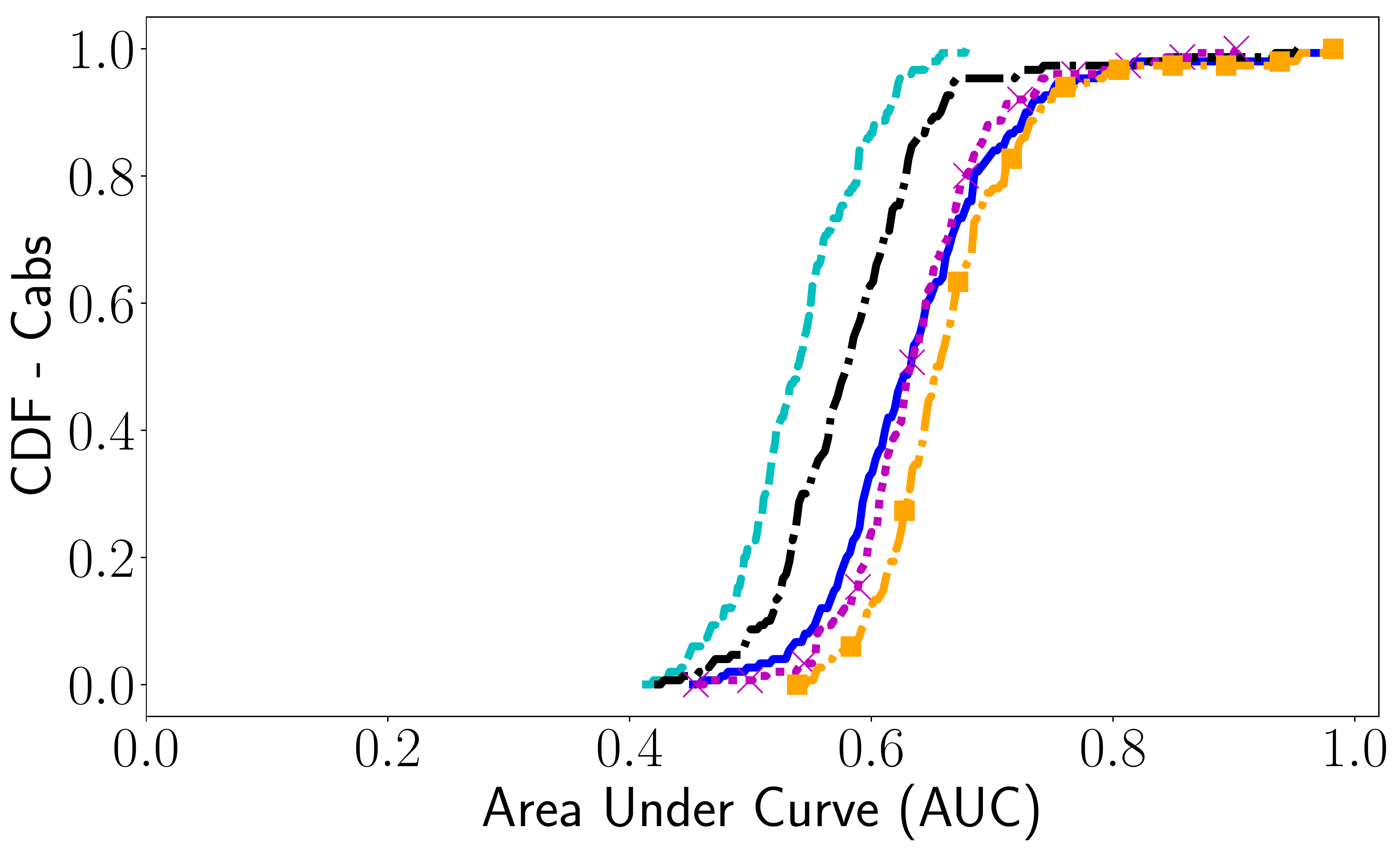}
        \caption{$m=50$}
        \label{fig:sfc-gr50-time}
    \end{subfigure}\\
    \begin{subfigure}[b]{0.23\textwidth}
        \includegraphics[width=\textwidth]{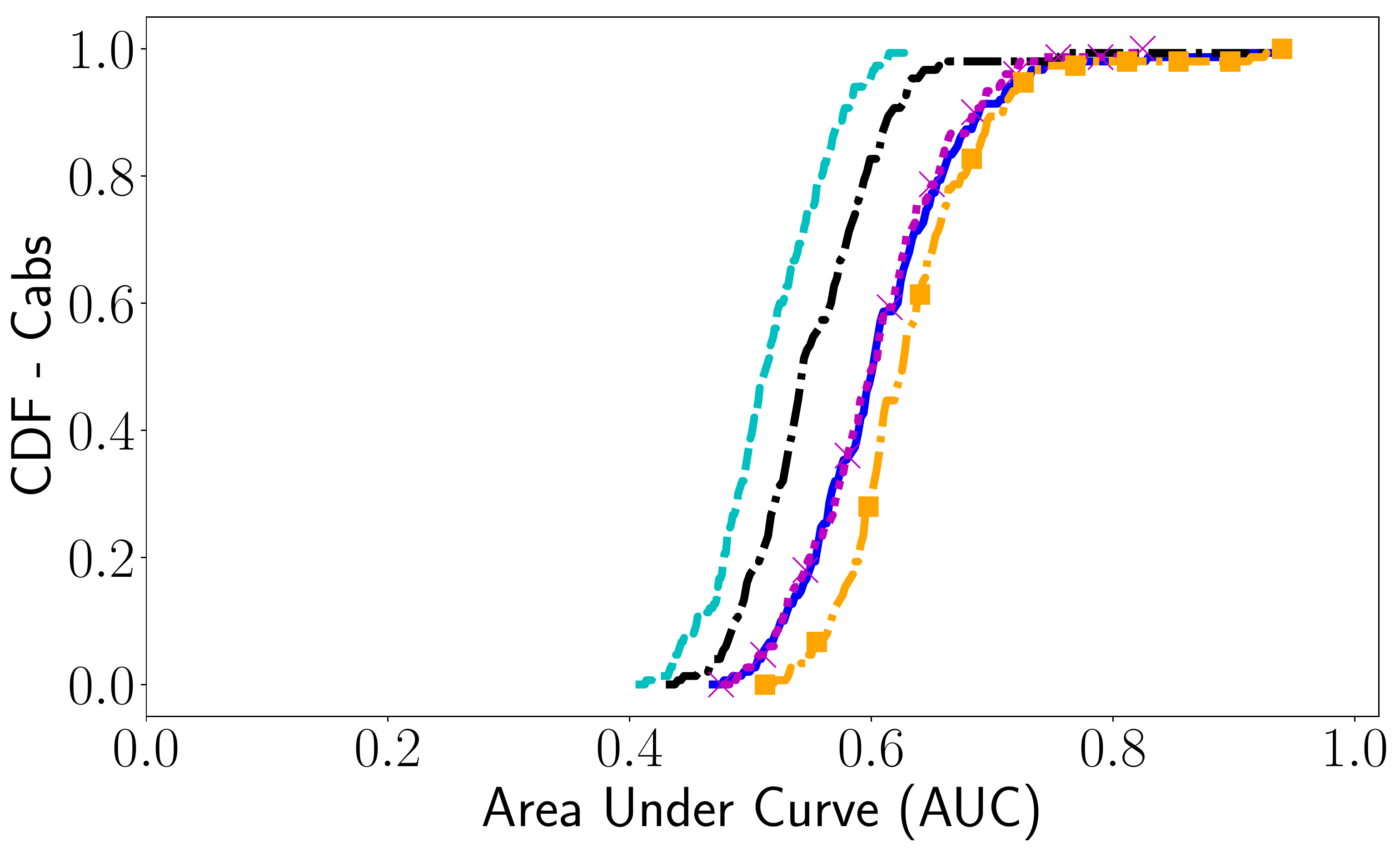}
        \caption{$m=100$}
        \label{fig:sfc-gr100-time}
    \end{subfigure}    
    ~
    \begin{subfigure}[b]{0.23\textwidth}
        \includegraphics[width=\textwidth]{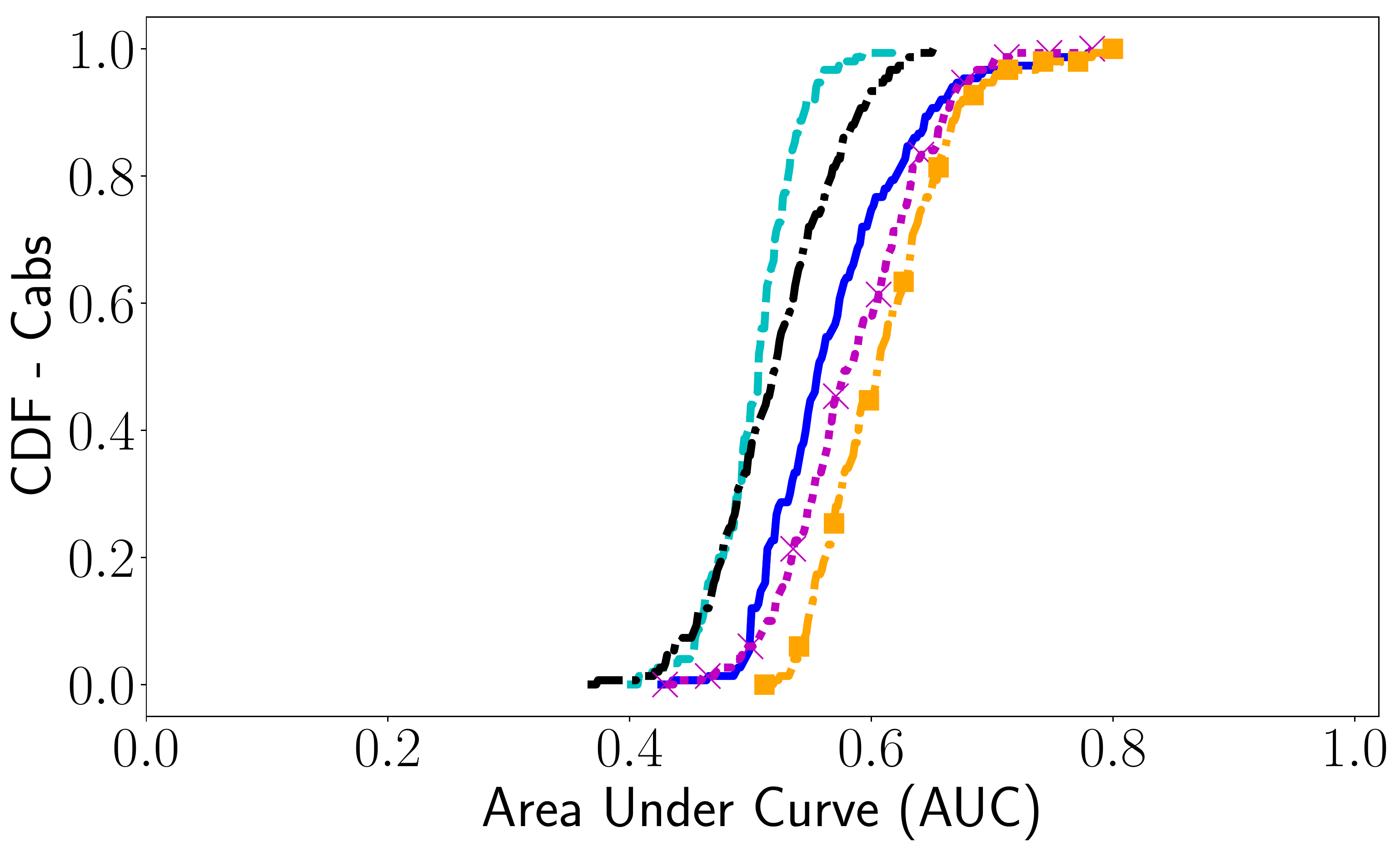}
        \caption{$m=300$}
        \label{fig:sfc-gr300-time}
    \end{subfigure}
    ~
    \begin{subfigure}[b]{0.23\textwidth}
        \includegraphics[width=\textwidth]{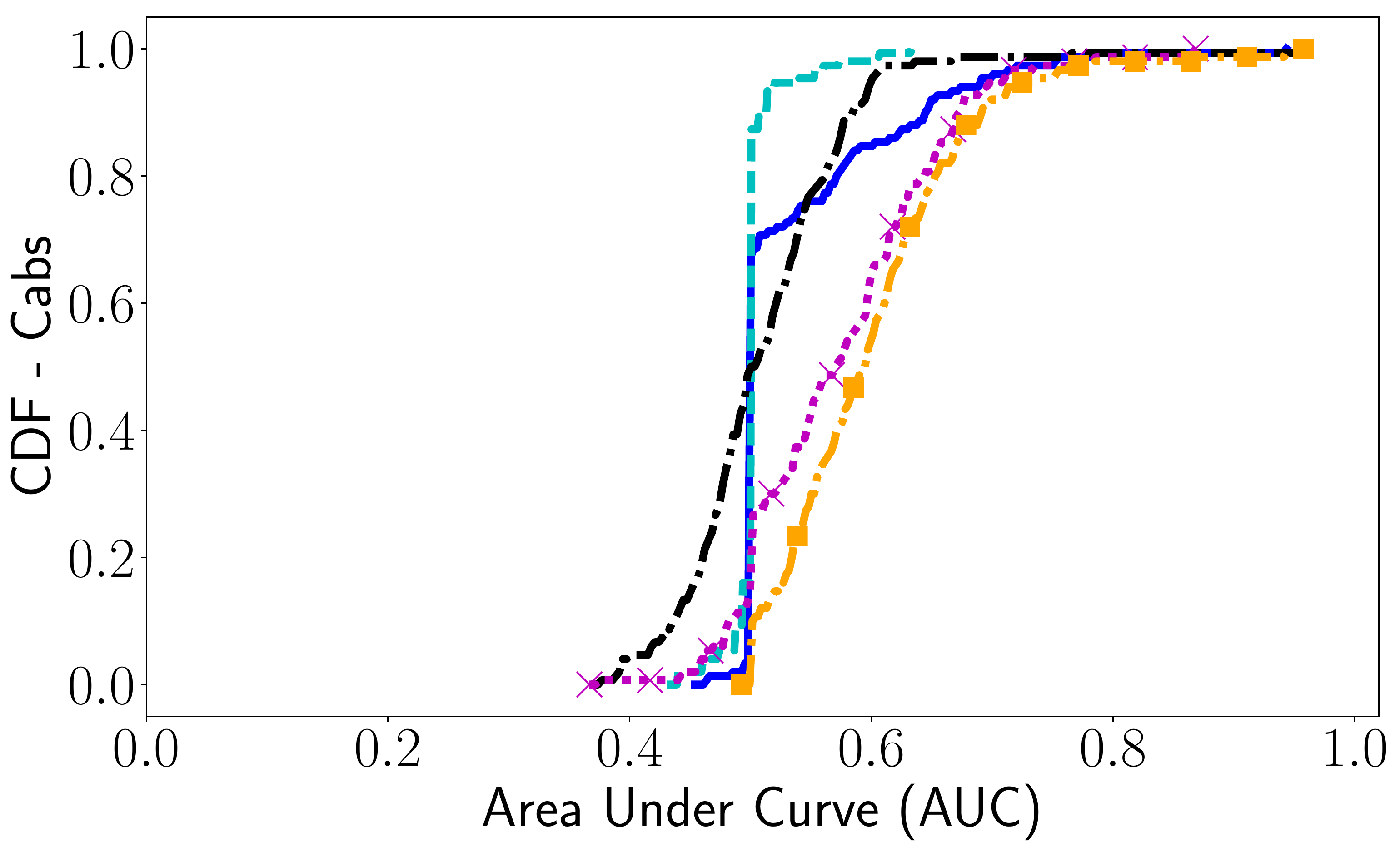}
        \caption{$m=500$}
        \label{fig:sfc-gr500-time}
    \end{subfigure}
	\vspace{-0.1cm}    
    \caption{{\em Same Groups as Released} prior (SFC, 67\%-33\% split, $\beta=150$, $|T_I|=168$) -- Adv's performance for different values of $m$.}
    \label{fig:sfc-1week-time}
	\vspace{-0.5cm}        
\end{figure*}

\subsection{Participation in Past Groups} 
Next, we simulate the setting where Adv's prior knowledge consists of aggregates of groups released in a past observation period, labeled as including data from the target user or not. As discussed in Section~\ref{sec:prior}, we consider two variants: when Adv's prior knowledge is built on either (a) the \textit{same} groups as or (b) \textit{different} groups than those used to compute the inference period aggregates.

\descr{\textit{Same} Groups as Released.} In this setting, we generate $D$ by computing the aggregates of $\beta=$ 150 randomly sampled unique user groups -- 75 that include and 75 that exclude the target -- and set the corresponding label of participation. We split $D$ over time to obtain the training and testing sets. As observation period, i.e., where Adv builds her prior knowledge, we consider the first 3 weeks for TFL (i.e., $|T_O|=$ 3 $\cdot$ 168 $=$ 504 hourly timeslots) and the first 2 weeks for SFC ($|T_O|=$ 336). In both cases, the inference period is the last week of data, thus $|T_I|=$ 168 hourly timeslots, yielding a 75\%-25\% split for TFL, and a 67\%-33\% split for SFC. 
Finally, we train the classifiers with features extracted from the aggregates of {\em each week} in the training set, and test them on those extracted from the aggregates of each group in the test set.

\descrit{TFL Dataset.} Fig.~\ref{fig:tfl-1week-time} shows the classifiers' performance for different aggregation group sizes ($m$). In this experiment, there is no limitation from the prior, thus we can consider groups as large as the dataset. As expected, we observe that for group sizes up to 100 (Figs.~\ref{fig:tfl-gr5}--\ref{fig:tfl-gr100}), membership inference is very accurate (all classifiers yield mean AUC scores over 0.9). Interestingly, as the groups grow to 1,000 commuters (Fig.~\ref{fig:tfl-gr1000}), LR, RF and MLP still yield very high AUC scores (0.99 on average), while k-NN slightly decreases (0.86). For groups of 9,500 commuters (Fig.~\ref{fig:tfl-gr9500}), MLP clearly outperforms the other classifiers yielding an AUC score of 0.99 compared to 0.81 for LR, 0.62 for k-NN and 0.61 for RF. Overall, this experiment indicates that when mobility patterns are regular, as the ones of commuters, an adversary with prior knowledge about specific groups can successfully infer membership in the future if groups are maintained, even if they are large. %

Fig.~\ref{fig:tfl-1week-pl} reports the privacy loss (PL) when the adversary picks the best classifier for each user. We see that, {\em independently} of the group size, commuters lose a huge amount of privacy when they are aggregated in groups for which the adversary has prior knowledge. The results reinforce the previous intuition: the effect of regularity on aggregates is very strong, and makes commuters very susceptible to membership inference attacks.

\descrit{SFC Dataset.} Fig.~\ref{fig:sfc-1week-time} illustrates the performance of the classifiers for variable aggregation group size on the SFC dataset. Recall that this is smaller than TFL, as it only contains 534 cabs. We observe that the lack of regularity in cabs movement has a great impact on the ability of an adversary to infer membership, even when the groups are maintained over time. For small groups ($m=$ 5 or 10), the classifiers' AUC ranges between 0.76 and 0.64, as opposed to 0.9 or more in TFL, with MLP now yielding the best results. As groups become larger (Figs.~\ref{fig:sfc-gr50-time}--\ref{fig:sfc-gr300-time}), irregularity has a bigger effect and, unexpectedly, performance drops further. Already for $m=$ 100, RF and k-NN perform similar to the random guess baseline, and LR's AUC drops to 0.52 when group size reaches $m=$ 500. MLP, however, is still somewhat better than random (0.57 mean AUC). Overall, if the adversary picks the best classifier for each cab (orange line), she can infer membership for half the cabs with AUC score larger than 0.6.

In terms of PL, Fig.~\ref{fig:sfc-1week-time-pl} shows that cabs lose privacy when they are aggregated in small groups. However, since cabs, as well as the groups they are aggregated in, are not as regular as TFL commuters, the loss drops drastically with larger groups (e.g., mean PL is 0.2 for groups of 500 cabs). In other words, irregularity makes inferring membership harder for the adversary. However, even though on average PL decreases, we observe that, for $m=$ 500, some instances of our experiment exhibit larger privacy loss than for $m=$ 300. This stems from the small size of the cab population. As there are only 534 cabs, when grouping them in batches of 500 elements, there inevitably is a big overlap across groups, which effectively creates a somewhat ``artificial'' regularity that increases the performance of the attack.

\begin{figure*}[h]
  \centering
    \begin{subfigure}[b]{0.245\textwidth}
        \includegraphics[width=\textwidth]{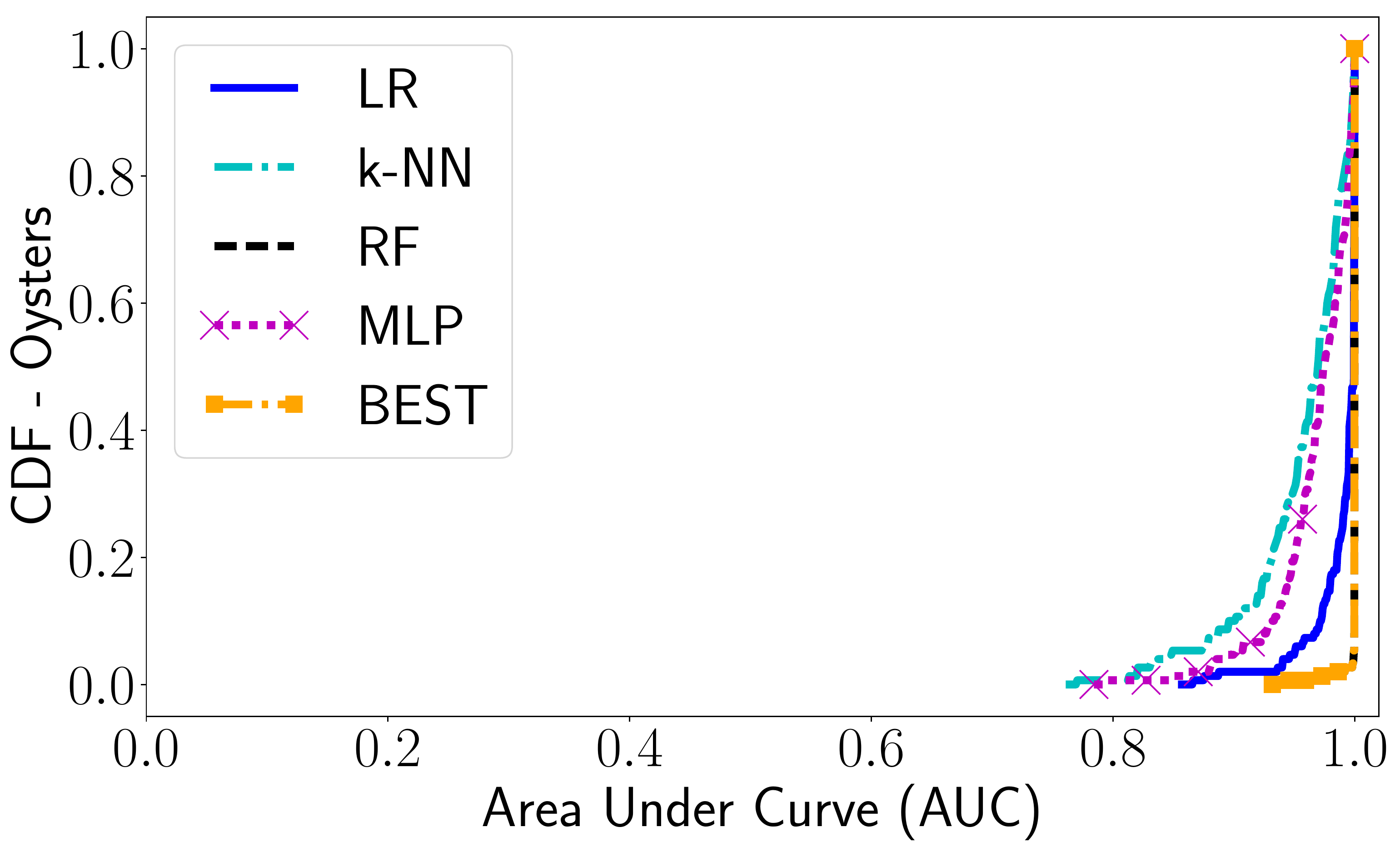}
        \caption{$m=5$}
        \label{fig:tfl-u-time-gr5}
    \end{subfigure}
	~
    \begin{subfigure}[b]{0.245\textwidth}
        \includegraphics[width=\textwidth]{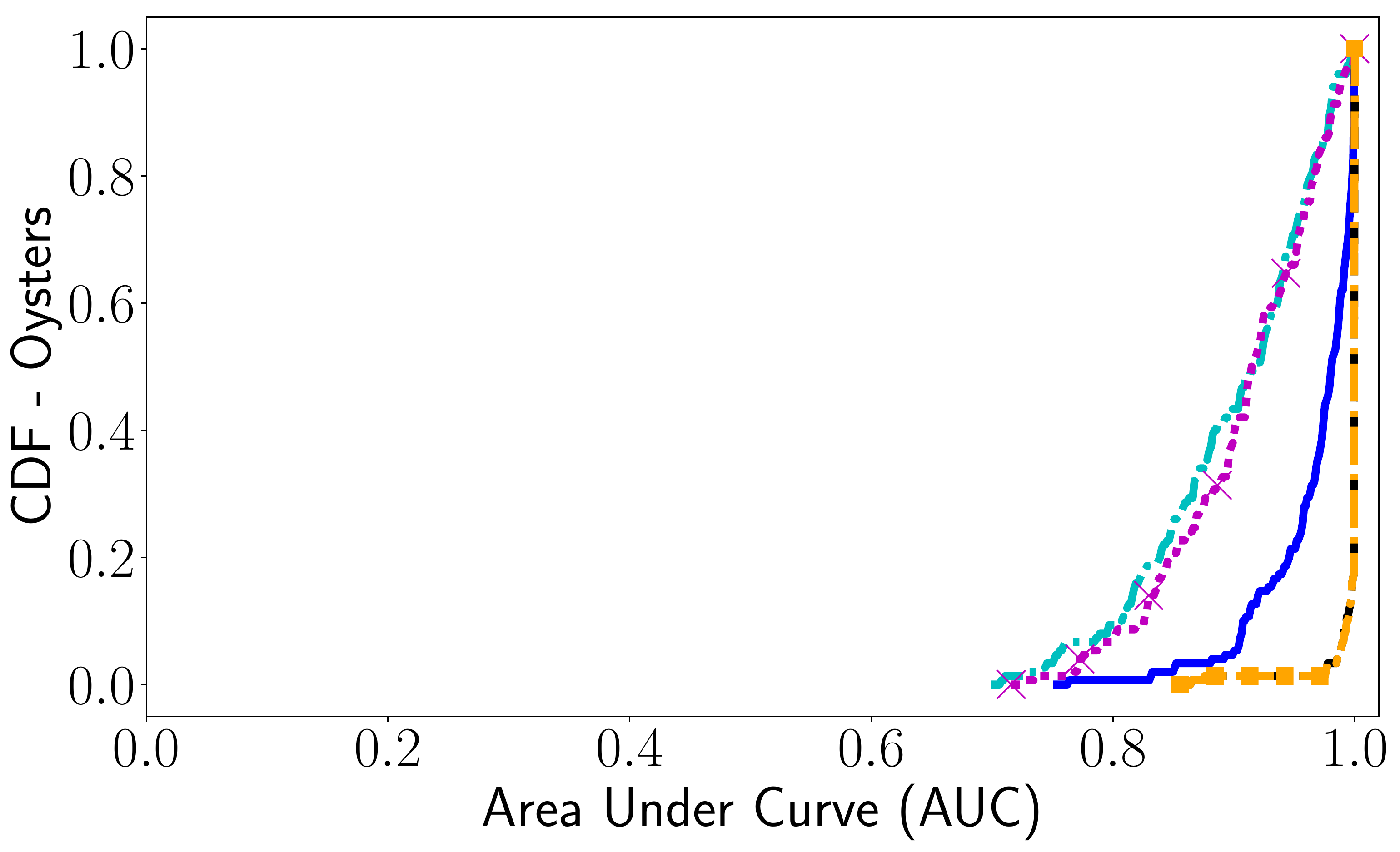}
        \caption{$m=10$}
        \label{fig:tfl-u-time-gr10}
    \end{subfigure}
    ~
    \begin{subfigure}[b]{0.245\textwidth}
        \includegraphics[width=\textwidth]{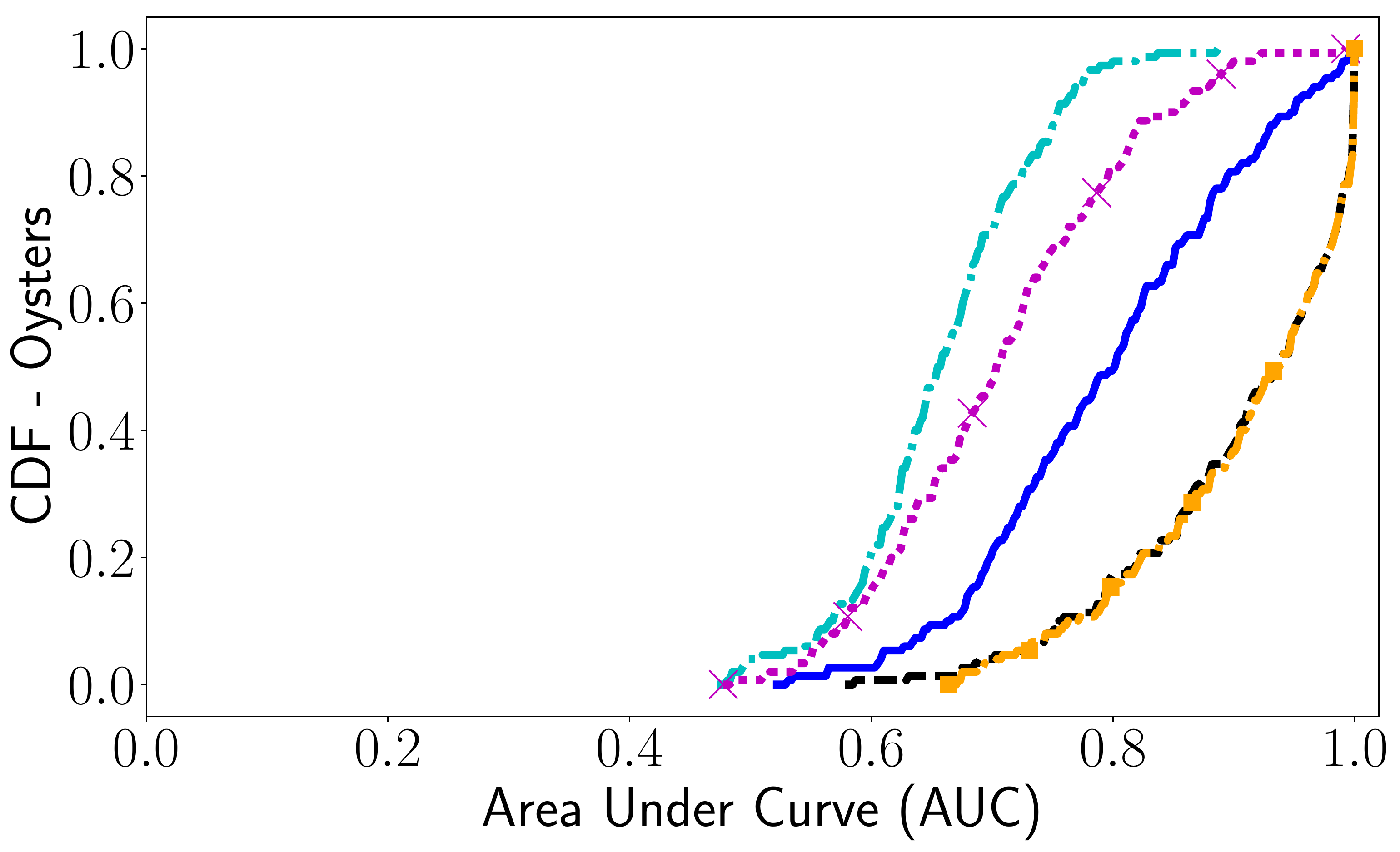}
        \caption{$m=50$}
        \label{fig:tfl-u-time-gr50}
    \end{subfigure}
	~
    \begin{subfigure}[b]{0.245\textwidth}
        \includegraphics[width=\textwidth]{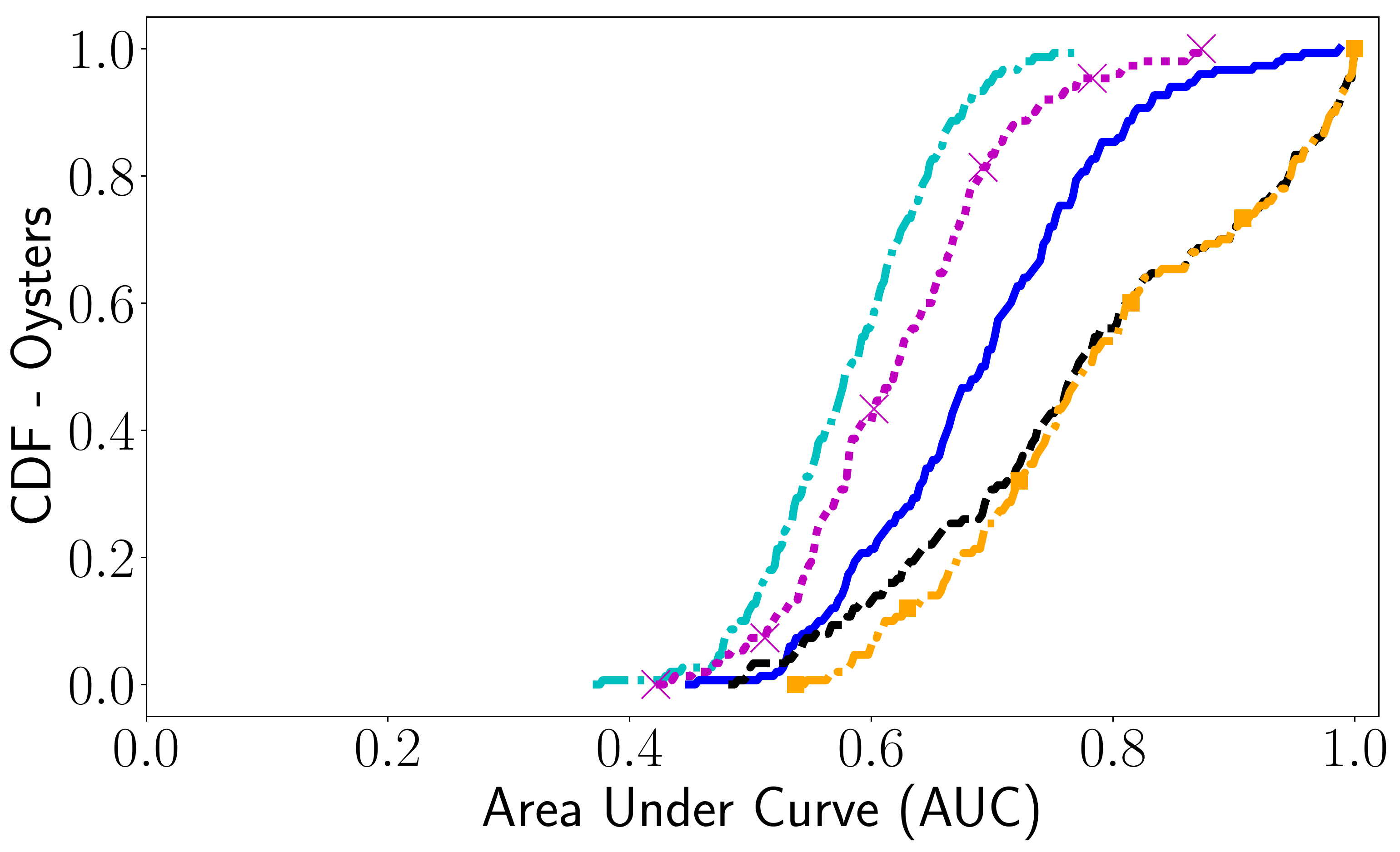}
        \caption{$m=100$}
        \label{fig:tfl-u-time-gr100}
    \end{subfigure}    
    ~
    \begin{subfigure}[b]{0.245\textwidth}
        \includegraphics[width=\textwidth]{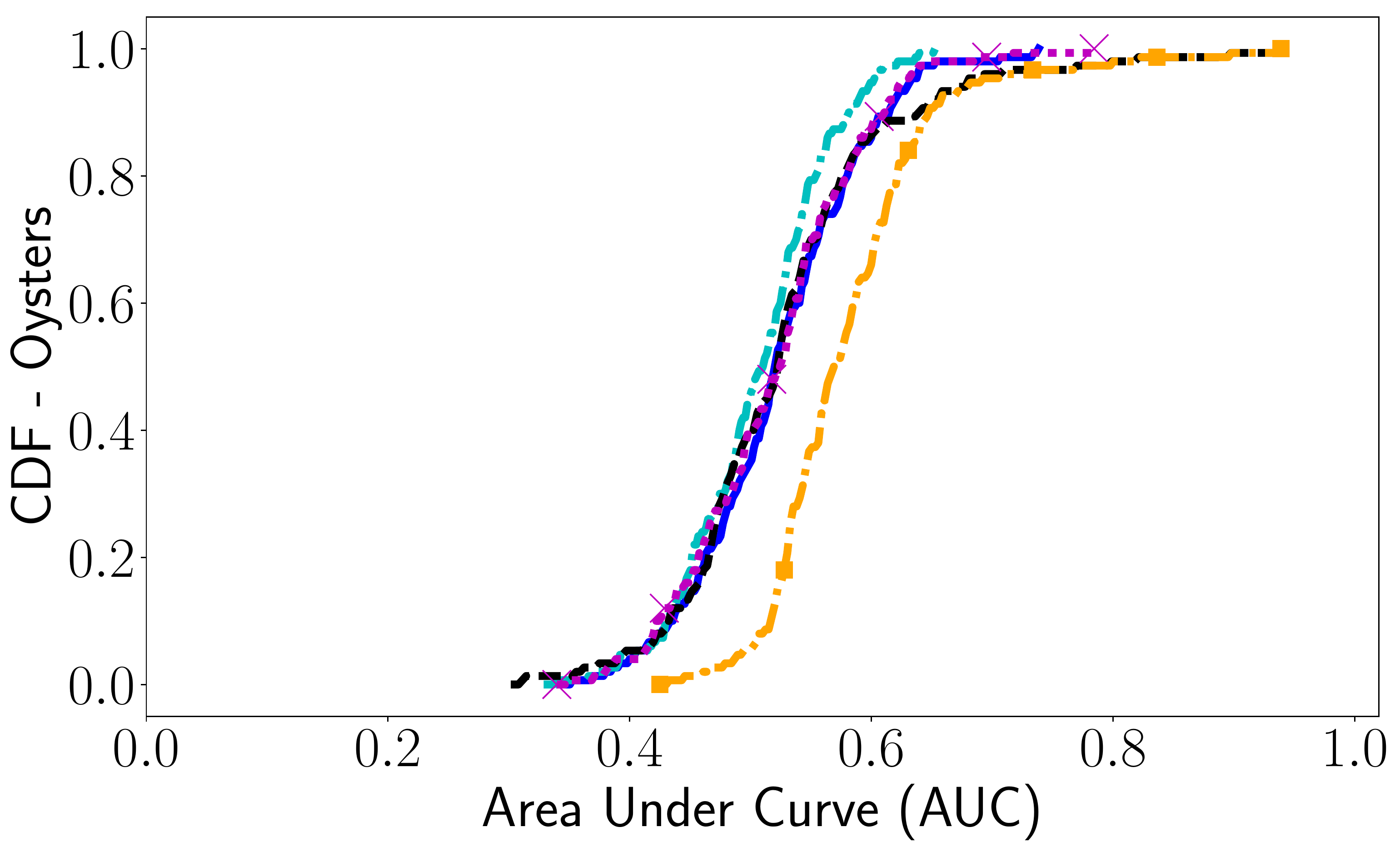}
        \caption{$m=1,000$}
        \label{fig:tfl-u-time-gr1000}
    \end{subfigure}
    ~
    \begin{subfigure}[b]{0.245\textwidth}
        \includegraphics[width=\textwidth]{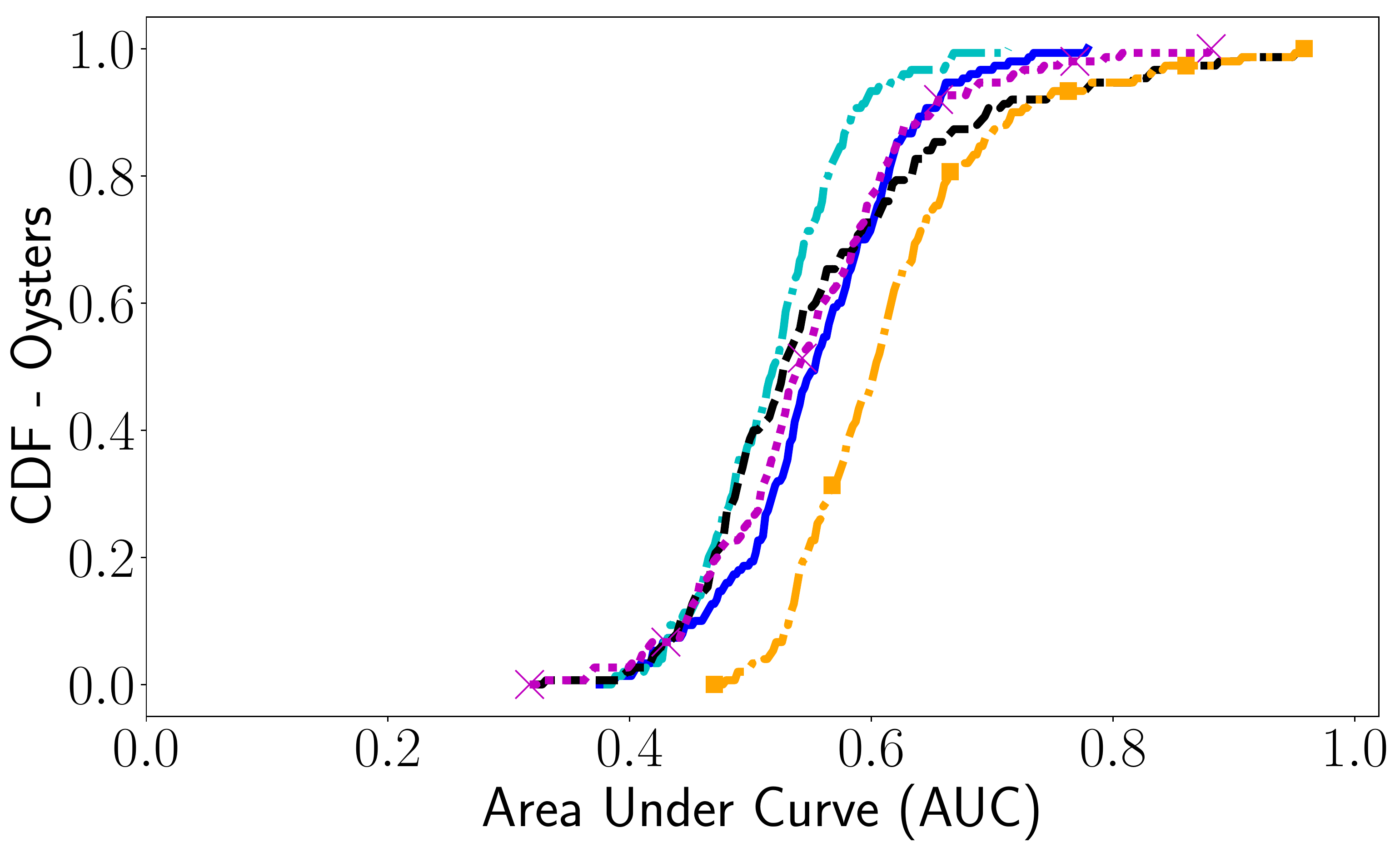}
        \caption{$m=9,500$}
        \label{fig:tfl-u-time-gr9500}
    \end{subfigure}
    \vspace{-0.3cm}
 \caption{{\em Different Groups than Released} prior (TFL, 75\%-25\% split, $\beta{=}300$, $|T_I|{=}168$) -- Adv's performance for different values of $m$.}   
    \label{fig:tfl-1week-u-time}
    \vspace{-0.2cm}
\end{figure*}
\begin{figure*}[h]
\centering
\begin{subfigure}[b]{0.4\textwidth}
\centering
\includegraphics[width=0.82\textwidth]{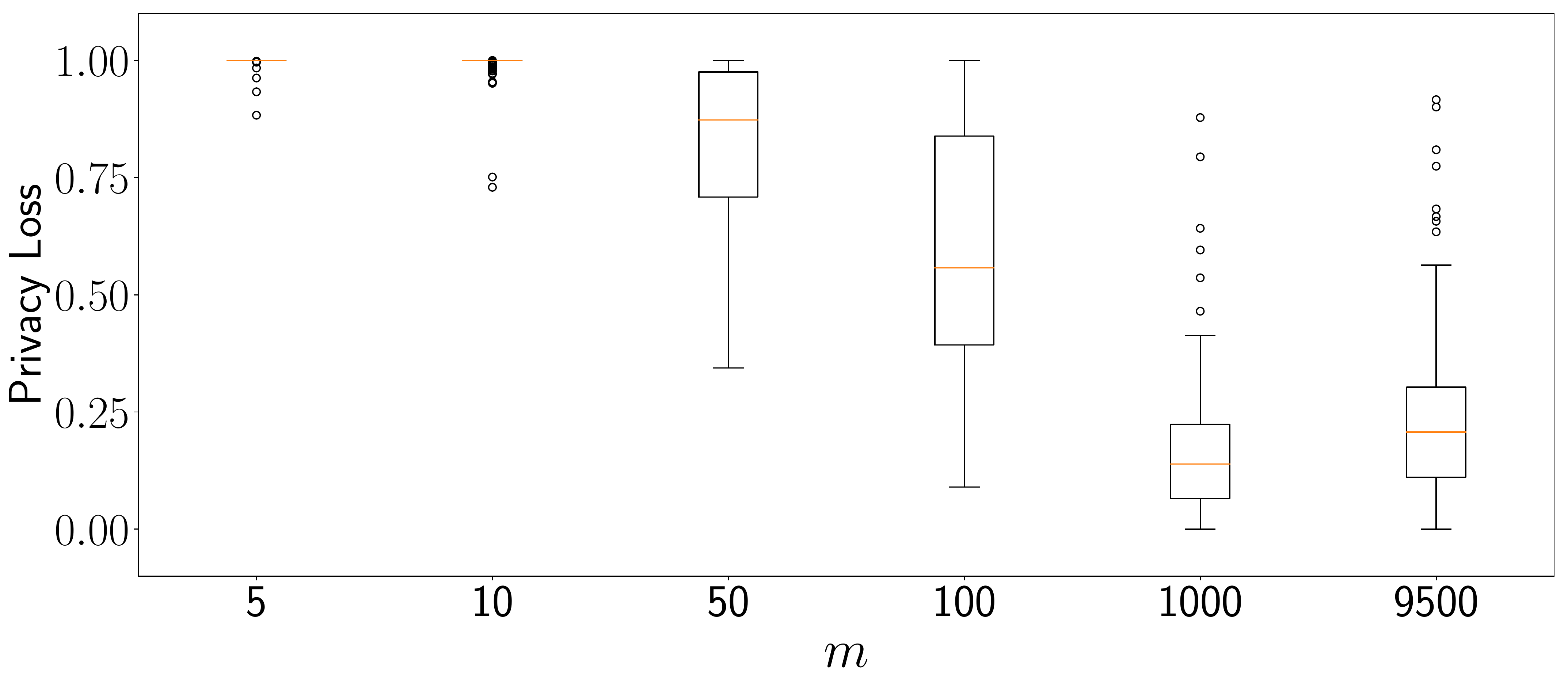}
\caption{TFL, $75\%-25\%$ split, $\beta=300$, $|T_I|=168$}
\label{fig:tfl-1week-u-time-pl}
\end{subfigure}
\centering
\begin{subfigure}[b]{0.4\textwidth}
\centering
\includegraphics[width=0.82\textwidth]{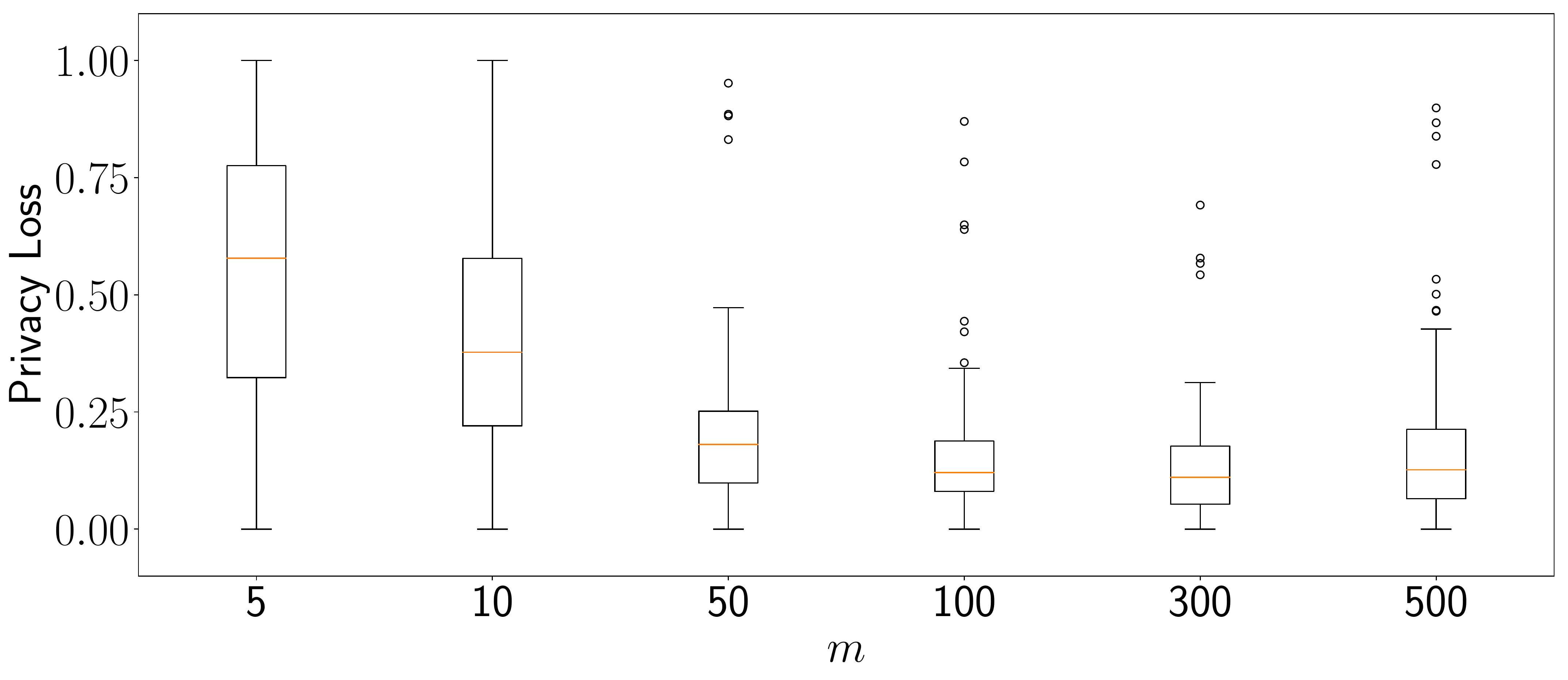}
\caption{SFC $67\%-33\%$ split, $\beta=300$, $|T_I|=168$}
\label{fig:sfc-1week-u-time-pl}
\end{subfigure}
\vspace{-0.3cm}
\caption{{\em Different Groups than Released} prior - Privacy Loss (PL) for different values of $m$.}
\vspace{-0.2cm}
\end{figure*}
\begin{figure*}[h]
  \centering
    \begin{subfigure}[b]{0.245\textwidth}
        \includegraphics[width=\textwidth]{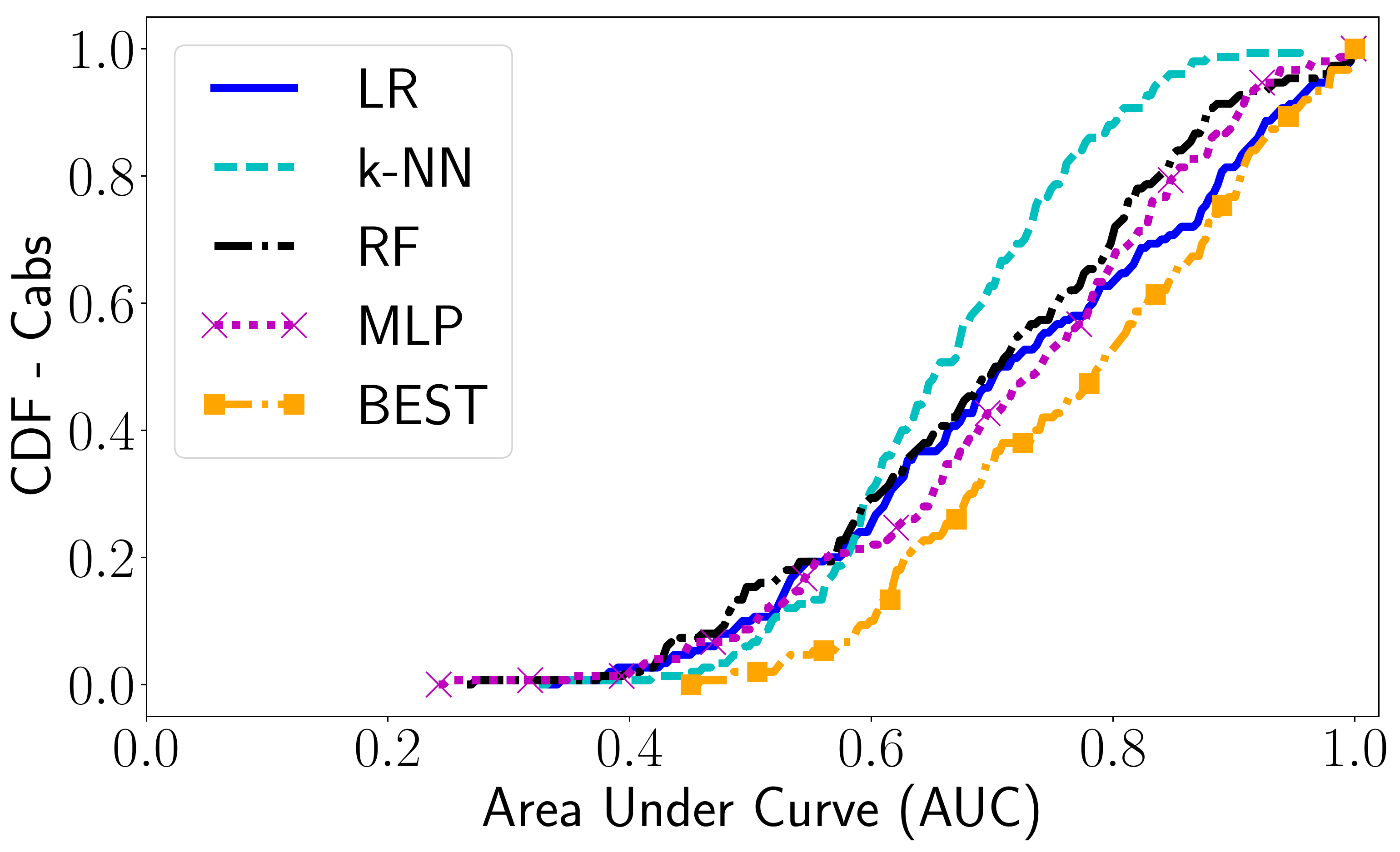}
        \caption{$m=5$}
        \label{fig:sfc-u-time-gr5}
    \end{subfigure}
	~
    \begin{subfigure}[b]{0.245\textwidth}
        \includegraphics[width=\textwidth]{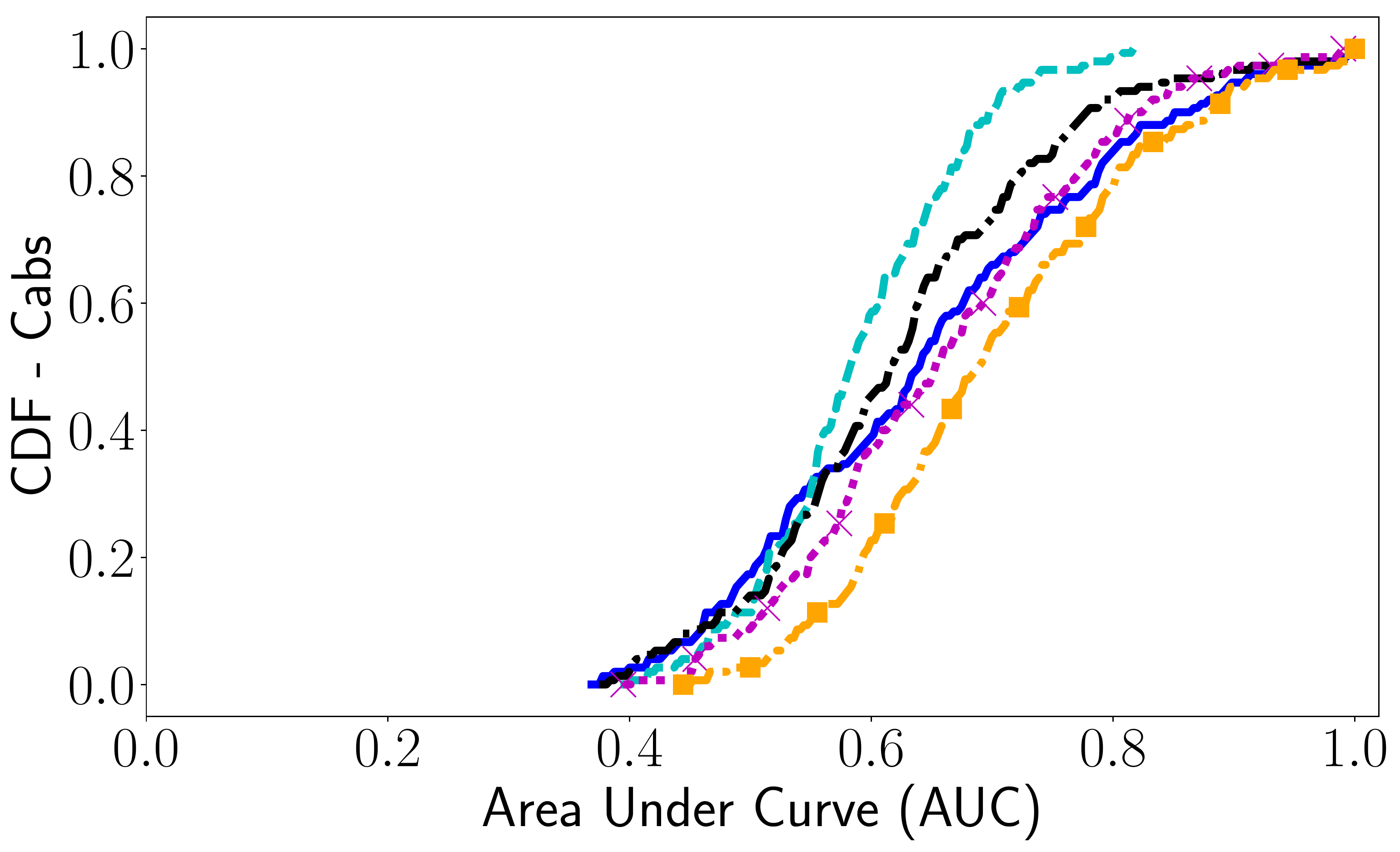}
        \caption{$m=10$}
        \label{fig:sfc-u-time-gr10}
    \end{subfigure}
    ~
    \begin{subfigure}[b]{0.245\textwidth}
        \includegraphics[width=\textwidth]{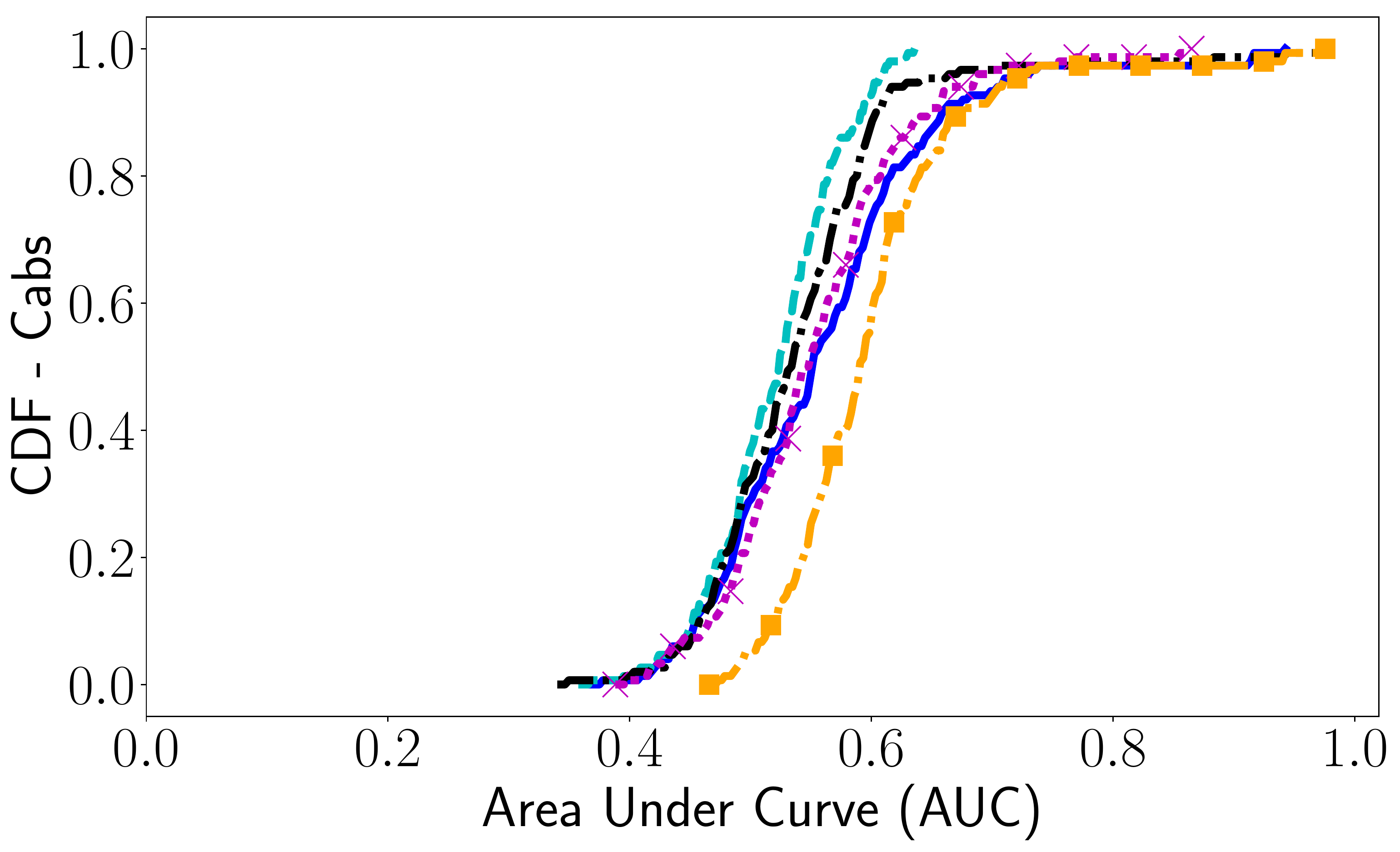}
        \caption{$m=50$}
        \label{fig:sfc-u-time-gr50}
    \end{subfigure}
	~
    \begin{subfigure}[b]{0.245\textwidth}
        \includegraphics[width=\textwidth]{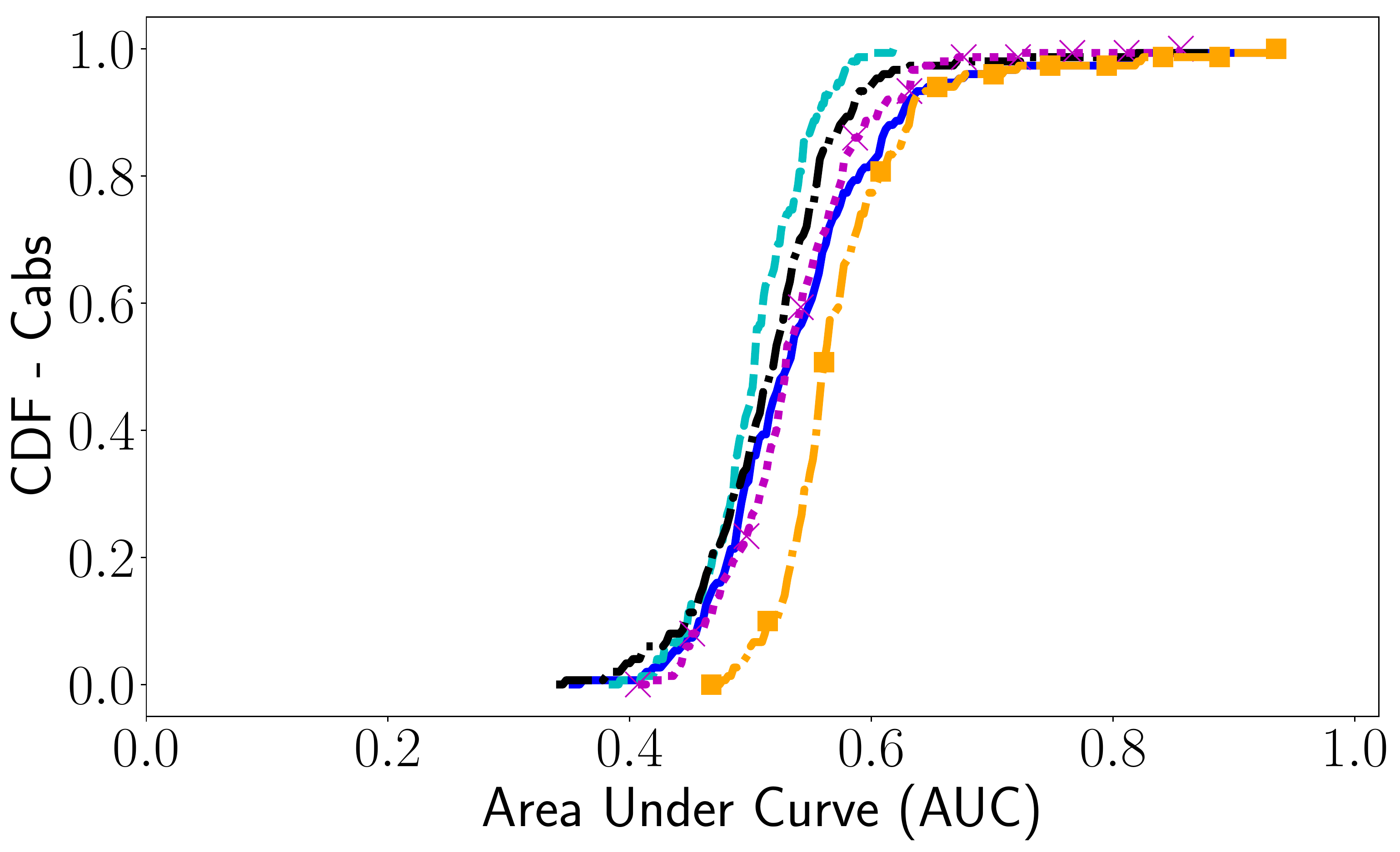}
        \caption{$m=100$}
        \label{fig:sfc-u-time-gr100}
    \end{subfigure}    
    ~
    \begin{subfigure}[b]{0.245\textwidth}
        \includegraphics[width=\textwidth]{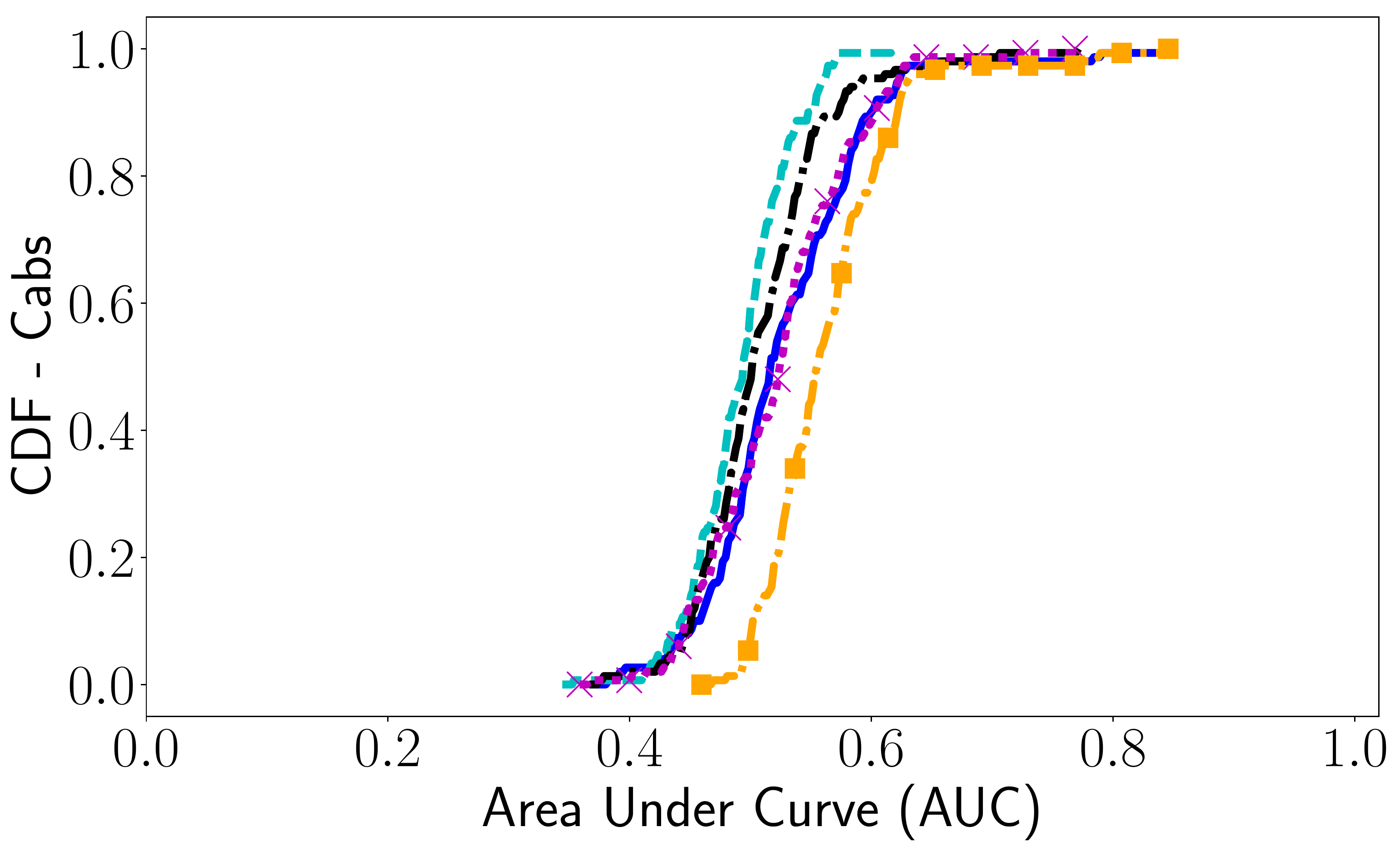}
        \caption{$m=300$}
        \label{fig:sfc-u-time-gr300}
    \end{subfigure}
    ~
    \begin{subfigure}[b]{0.245\textwidth}
        \includegraphics[width=\textwidth]{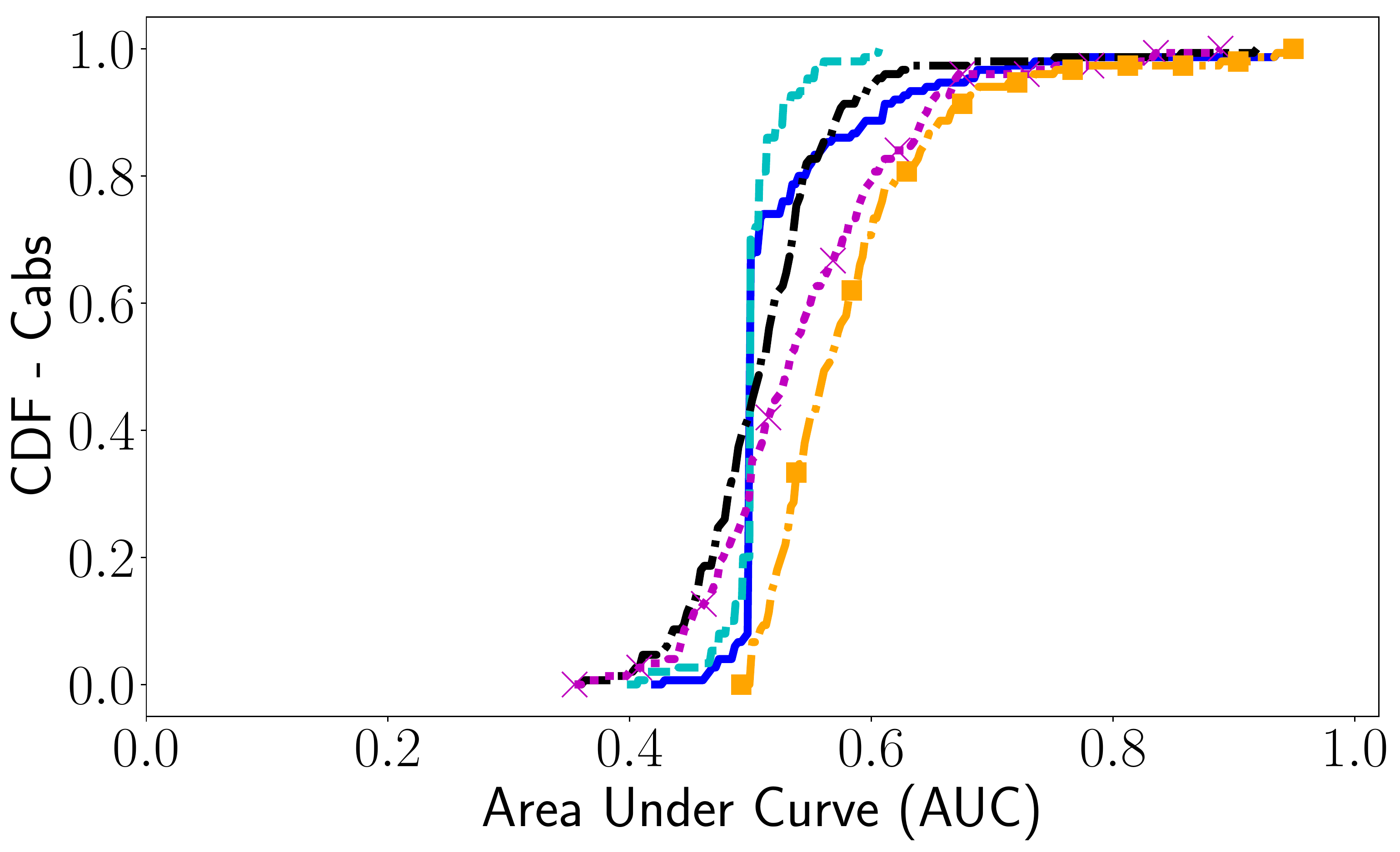}
        \caption{$m=500$}
        \label{fig:sfc-u-time-gr500}
    \end{subfigure}
    \vspace{-0.1cm}   
\caption{{\em Different Groups than Released} prior (SFC 67\%-33\% split, $\beta{=}300$, $|T_I|{=}168$) -- Adv's performance for different values of $m$.}
    \label{fig:sfc-1week-u-time}
    \vspace{-0.4cm}
\end{figure*}

\begin{figure*}[t]
	\centering
    \begin{subfigure}[t]{0.314\textwidth}
        \includegraphics[width=1.0\textwidth]{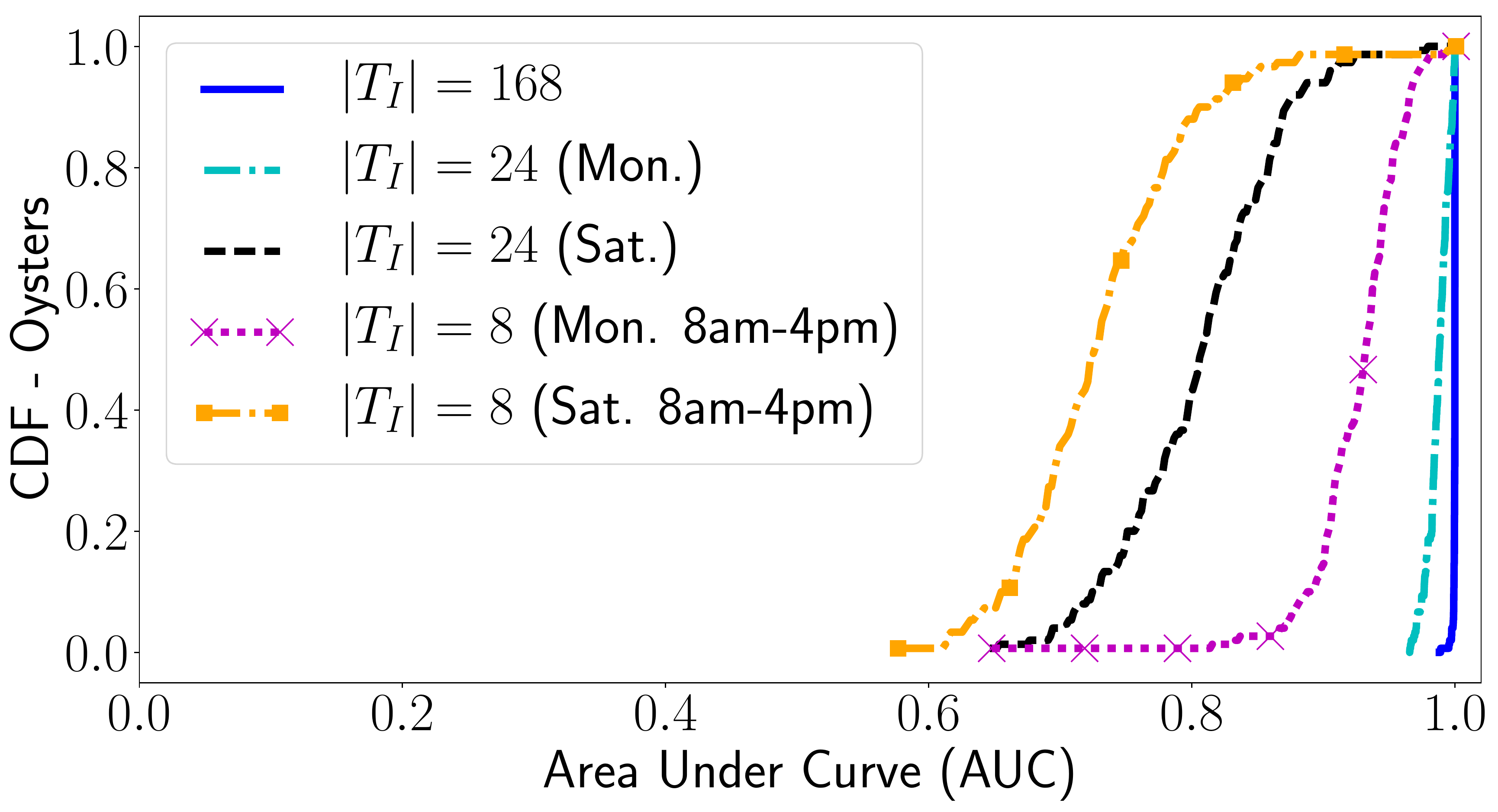}
        \caption{TFL (75\%-25\% split, $m=\text{1,000}$)}
        \label{fig:tfl-gr1000-dur}
    \end{subfigure}
    ~~~
	\begin{subfigure}[t]{0.314\textwidth}
		\includegraphics[width=1.0\textwidth]{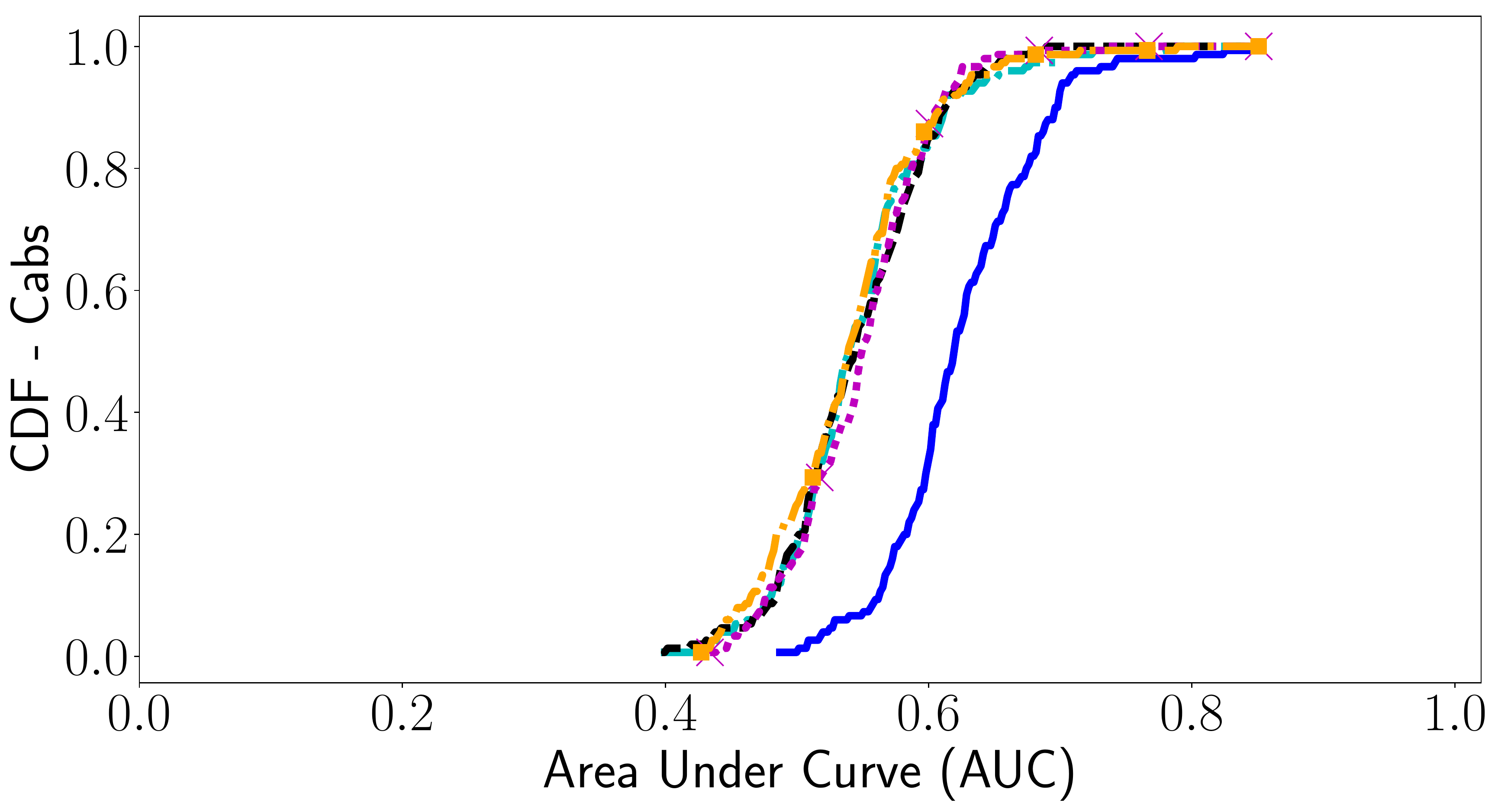}
		\caption{SFC (67\%-33\% split, $m=\text{100}$)}
		\label{fig:sfc-gr100-dur}
	\end{subfigure}\\
    \begin{subfigure}[t]{0.314\textwidth}
		\includegraphics[width=1.0\textwidth]{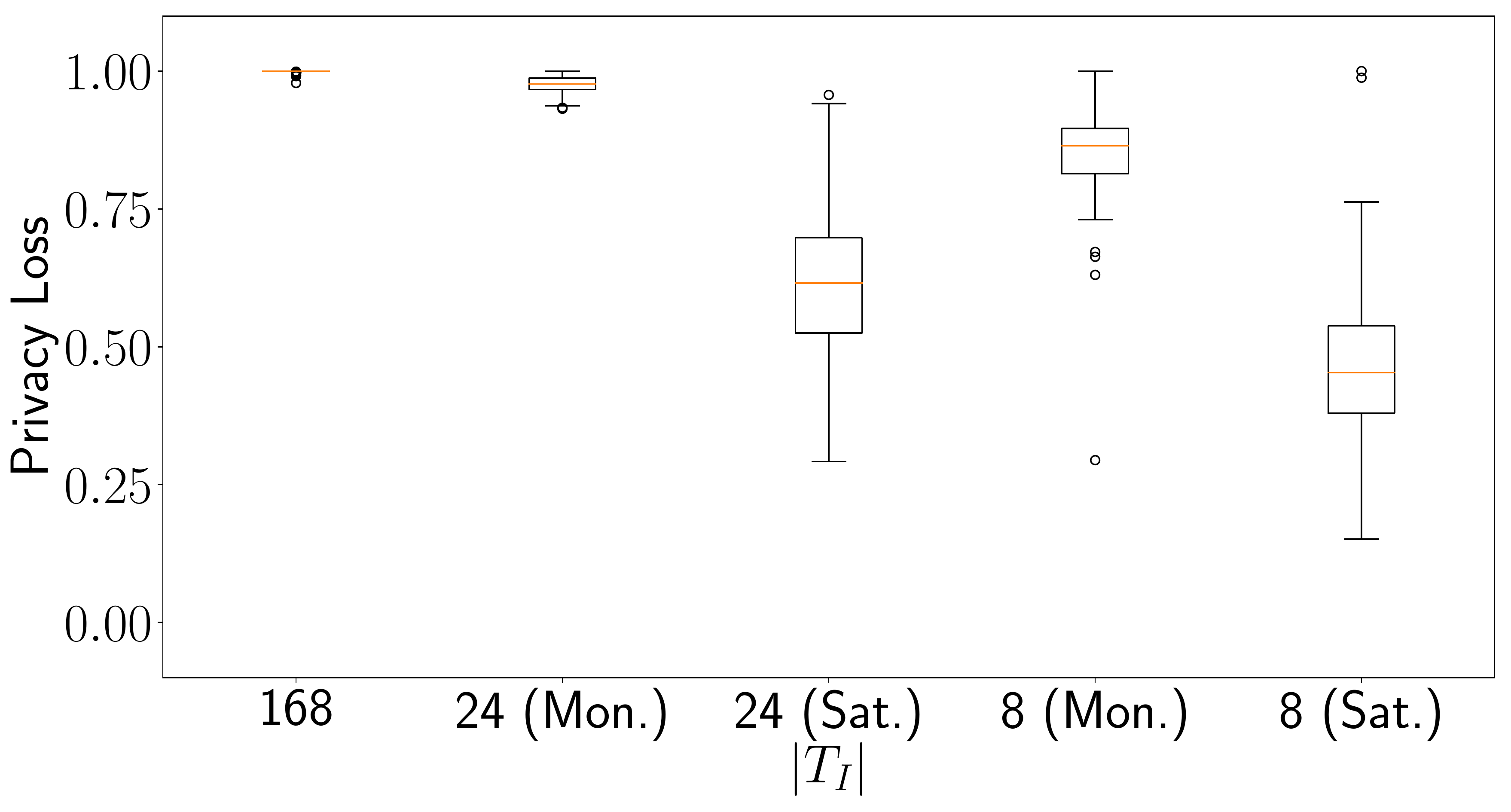}
		\caption{TFL (75\%-25\% split, $m=\text{1,000}$)}	
		\label{fig:tfl-gr1000-dur-pl}
	\end{subfigure}	
	~~~
	\begin{subfigure}[t]{0.314\textwidth}
		\includegraphics[width=1.0\textwidth]{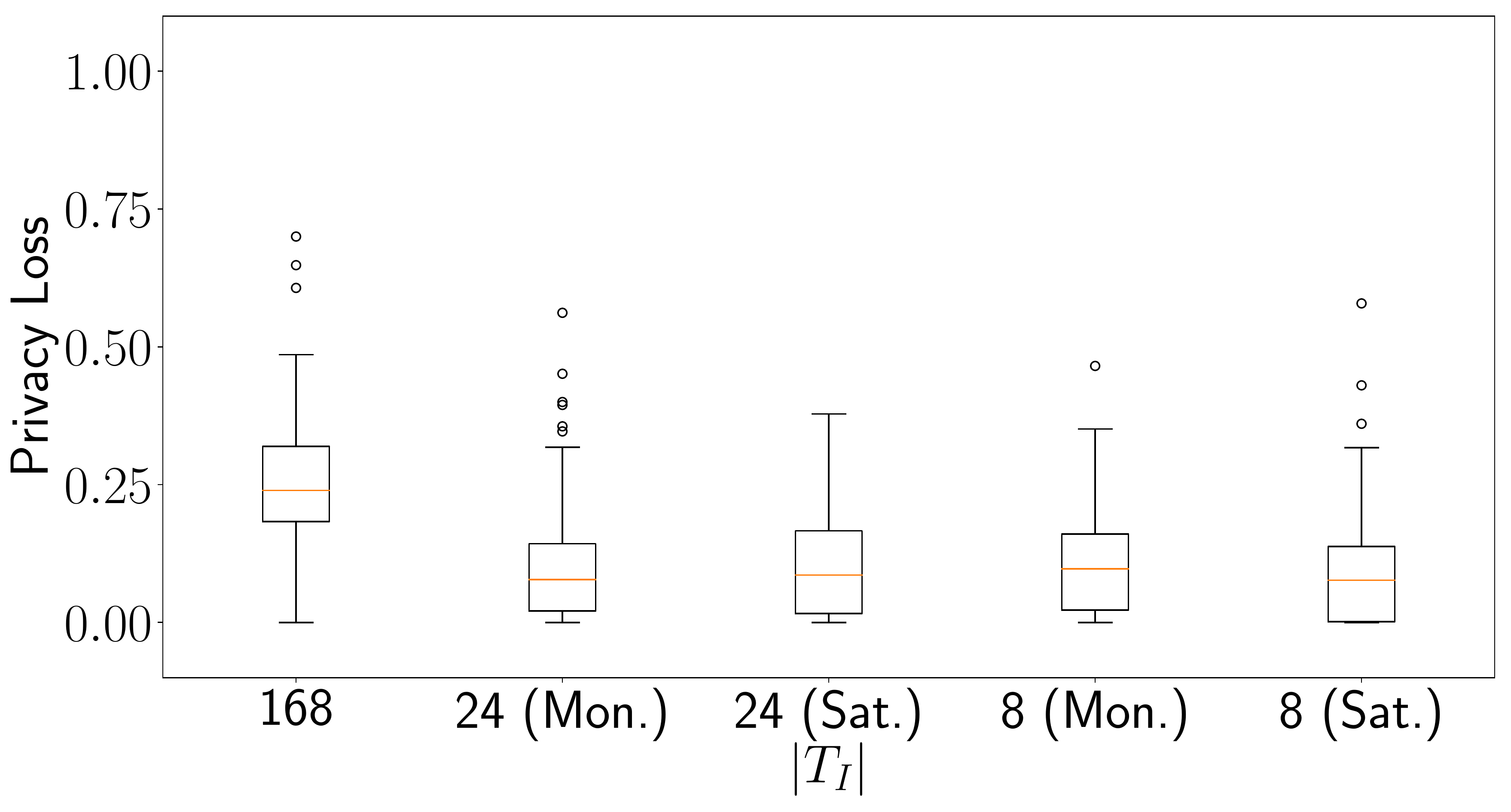}
		\caption{SFC (67\%-33\% split, $m=\text{100}$)}
		\label{fig:sfc-gr100-dur-pl}
	\end{subfigure}
	\vspace{-0.1cm}	
	\caption{{\em Same Groups as Released} prior ($\beta{=}150$) - Adv's performance for variable inference period length ($|T_I|$), on (a) TFL and (b) SFC, and Privacy Loss on (c) TFL and (d) SFC.}
	\vspace{-0.5cm}
\end{figure*}

\descr{\textit{Different} Groups than Released.} 
In this setting, for each target user, we design a balanced experiment by generating a dataset $D$ with the aggregates of 400 \emph{unique} randomly sampled groups -- half including the target and half not -- and set the corresponding participation label (in/out). 
Once again, the experiment size is chosen to provide enough data for our classifiers to learn patterns, while keeping the computation time reasonable on commodity hardware. To simulate the difference in groups between observation and inference period, we first perform a 75\%-25\% stratified random split of $D$, whereby we keep 300 groups for training and 100 for testing. Then, for TFL, we define the observation period to be the first 3 weeks of data (i.e., $|T_O|=$ 504) while for SFC the first 2 weeks ($|T_O| =$ 336) and in both cases, the inference period is the last week (i.e., $|T_I|=$ 168). We then split the training and testing sets according to time: from the training set, we keep {\em only} the aggregates of the observation period, while, from the testing set, {\em only} those from the inference period (i.e., overall, we perform a 75\%-25\% split for TFL and 67\%-33\% for SFC). That is, we let the adversary obtain knowledge for 300 user groups (i.e., $\beta=$ 300), half including her target, whose aggregates are generated during the observation period. Finally, we train the classifiers on the features extracted from the aggregates of the groups in the \emph{training} set for each week of the {\em observation period}, and test them against those extracted from the groups in the \emph{testing} set (during the {\em inference} period).

\descrit{TFL Dataset.} Fig.~\ref{fig:tfl-1week-u-time} illustrates the classifiers' performance for different aggregation group sizes (up to 9,500 commuters, since there are no restrictions). Once again, for small groups ($m=$ 5 or 10) membership can be easily inferred (AUC $>0.89$ for all classifiers).
As $m$ starts increasing, we first observe a small drop in the adversarial performance, with RF achieving mean AUC of 0.89 and 0.78 for groups of 50 and 100 commuters, resp. This indicates that regularity still helps membership inference in small groups even when these groups change. %
However, when $m$ reaches 1,000 all the classifiers perform, on average, similar to the baseline indicating that the effect of regularity dilutes. Interestingly, for $m=$ 9,500, we note a small increase in the classifiers' AUC scores due to the big user overlap across training and testing groups, i.e., the different-groups prior becomes more similar to the same-groups prior.

This effect can also be observed in terms of PL (Fig.~\ref{fig:tfl-1week-u-time-pl}). Membership inference is quite effective for groups of size up to 100, where commuters suffer a privacy loss of at least 0.59. However, when data of more commuters is aggregated, mean PL decreases to 0.17 for groups of 1,000, and it slightly increases to 0.22 when $m=$ 9,500. Overall, we note that the privacy loss is smaller in this setting, however, this is not surprising, since this is a much weaker adversarial setting than the previous ones.

\descrit{SFC Dataset.} Similar to the experiment with the same groups prior, we observe in Fig.~\ref{fig:sfc-1week-u-time} that the classifiers perform worse for SFC than TFL, due to the lack of regularity. Already for small groups ($m=$ 5) the mean AUC drops to $0.71$ for the best classifiers, LR and MLP. With larger groups, the performance is significantly lower, and all classifiers converge towards the random guess baseline. When $m=$ 500, MLP and LR yield slightly better results and membership inference can be achieved with AUC larger than 0.6 for only a small percentage of cabs (about 20\%).

From Fig.~\ref{fig:sfc-1week-u-time-pl}, we see that, due to the weaker prior, PL values are smaller across the board compared to the previous setting. Overall, PL decreases with increasing aggregation group size, ranging from mean PL of 0.54 with $m=$ 5 to 0.12 for $m=$ 300. Similar to the TFL case, we observe a small increase for groups of 500 cabs. The reason is the same, i.e., the user overlaps between training and testing groups slightly improve the effectiveness of the membership inference attack.

\subsection{Length of Inference Period} 
In the previous experiments, we have studied the effect of the size of the aggregation groups ($m$) on the success of membership inference, for various types of adversarial prior knowledge. In this section, we examine the {\em effect of the inference period length}, i.e., $|T_I|$. We consider lengths of 1 week (168 hourly timeslots), 1 day (24 timeslots), and 8 hours (8 timeslots). In particular, for the last two, we also consider working vs weekend days to account for the difference in mobility behavior.
Due to space limitations, we only report experiments in the setting where Adv has prior knowledge about the exact groups that are released by Ch -- i.e., prior (2a) in Section~\ref{sec:prior} -- and fix the group size to 1,000 commuters for TFL and to 100 cabs for SFC. 
For each target user, we create a dataset of $\beta=$ 150 random unique groups, half of which include the user and half of which do not, and split their aggregates in training and testing sets according to time following a 75\%-25\% split for TFL and a 67\%-33\% for SFC. We choose RF as classifier for TFL, and MLP for SFC, since they yield the best AUC scores in this setting, as shown in Figs.~\ref{fig:tfl-gr1000} and~\ref{fig:sfc-gr100-time}. 
For each $|T_I| \in \{\text{8, 24, 168}\}$, we train the classifiers on aggregates of that length, for {\em each week} in the training set (observation period), and evaluate them against the corresponding aggregates in the test set (inference period).

Fig.~\ref{fig:tfl-gr1000-dur} reports the results on the TFL dataset: as the number of points in the inference period $|T_I|$ decreases, the adversarial performance degrades as there is less information about mobility patterns to be exploited. Also, there is indeed a difference between working days and weekends. Mean AUC is 0.97 when training and testing on a Monday, and 0.8 on a Saturday. 
This seems to be due to regularity, as commuters' regular patterns during the week make them more susceptible to membership inference than sporadic/leisure activities over weekends.
This is confirmed by the classifier's performance for $|T_I|=$ 8, as we obtain much better results when the inference period is set to Monday 8am--4pm (AUC $=$ 0.91) than on Saturday during the same timeframe (AUC $=$ 0.72).

Once again, the lack of regularity affects negatively Adv's performance when attacking the SFC dataset (Fig.~\ref{fig:sfc-gr100-dur}). As for the length of the inference period, our results confirm that the inference task becomes harder with fewer points in time: mean AUC drops from 0.62 to 0.54 when $|T_I|$ goes from 1 week to 1 day. However, as cabs are never regular (their movements are mandated by client demand), we do not observe significant difference between working days and weekends, nor when considering full days vs 8h slots.

The Privacy Loss (PL) exhibits similar trends. For TFL (see Fig.~\ref{fig:tfl-gr1000-dur-pl}), the highest loss is observed when more points are available (0.98 on average when $|T_I|$ is 1 week), while the loss is reduced as the length of the inference period decreases and the adversary has less information. Also, we see how regularity in working days results in better membership inference attacks than during weekends, i.e., 
mean PL is 0.96 and 0.85 on Mondays vs 0.61 and 0.46 on Saturdays for $|T_I|$ set at 24 and 8 hours, respectively.
Finally, Fig.~\ref{fig:sfc-gr100-dur-pl} %
highlights smaller PL for the SFC cabs, for all period lengths, with a maximum mean PL of 0.25 when $|T_I|$ is 1 week, down to 0.1 and 0.09 for 1 day and 8 hours, respectively. There is no significant difference between Mondays and Saturdays, confirming that regularity has a strong influence on the problem.

\subsection{Raw Aggregates Evaluation -- Take-Aways}
\label{sec:take-aways}

Overall, our evaluation showcases the effectiveness of modeling membership inference attacks on aggregate location time-series as a classification task, vis-\`a-vis different datasets and priors. We show that an adversary can build a machine learning model by extracting features from known aggregate location time-series and use it to guess whether a target user has contributed to a set of previously unseen aggregates. Our results evidence that the risks stemming from such attacks are significant, with the actual level of privacy leakage depending on the adversary's prior knowledge, the characteristics of the data, as well as the group size and timeframe on which aggregation is performed.

We find that, up to certain aggregation group sizes, membership inference is very successful when the adversary knows the actual locations of a subset of users (including her target), or when she knows past aggregates for the same groups on which she tries to perform the inference. In the least restrictive setting, where the past groups known to the adversary are different than those whose statistics are released, privacy leakage is relatively small, but still non-negligible.

Moreover, the characteristics of the data used for aggregation also influence the adversarial performance: in general, privacy leakage on the dataset containing mobility of commuters (TFL) is larger than on the one including cab traces (SFC). This highlights that regularity in users' movements, as well as sparseness of the location signal, significantly ease the membership inference task.

Finally, the number of users that contribute to aggregation also has a profound effect on the adversarial performance. Unsurprisingly, membership inference is very successful when aggregation is performed over small groups, while users generally enjoy more privacy in larger groups. A notable exception is the TFL case where, due to the regularity of commuters, membership inference attacks are still very effective even for large groups. %
Also, the length, as well as the time semantics, of the inference period play a very important role. Inference is easier if the aggregates of longer periods are released (i.e., more information is available to extract patterns), and at times when mobility patterns are likely to be more regular (e.g., workdays or mornings).%

\section{Evaluating DP Defenses}
\label{sec:counter}
In this section, we evaluate the effectiveness of available defense mechanisms to prevent privacy leakage from membership inference attacks on aggregate location time-series.

\subsection{Differential Privacy (DP)}
\label{sec:dp-prelim}
The established framework to define private functions that are free from inferences is Differential Privacy (DP)~\cite{dwork2008differential}.
Applying differentially private mechanisms to a dataset ensures that only a bounded amount of information is disclosed upon its release. This can mitigate membership inference attacks, as DP's indistinguishability-based definition guarantees that the outcome of any computation on a dataset is insensitive to the inclusion of any data record (user) in the dataset. 

Formally, we say that a randomized algorithm $\mathcal{M}$ is 
\textit{$(\epsilon, \delta)$-differentially private} if for all datasets $D_1$ and $D_2$ that differ on at most one element, and all $\mathcal{S} \subseteq \text{Range}(\mathcal{M})$:
\begin{equation}
\Pr[\mathcal{M}(D_1) \in \mathcal{S}] \leq e^{\epsilon} \times \Pr[\mathcal{M}(D_2) \in \mathcal{S}] + \delta
\end{equation}
where the probability is calculated over the coin tosses of $\mathcal{M}$ and $\text{Range}(\mathcal{M})$ denotes the set of possible outcomes of $\mathcal{M}$. In other words, the outcome of $\mathcal{M}$ has a very small dependence on the members of the dataset. If $\delta=0$, we say that $\mathcal{M}$ is $\epsilon$-differentially private~\cite{dwork2008differential}.

We also review the notion of {\em sensitivity}, which captures how much one record affects the output of a function.
Formally, for any function $f: \mathcal{D} \rightarrow \mathbb{R}^d$, the sensitivity of $f$ is: 
\begin{equation}
\Delta f_{p} = \underset{D_1, D_2}{\text{max}} \parallel f(D_1) - f(D_2) \parallel_{p}
\end{equation}
for all datasets $D_1, D_2$ differing on at most one element, with $\parallel \cdot \parallel_{p}$ indicating the $\ell_{p}$ norm.

\descrit{Laplacian Mechanism (LPA):} The most common used mechanism to achieve DP is to randomize the aggregate statistics to be released using random noise independently drawn from the Laplace distribution. For a function $f: \mathcal{D} \rightarrow \mathbb{R}^d$, a randomized mechanism $\mathcal{M}$ defined as $\mathcal{M}(X) = f(X) + \text{Lap}(0, \sigma)^{d}$ guarantees $\epsilon$-DP if $\sigma \geq \frac{\Delta f_{1}}{\epsilon}$. 
A {\em weaker} version of LPA has been proposed for time-series, which perturbs the counts of a time-series with noise distributed according to Lap($1/\epsilon$)~\cite{chan2011private}. %
However, this mechanism only provides \textit{event-level privacy}~\cite{dwork2010differential}, i.e., in our setting, it can only protect single location visits. We include it as a {\em baseline} for our evaluation, since we do not expect it to perform well for full time-series.

\descrit{Gaussian Mechanism (GSM):} Another mechanism consists in perturbing the statistics with random noise drawn from the Gaussian distribution $\mathcal{N}$. Given $ f: \mathcal{D} \rightarrow \mathbb{R}^d$, a randomized mechanism $\mathcal{M}$ defined as $\mathcal{M}(X) = f(X) + \mathcal{N}(0, \sigma^{2})^d$ provides $(\epsilon, \delta)$-DP when $\sigma \geq \frac{\sqrt{2 \cdot ln(2 / \delta)}}{\epsilon} \cdot \Delta f_{2}$~\cite{dwork2006our}. Note that this is a weaker privacy guarantee than the one offered by LPA.

\descrit{Fourier Perturbation Algorithm (FPA):} We then consider differentially private mechanisms proposed specifically for  time-series settings. One is FPA~\cite{rastogi2010differentially}, which performs the noise addition on the compressed frequency domain: a time-series is compressed using the Discrete Fourier Transform (DFT) and the first $\kappa$ Fourier coefficients $F_{\kappa}$ are kept. Then $F_{\kappa}$ is perturbed with noise distributed according to $\text{Lap}(\sqrt{\kappa} \cdot \Delta f_{2} / \epsilon)$ and padded with zeros to the size of the original time-series. Finally, the inverse DFT is applied to obtain the perturbed time-series. 
As per~\cite{rastogi2010differentially}, FPA provably guarantees $\epsilon$-DP.

\descrit{Enhanced Fourier Perturbation Algorithm with Gaussian Noise (EFPAG):} EFPAG~\cite{acs2014case} improves FPA by choosing the number of coefficients ($\kappa$) to be perturbed probabilistically, and using the exponential mechanism to assign larger probability to values that minimize the root-sum-squared error between the input time-series and its noisy version. Then, rather than DFT, it uses the Discrete Cosine Transform (DCT) and employs Gaussian noise instead of Laplacian to achieve better accuracy. As a result, EFPAG guarantees $(\epsilon, \delta)$-DP~\cite{acs2014case}.

\vspace{-0.1cm}
\subsection{Experimental Design}
\label{sec:dp-setup}
\descr{Intuition.} Our evaluation of membership inference on raw aggregate location time-series showed that releasing them poses a significant privacy threat for users whose times-series are aggregated, and more so in settings where aggregation is performed over small groups. 
In this section, we present experiments aiming to evaluate the effectiveness of differentially private mechanisms in defending against such inferences. 
Note that we opt to evaluate them over large groups, %
since we expect (and have also verified experimentally) that, for small groups, the loss of utility incurred by DP-based mechanisms is prohibitively high. This is because the {\em sensitivity} of the location aggregation function, which directly affects the amount of noise to be employed, does not depend on the group size ($m$). As such, the aggregate time-series of groups with few users, which naturally have small counts, are affected more by the noise required by the DP-based mechanism.%

As the large group size gives the defense mechanism an ``advantage'' in terms of utility, and since DP provides protection against arbitrary risks~\cite{dwork2008differential}, we set to run our experiments considering a worst-case adversary that obtains \textit{perfect} prior knowledge for the users, i.e., she knows the inference period aggregates of the groups that are released by the challenger as well as the target user's membership in these groups. %
With this knowledge, Adv is able to train an accurate machine learning classifier that, upon release of raw statistics, always guesses correctly the target user's membership -- i.e., achieves an AUC score of $1$. %
In other words, we evaluate the privacy/utility trade-off of differentially private mechanisms considering the best setting for utility and the worst one for privacy.

\descr{Experiments.} We slightly modify the DG game in Fig.~\ref{fig:dg}, such that Ch applies a differentially private mechanism on the aggregates, before sending her \textit{challenge} to Adv. We evaluate the gain in privacy offered by the mechanisms on two cases, depending on whether Adv's classifier (which, once again, instantiates the distinguishing function) is trained on (1) the raw aggregates of the groups to be released; or (2) noisy aggregates of the groups to be released using the defense mechanism under examination. 

In both cases, %
testing is done on the aggregates of the released groups, perturbed with the defense mechanism (using \textit{fresh} noise). The first scenario represents a \textit{passive} adversary that attempts to infer user membership on the noisy aggregates, exploiting only the raw aggregate information from her prior knowledge. The second one, represents a strategic \textit{active} adversary that tries to mimic the behavior of the defender during training, knowing the parameters of the defense mechanism (i.e., $\epsilon$ and the sensitivity of the aggregation function denoted as $\Delta$), but not knowing the concrete values used in the defender's perturbation. We follow the same procedure as in Section~\ref{sec:setup}, i.e., we extract features from the aggregate location time-series of the user groups, and use RFE to reduce the number of features to the number of samples.

\descr{Settings.} We run membership inference attacks against the 150 users sampled from each dataset (cf. Section~\ref{sec:datasets}). We take as observation/inference period the first week in each dataset, i.e. $|T_O| = |T_I| = 168$. 
Aiming to examine a favorable setting for the utility of DP-based mechanisms, we construct large user groups: we set $m=\text{9,500}$ for TFL, and $m=\text{500}$ for SFC. Then, we generate the dataset $D$ by randomly sampling 200 and 400 elements for TFL and SFC, respectively, half including the target and half not. We pick a different number of groups for practical reasons: the TFL dataset has six times more ROIs than the SFC one, and this makes feature extraction significantly more expensive. 
As classifier, we employ MLP, which performed well overall %
in the previous experiments. 

To configure the perturbation mechanisms, we calculate the sensitivity $\Delta$ for the users in each dataset (i.e., the maximum number of ROIs reported by an oyster/cab in the inference week), obtaining $\Delta=$ 207 for TFL and $\Delta=$ 2,685 for SFC. We consider $\epsilon$ values of DP in the range $\{ 0.01, 0.1, 1.0, 10.0\}$, and set $\delta=$ 0.1 for GSM and EFPAG. For FPA, we empirically find the best value for $\kappa$ in terms of utility, setting $\kappa=$ 25 for TFL, and $\kappa=$ 20 for SFC.

\descr{Metrics.} Our evaluation uses the following privacy and utility metrics to capture the amount of privacy gained compared to a setting where the DG game is played on raw aggregates, and the utility lost due to the noise addition.

\descrit{Privacy Gain (PG):}
We define PG as the relative decrease in a classifier's AUC score when tested on perturbed aggregates ($\text{AUC}_{A'}$) versus raw aggregates ($\text{AUC}_{A}$):
\begin{equation}
\text{PG} = \begin{cases}
\frac{\text{AUC}_{A} - \text{AUC}_{A'}}{\text{AUC}_{A} - 0.5} ~~~~~\text{if} ~ \text{AUC}_{A} > \text{AUC}_{A'} \geq 0.5  \\
0 ~~~~~~~~~~~~~~~~~~\text{otherwise} \\
\end{cases}
\end{equation}
PG is a value between $0$ and $1$, which captures the decrease in adversarial performance, i.e., how much the adversary's inference power deteriorates towards the random guess baseline, when a defense mechanism is implemented. 

\descrit{Mean Relative Error (MRE):}
We evaluate the utility loss as a result of using DP-based defense mechanisms by means of the standard MRE metric, computed between the raw aggregate time series $Y$, of length $n$, and its perturbed version $Y'$:
\begin{equation}
\text{MRE}(Y, Y') = \frac{1}{n} \sum_{i=1}^{n} \frac{ | Y'_{i} - Y_{i} | }{ \text{max}(\gamma, Y_{i}) }
\end{equation}
where $\gamma$ is a sanity bound mitigating the effect of very small counts. Following previous work~\cite{acs2014case}, we set $\gamma$ to $0.1\%$ of $\sum_{i=1}^{n} Y_{i}$. We compute the MRE over the aggregate time-series for all ROIs ($s \in S$) in our datasets and report the mean value. %

\subsection{Results}

\descr{Utility.} We first report the utility measured as per the Mean Relative Error (MRE) of the TFL and SFC perturbed aggregates, for each mechanism and different values of $\epsilon$, in Tables~\ref{table:tfl-mre} and~\ref{table:sfc-mre}.
Naturally, as $\epsilon$ increases, so does utility. %
Note that LPA($\Delta$/$\epsilon$) incurs the highest MRE, with the noisy aggregate values being 8 (resp., 41) times less accurate than raw aggregates on TFL (resp., SFC) data, in the most relaxed privacy setting ($\epsilon=10$). With GSM, utility does not improve much. On the other hand, FPA and EFPAG achieve better results, e.g., MRE is under 1.1 for values of $\epsilon \geq 1$. Finally, note that LPA($1 / \epsilon$) achieves the best utility, but it provides poor privacy protection against membership inference attacks, as shown below.

\begin{table}[t]
\centering
\footnotesize
\begin{tabular}{  l  r  r  r  r  }
\toprule
\multicolumn{1}{r}{\boldmath$\epsilon$}  & \textbf{0.01} & \textbf{0.1} & \textbf{1.0} & \textbf{10} \\
\midrule
\textbf{LPA}(\boldmath$\Delta / \epsilon$) & 3056.1 & 812.6 & 81.7 & 8.2 \\
\textbf{GSM} & 753.2 & 75.8 & 7.4 & 0.75 \\
\textbf{FPA} & 67.2 & 6.1 & 0.7 & 0.03 \\
\textbf{EFPAG} & 36.8 & 3.6 & 0.4 & 0.03 \\
\textbf{LPA}(1 / \boldmath$\epsilon$) & 38.5 & 3.7 & 0.3 & 0.002 \\
\bottomrule
\end{tabular}
\vspace{-0.1cm}
\caption{MRE of aggregate location time-series with different differentially private mechanisms and parameter $\epsilon$ (TFL).}
\label{table:tfl-mre}
\vspace{-0.2cm}
\end{table}

\descr{Privacy.} We now evaluate the Privacy Gain (PG) provided by the different DP-based mechanisms, distinguishing between the two settings introduced in Section~\ref{sec:dp-setup}.
  
\begin{figure*}[t]
\centering
	\begin{subfigure}[b]{0.4\textwidth}
		\includegraphics[width=1.0\textwidth]{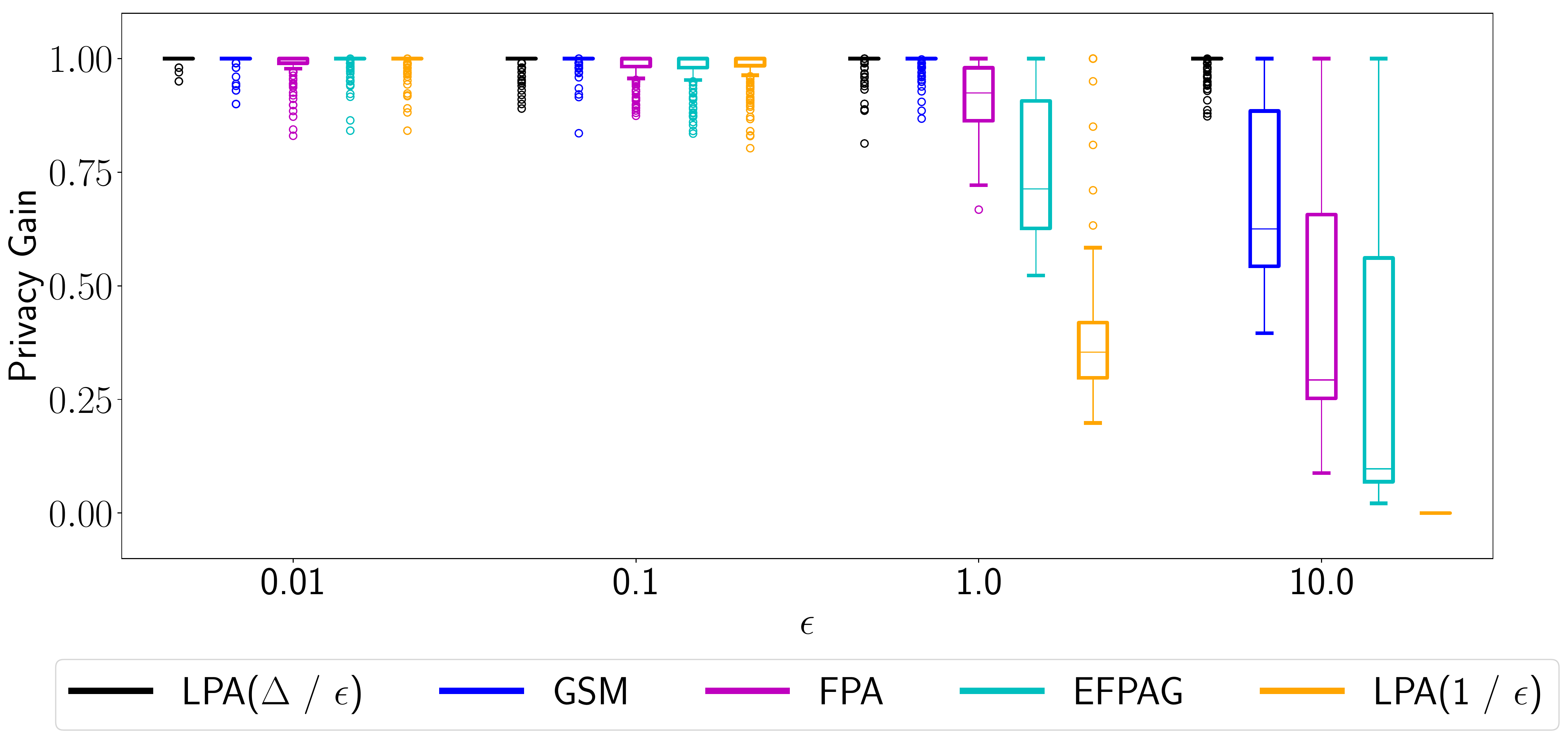}
		\caption{TFL ($m=9,500$, $|T_I|=168$)}
		\label{fig:tfl-pg}
	\end{subfigure}
	~
	\begin{subfigure}[b]{0.4\textwidth}
		\includegraphics[width=1.0\textwidth]{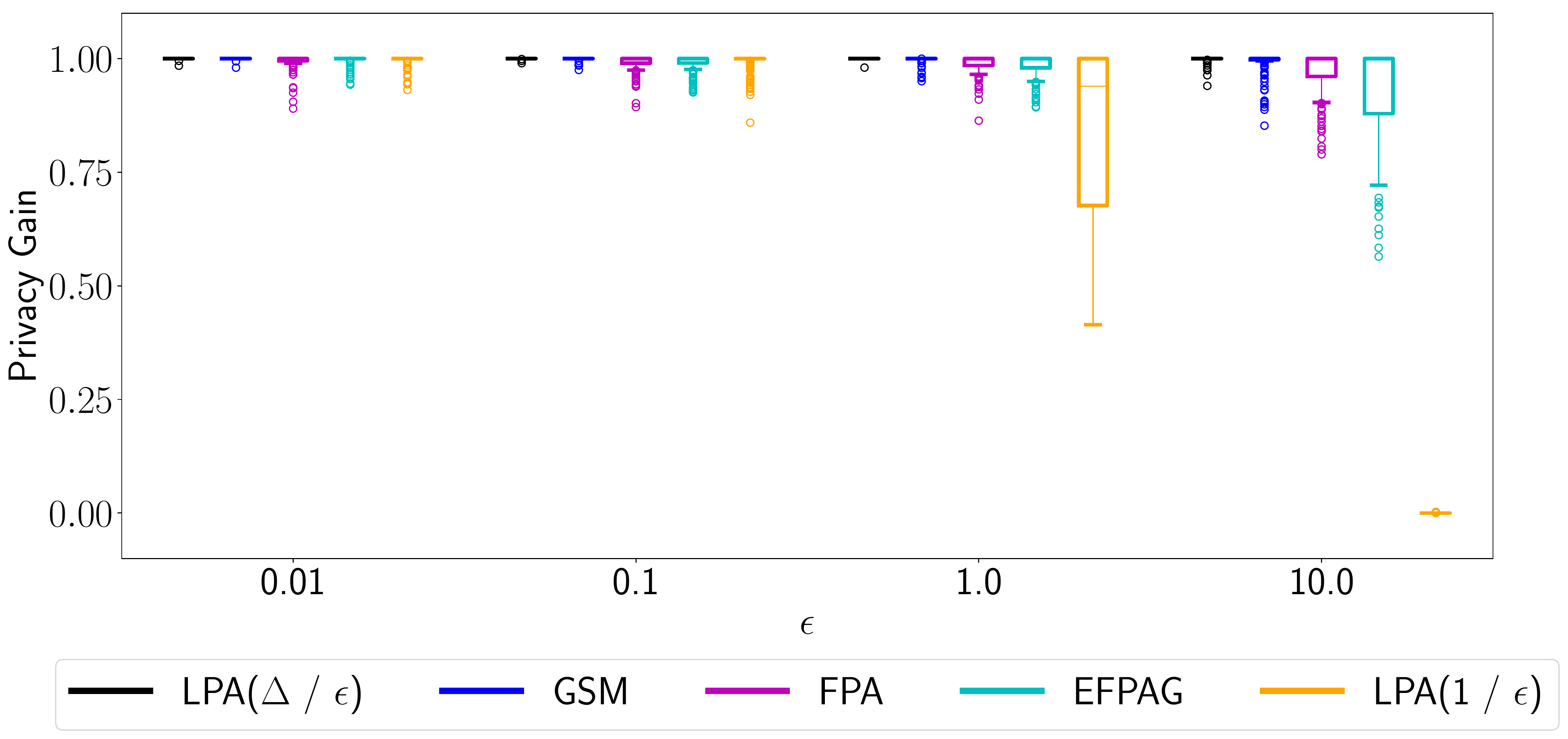}
		\caption{SFC ($m=500$, $|T_I|=168$)}
		\label{fig:sfc-pg}
	\end{subfigure}
\vspace{-0.3cm}	
\caption{Privacy Gain (PG) achieved by differentially private mechanisms with different values of $\epsilon$, against a MLP classifier trained on raw aggregates and tested on noisy aggregates.}
\label{fig:tfl-sfc-pg}
\vspace{-0.2cm}
\end{figure*}

\descrit{Train on Raw / Test on Noisy Aggregates.} Fig.~\ref{fig:tfl-sfc-pg} plots the PG achieved by various mechanisms against a MLP classifier trained on raw aggregates and tested on perturbed ones.
For TFL (Fig.~\ref{fig:tfl-pg}), we observe that for low $\epsilon$ values (up to 0.1) all mechanisms provide excellent privacy protection, achieving Privacy Gain (PG) close to 1. However, this protection comes with poor utility, as shown in Table~\ref{table:tfl-mre}. As $\epsilon$ increases to $1$, LPA($\Delta / \epsilon$) and GSM still provide good protection, while we observe a small drop in PG for the mechanisms that achieve MRE $<$ 1. In particular, FPA now yields a mean PG of 0.9, while EFPAG and LPA($1/\epsilon$) 0.75 and 0.38, resp. When $\epsilon=10$ and the utility of FPA and EFPAG is good, the decrease in PG is bigger (0.45 and 0.3, resp.), as expected.

With SFC data (Fig.~\ref{fig:sfc-pg}), we find that PG for all the mechanisms stays high for values of $\epsilon$ up to $1$. This is reasonable, since the sensitivity, and thus, the magnitude of noise required, is much larger on SFC compared to TFL ($\Delta=\text{2,685}$ vs 207). However, as seen from Table~\ref{table:sfc-mre}, utility is quite poor in these settings. 
With $\epsilon=10$, mean PG is almost $1$ for LPA($\Delta/\epsilon$) and GSM, i.e., users are well protected against membership inference attacks. Meanwhile, PG slightly drops for FPA and EFPAG ($0.96$ and $0.92$ on average) while their utility is higher.
Unsurprisingly, LPA($1/\epsilon$) achieves negligible privacy gain in this setting.

\begin{figure*}[t]
\centering
	\begin{subfigure}[b]{0.4\textwidth}
	\includegraphics[width=1.0\textwidth]{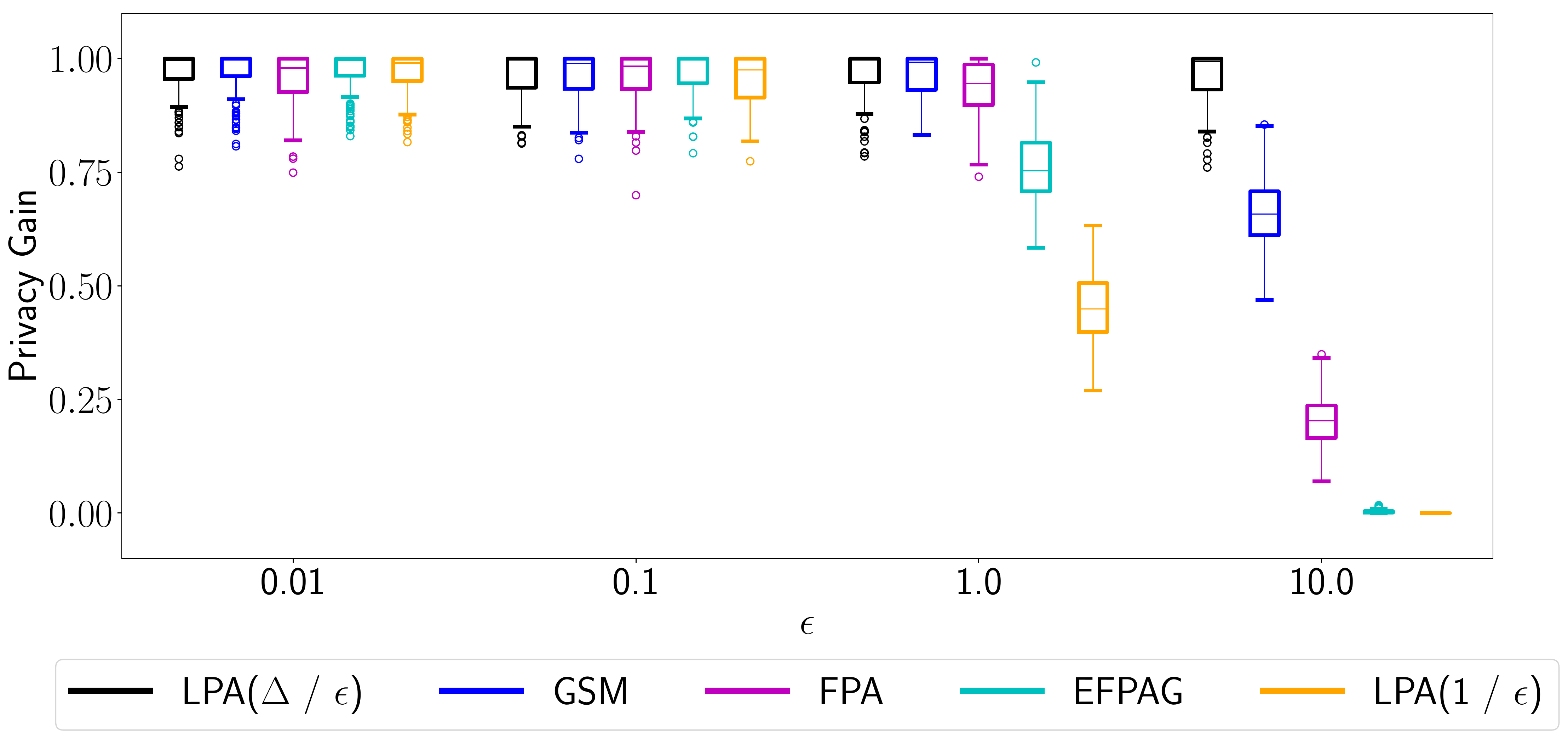}
		\caption{TFL ($m=9,500$, $|T_I|=168$)}
		\label{fig:tfl-dp-tr-ts}
	\end{subfigure}
	~	
	\begin{subfigure}[b]{0.4\textwidth}	
		\includegraphics[width=1.0\textwidth]{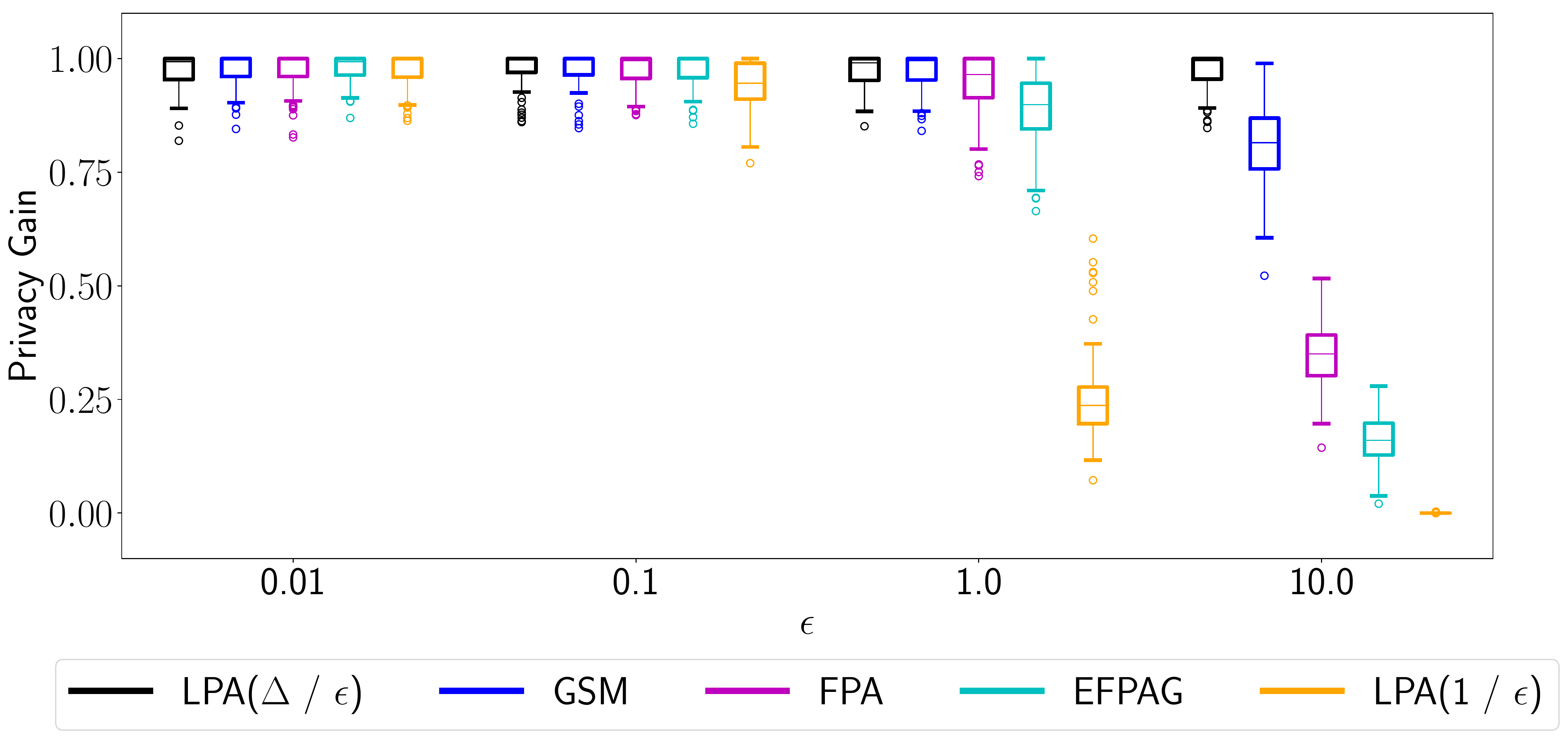}
		\caption{SFC ($m=500$, $|T_I|=168$)}
		\label{fig:sfc-dp-tr-ts}
	\end{subfigure}
\vspace{-0.3cm}	
\caption{Privacy Gain (PG) achieved by differentially private mechanisms with different values of $\epsilon$, against a MLP classifier trained and tested on noisy aggregates.}
\label{fig:dp-tr-ts-pg}
\vspace{-0.4cm}
\end{figure*}

\descrit{Train on Noisy / Test on Noisy Aggregates.} Fig.~\ref{fig:dp-tr-ts-pg} reports the PG results when the MLP classifier is trained on noisy aggregates. Interestingly, the protection of the mechanisms decreases much faster for increasing values of $\epsilon$.
For TFL (Fig.~\ref{fig:tfl-dp-tr-ts}), we observe that for values of $\epsilon \leq 1$, the PG decreases only slightly compared to the previous setting, where training was done on raw aggregates (Fig.~\ref{fig:tfl-pg}). That is, the DP-based mechanisms still provide good protection against membership inference. However, when $\epsilon=10$, we notice a notable decrease in PG, with FPA and EFPAG. More precisely, FPA now achieves 0.2 mean PG (vs 0.45 in the previous setting) and EFPAG provides negligible protection against membership inference (compared to 0.3 in Fig.~\ref{fig:tfl-pg}, $\epsilon=10$).

Similarly, with SFC (Fig.~\ref{fig:sfc-dp-tr-ts}), mean PG remains high for $\epsilon \leq 1$ for all mechanisms, except for LPA($1/\epsilon$). For $\epsilon=10$, there is a significant decline in PG with GSM, FPA, and EFPAG. In particular, GSM now yields 0.8 mean PG, while FPA and EFPAG 0.32 and 0.15, respectively. This corresponds to a significant drop in privacy protection (20\%, 66\% and 83\% for GSM, FPA, and EFPAG) compared to the setting where training was done on raw aggregates (cf.~Fig.~\ref{fig:sfc-pg}).

\begin{table}[t]
\centering
\footnotesize
\begin{tabular}{  l  rrrr  }
\toprule
\multicolumn{1}{r}{\boldmath$\epsilon$} & \textbf{0.01} & \textbf{0.1} & \textbf{1.0} & \textbf{10} \\
\midrule
\textbf{LPA}(\boldmath$\Delta / \epsilon$) & 131.9 & 129.3 & 114.4 & 41.9 \\
\textbf{GSM} & 129.6 & 94.7 & 14.1 & 1.4 \\
\textbf{FPA} & 85.9 & 11.3 & 1.1 & 0.11 \\
\textbf{EFPAG} & 57.9 & 6.1 & 0.6 & 0.04 \\
\textbf{LPA}(1 / \boldmath$\epsilon$) & 24.7 & 2.5 & 0.2 & 0.001 \\ %
\bottomrule
\end{tabular}
\vspace{-0.1cm}
\caption{MRE of aggregate location time-series with different differentially private mechanisms and parameter $\epsilon$ (SFC).}
\label{table:sfc-mre}
\vspace{-0.2cm}
\end{table}

\subsection{DP Evaluation -- Take-Aways} 
The experiments presented in this section evaluate the performance of differentially private mechanisms against membership inference on aggregate location time-series, both in terms of privacy and utility. %
Considering an advantageous setting for utility, but worst-case for privacy, we find that differentially private mechanisms can be effective at preventing inferences, but with some important caveats.
In particular, our results show that a \textit{passive} adversary who trains a classifier on raw aggregate location data is not very successful at inferring membership on noisy aggregates. 
However, when we consider a \textit{strategic} adversary that mimics the behavior of the defender, and trains a classifier on noisy aggregates, we find that the actual privacy gain offered from the DP-based mechanisms is significantly reduced, and also decreases much faster with increasing $\epsilon$ values. This should draw the attention of the research community as advances in deep learning, might lead to stronger attacks against defense mechanisms based on perturbation (e.g., by achieving noise filtering).

Among the defense mechanisms considered, we observe that the straightforward application of LPA and GSM %
yields very poor utility. This is not surprising, as previous work highlights the difficulty of releasing private statistics under continual observation~\cite{chan2011private,dwork2010differential,kellaris2014differentially}.
Mechanisms specifically proposed for time-series settings (i.e., FPA and EFPAG) yield much better utility, at the cost of reduced privacy. This shows that achieving an optimal trade-off between privacy and utility in the settings we consider is still a challenging task.

Finally, our analysis also shows how dataset characteristics affect the performance of differentially private mechanisms too. Specifically, the privacy gain on a sparser dataset (TFL) decreases faster with growing $\epsilon$, compared to a denser one (SFC). This is not surprising, taking into account the scale difference between the sensitivity of the aggregation in each case (recall that $\Delta=\text{207}$ on TFL and 2,685 on SFC).

\vspace{-0.1cm}
\section{Conclusion}
\label{sec:conclusion}
Location privacy has been a prolific research area over the past few years, with a number of attacks and defenses having been proposed on mobility profiles and users' locations.
Although this line of research has improved our understanding about protecting users against the disclosure of sensitive information, to the best of our knowledge, little work has focused on the privacy threats that the availability of aggregate location time-series may pose for individuals contributing to the aggregates.

This paper presented the first evaluation of membership inference in the context of location data. 
We formalize this inference as a distinguishability game in which an adversary has to guess whether or not a target user's location data has been used to compute a given set of aggregates. Instantiating the distinguishing function as a machine learning classifier, we quantify the inference power of an adversary with various types of prior knowledge on two real datasets with different characteristics. We show that, membership inference is very accurate when groups are small, and that users that have regular habits are easier to classify correctly than those performing sporadic movements.

We also evaluate the extent to which defense mechanisms based on differential privacy can prevent membership inference. We find that they are quite effective if the adversary trains the classifier on raw aggregates, though they entail a significant loss in utility. However, they are much less effective if the adversary mimics the behavior of the perturbation mechanism by training her classifier on noisy aggregates.

We remark that our methodology can be used to evaluate membership inference attacks, as well as defenses, in real-world settings.
We also hope that our techniques can be leveraged by providers to test the quality of privacy protection before data release or by regulators aiming to detect violations.

\descr{Acknowledgments.} We wish to thank Mirco Musolesi, Gordon Ross, and Claude Castelluccia for useful feedback and comments on our work. This research is partially supported by a Xerox University Affairs Committee grant on ``Secure Collaborative Analytics'' and the EU H2020 project ``NEXTLEAP'' (Grant Agreement No. 688722).

{\small
\bibliographystyle{abbrv}
\bibliography{bibfile}
}

\appendices
\section{Machine Learning Classifiers}
\label{sec:appendix}

We now briefly review the machine learning classifiers used throughout the paper. %

\descr{Logistic Regression (LR).} LR is a linear model where the probabilities describing the possible outcomes of a single trial are modeled via a logistic (logit) function. The parameters of the model are estimated with maximum likelihood estimation, using an iterative algorithm.

\descr{Nearest Neighbors (k-NN).} k-NN performs classification with a simple majority vote of the nearest neighbors of each data point: a query point is assigned the data class which has the most representatives within the nearest neighbors of the point.

\descr{Random Forest (RF).} RF is an ensemble learning method which constructs a number of decision trees during training and outputs the majority class voted by the individual trees during testing. With RF, each tree in the ensemble is built from a sample drawn with replacement from the training set. When splitting a node during the construction of the tree, the split that is picked is the best split among a random subset of the features. As a result of this randomness, the bias of the forest usually slightly increases but, due to averaging, its variance also decreases.

\descr{Multi-Layer Perceptron (MLP).} MLP is a kind of artificial neural network, consisting of at least three layers of nodes: the input, the hidden and the output layers. Except for the input nodes, each other node is a neuron that uses a non-linear activation function. MLP utilizes a supervised learning technique called back propagation for training and its multiple layers along with the non-linear activation allow it to distinguish data that is not linearly separable.

\end{document}